\documentstyle[11pt,aaspp4]{article}

\newcommand{\kms}{\mbox{ km~s$^{-1}$}}

\begin{document}

\def\PsfigVersion{1.10}
\def\setDriver{\DvipsDriver} 
\ifx\undefined\psfig\else \fi
%

\let\LaTeXAtSign=\@
\let\@=\relax
\edef\psfigRestoreAt{\catcode`\@=\number\catcode`@\relax}
\catcode`\@=11\relax
\newwrite\@unused
\def\ps@typeout#1{{\let\protect\string\immediate\write\@unused{#1}}}

\def\DvipsDriver{
	\ps@typeout{psfig/tex \PsfigVersion -dvips}
\def\PsfigSpecials{\DvipsSpecials} 	\def\ps@dir{/}
\def\ps@predir{} }
\def\OzTeXDriver{
	\ps@typeout{psfig/tex \PsfigVersion -oztex}
	\def\PsfigSpecials{\OzTeXSpecials}
	\def\ps@dir{:}
	\def\ps@predir{:}
	\catcode`\^^J=5
}


\def\figurepath{./:}
\def\psfigurepath#1{\edef\figurepath{#1:}}

\def\DoPaths#1{\expandafter\EachPath#1\stoplist}
\def\leer{}
\def\EachPath#1:#2\stoplist{
  \ExistsFile{#1}{\SearchedFile}
  \ifx#2\leer
  \else
    \expandafter\EachPath#2\stoplist
  \fi}
%
%
\def\ps@dir{/}
\def\ExistsFile#1#2{%
   \openin1=\ps@predir#1\ps@dir#2
   \ifeof1
       \closein1
   \else
       \closein1
        \ifx\ps@founddir\leer
           \edef\ps@founddir{#1}
        \fi
   \fi}
%
%
\def\get@dir#1{%
  \def\ps@founddir{}
  \def\SearchedFile{#1}
  \DoPaths\figurepath
}

%
%
\def\@nnil{\@nil}
\def\@empty{}
\def\@psdonoop#1\@@#2#3{}
\def\@psdo#1:=#2\do#3{\edef\@psdotmp{#2}\ifx\@psdotmp\@empty \else
    \expandafter\@psdoloop#2,\@nil,\@nil\@@#1{#3}\fi}
\def\@psdoloop#1,#2,#3\@@#4#5{\def#4{#1}\ifx #4\@nnil \else
       #5\def#4{#2}\ifx #4\@nnil \else#5\@ipsdoloop #3\@@#4{#5}\fi\fi}
\def\@ipsdoloop#1,#2\@@#3#4{\def#3{#1}\ifx #3\@nnil 
       \let\@nextwhile=\@psdonoop \else
      #4\relax\let\@nextwhile=\@ipsdoloop\fi\@nextwhile#2\@@#3{#4}}
\def\@tpsdo#1:=#2\do#3{\xdef\@psdotmp{#2}\ifx\@psdotmp\@empty \else
    \@tpsdoloop#2\@nil\@nil\@@#1{#3}\fi}
\def\@tpsdoloop#1#2\@@#3#4{\def#3{#1}\ifx #3\@nnil 
       \let\@nextwhile=\@psdonoop \else
      #4\relax\let\@nextwhile=\@tpsdoloop\fi\@nextwhile#2\@@#3{#4}}
%
\ifx\undefined\fbox
\newdimen\fboxrule
\newdimen\fboxsep
\newdimen\ps@tempdima
\newbox\ps@tempboxa
\fboxsep = 3pt
\fboxrule = .4pt
\long\def\fbox#1{\leavevmode\setbox\ps@tempboxa\hbox{#1}\ps@tempdima\fboxrule
    \advance\ps@tempdima \fboxsep \advance\ps@tempdima \dp\ps@tempboxa
   \hbox{\lower \ps@tempdima\hbox
  {\vbox{\hrule height \fboxrule
          \hbox{\vrule width \fboxrule \hskip\fboxsep
          \vbox{\vskip\fboxsep \box\ps@tempboxa\vskip\fboxsep}\hskip 
                 \fboxsep\vrule width \fboxrule}
                 \hrule height \fboxrule}}}}
\fi
%
%
\newread\ps@stream
\newif\ifnot@eof       
\newif\if@noisy        
\newif\if@atend        
\newif\if@psfile       
%
%
{\catcode`\%=12\global\gdef\epsf@start{
\def\epsf@PS{PS}
\def\epsf@getbb#1{%
%
%
\openin\ps@stream=\ps@predir#1
\ifeof\ps@stream\ps@typeout{Error, File #1 not found}\else
%
%
   {\not@eoftrue \chardef\other=12
    \def\do##1{\catcode`##1=\other}\dospecials \catcode`\ =10
    \loop
       \if@psfile
	  \read\ps@stream to \epsf@fileline
       \else{
	  \obeyspaces
          \read\ps@stream to \epsf@tmp\global\let\epsf@fileline\epsf@tmp}
       \fi
       \ifeof\ps@stream\not@eoffalse\else
%
%
       \if@psfile\else
       \expandafter\epsf@test\epsf@fileline:. \\%
       \fi
%
%
          \expandafter\epsf@aux\epsf@fileline:. \\%
       \fi
   \ifnot@eof\repeat
   }\closein\ps@stream\fi}%
%
%
\long\def\epsf@test#1#2#3:#4\\{\def\epsf@testit{#1#2}
			\ifx\epsf@testit\epsf@start\else
\ps@typeout{Warning! File does not start with `\epsf@start'.  It may not be a PostScript file.}
			\fi
			\@psfiletrue} 
%
%
{\catcode`\%=12\global\let\epsf@percent=
%
%
%
\long\def\epsf@aux#1#2:#3\\{\ifx#1\epsf@percent
   \def\epsf@testit{#2}\ifx\epsf@testit\epsf@bblit
	\@atendfalse
        \epsf@atend #3 . \\%
	\if@atend	
	   \if@verbose{
		\ps@typeout{psfig: found `(atend)'; continuing search}
	   }\fi
        \else
        \epsf@grab #3 . . . \\%
        \not@eoffalse
        \global\no@bbfalse
        \fi
   \fi\fi}%
%
%
\def\epsf@grab #1 #2 #3 #4 #5\\{%
   \global\def\epsf@llx{#1}\ifx\epsf@llx\empty
      \epsf@grab #2 #3 #4 #5 .\\\else
   \global\def\epsf@lly{#2}%
   \global\def\epsf@urx{#3}\global\def\epsf@ury{#4}\fi}%
%
%
\def\epsf@atendlit{(atend)} 
\def\epsf@atend #1 #2 #3\\{%
   \def\epsf@tmp{#1}\ifx\epsf@tmp\empty
      \epsf@atend #2 #3 .\\\else
   \ifx\epsf@tmp\epsf@atendlit\@atendtrue\fi\fi}


\chardef\psletter = 11 
\chardef\other = 12

\newif \ifdebug 
\newif\ifc@mpute 
\c@mputetrue 

\let\then = \relax
\def\r@dian{pt }
\let\r@dians = \r@dian
\let\dimensionless@nit = \r@dian
\let\dimensionless@nits = \dimensionless@nit
\def\internal@nit{sp }
\let\internal@nits = \internal@nit
\newif\ifstillc@nverging
\def \Mess@ge #1{\ifdebug \then \message {#1} \fi}

{ 
	\catcode `\@ = \psletter
	\gdef \nodimen {\expandafter \n@dimen \the \dimen}
	\gdef \term #1 #2 #3%
	       {\edef \t@ {\the #1}
		\edef \t@@ {\expandafter \n@dimen \the #2\r@dian}%
		\t@rm {\t@} {\t@@} {#3}%
	       }
	\gdef \t@rm #1 #2 #3%
	       {{%
		\count 0 = 0
		\dimen 0 = 1 \dimensionless@nit
		\dimen 2 = #2\relax
		\Mess@ge {Calculating term #1 of \nodimen 2}%
		\loop
		\ifnum	\count 0 < #1
		\then	\advance \count 0 by 1
			\Mess@ge {Iteration \the \count 0 \space}%
			\Multiply \dimen 0 by {\dimen 2}%
			\Mess@ge {After multiplication, term = \nodimen 0}%
			\Divide \dimen 0 by {\count 0}%
			\Mess@ge {After division, term = \nodimen 0}%
		\repeat
		\Mess@ge {Final value for term #1 of 
				\nodimen 2 \space is \nodimen 0}%
		\xdef \Term {#3 = \nodimen 0 \r@dians}%
		\aftergroup \Term
	       }}
	\catcode `\p = \other
	\catcode `\t = \other
	\gdef \n@dimen #1pt{#1} 
}

\def \Divide #1by #2{\divide #1 by #2} 

\def \Multiply #1by #2
       {{
	\count 0 = #1\relax
	\count 2 = #2\relax
	\count 4 = 65536
	\Mess@ge {Before scaling, count 0 = \the \count 0 \space and
			count 2 = \the \count 2}%
	\ifnum	\count 0 > 32767 
	\then	\divide \count 0 by 4
		\divide \count 4 by 4
	\else	\ifnum	\count 0 < -32767
		\then	\divide \count 0 by 4
			\divide \count 4 by 4
		\else
		\fi
	\fi
	\ifnum	\count 2 > 32767 
	\then	\divide \count 2 by 4
		\divide \count 4 by 4
	\else	\ifnum	\count 2 < -32767
		\then	\divide \count 2 by 4
			\divide \count 4 by 4
		\else
		\fi
	\fi
	\multiply \count 0 by \count 2
	\divide \count 0 by \count 4
	\xdef \product {#1 = \the \count 0 \internal@nits}%
	\aftergroup \product
       }}

\def\r@duce{\ifdim\dimen0 > 90\r@dian \then   
		\multiply\dimen0 by -1
		\advance\dimen0 by 180\r@dian
		\r@duce
	    \else \ifdim\dimen0 < -90\r@dian \then  
		\advance\dimen0 by 360\r@dian
		\r@duce
		\fi
	    \fi}

\def\Sine#1%
       {{%
	\dimen 0 = #1 \r@dian
	\r@duce
	\ifdim\dimen0 = -90\r@dian \then
	   \dimen4 = -1\r@dian
	   \c@mputefalse
	\fi
	\ifdim\dimen0 = 90\r@dian \then
	   \dimen4 = 1\r@dian
	   \c@mputefalse
	\fi
	\ifdim\dimen0 = 0\r@dian \then
	   \dimen4 = 0\r@dian
	   \c@mputefalse
	\fi
	\ifc@mpute \then
		\divide\dimen0 by 180
		\dimen0=3.141592654\dimen0
		\dimen 2 = 3.1415926535897963\r@dian 
		\divide\dimen 2 by 2 
		\Mess@ge {Sin: calculating Sin of \nodimen 0}%
		\count 0 = 1 
		\dimen 2 = 1 \r@dian 
		\dimen 4 = 0 \r@dian 
		\loop
			\ifnum	\dimen 2 = 0 
			\then	\stillc@nvergingfalse 
			\else	\stillc@nvergingtrue
			\fi
			\ifstillc@nverging 
			\then	\term {\count 0} {\dimen 0} {\dimen 2}%
				\advance \count 0 by 2
				\count 2 = \count 0
				\divide \count 2 by 2
				\ifodd	\count 2 
				\then	\advance \dimen 4 by \dimen 2
				\else	\advance \dimen 4 by -\dimen 2
				\fi
		\repeat
	\fi		
			\xdef \sine {\nodimen 4}%
       }}

\def\Cosine#1{\ifx\sine\UnDefined\edef\Savesine{\relax}\else
		             \edef\Savesine{\sine}\fi
	{\dimen0=#1\r@dian\advance\dimen0 by 90\r@dian
	 \Sine{\nodimen 0}
	 \xdef\cosine{\sine}
	 \xdef\sine{\Savesine}}}	      

\def\psdraft{
	\def\@psdraft{0}
}
\def\psfull{
	\def\@psdraft{100}
}

\psfull

\newif\if@scalefirst
\def\psscalefirst{\@scalefirsttrue}
\def\psrotatefirst{\@scalefirstfalse}
\psrotatefirst

\newif\if@draftbox
\def\psnodraftbox{
	\@draftboxfalse
}
\def\psdraftbox{
	\@draftboxtrue
}
\@draftboxtrue

\newif\if@prologfile
\newif\if@postlogfile
\def\pssilent{
	\@noisyfalse
}
\def\psnoisy{
	\@noisytrue
}
\psnoisy
\newif\if@bbllx
\newif\if@bblly
\newif\if@bburx
\newif\if@bbury
\newif\if@height
\newif\if@width
\newif\if@rheight
\newif\if@rwidth
\newif\if@angle
\newif\if@clip
\newif\if@verbose
\def\@p@@sclip#1{\@cliptrue}
\newif\if@decmpr
\def\@p@@sfigure#1{\def\@p@sfile{null}\def\@p@sbbfile{null}\@decmprfalse
   \openin1=\ps@predir#1
   \ifeof1
	\closein1
	\get@dir{#1}
	\ifx\ps@founddir\leer
		\openin1=\ps@predir#1.bb
		\ifeof1
			\closein1
			\get@dir{#1.bb}
			\ifx\ps@founddir\leer
				\ps@typeout{Can't find #1 in \figurepath}
			\else
				\@decmprtrue
				\def\@p@sfile{\ps@founddir\ps@dir#1}
				\def\@p@sbbfile{\ps@founddir\ps@dir#1.bb}
			\fi
		\else
			\closein1
			\@decmprtrue
			\def\@p@sfile{#1}
			\def\@p@sbbfile{#1.bb}
		\fi
	\else
		\def\@p@sfile{\ps@founddir\ps@dir#1}
		\def\@p@sbbfile{\ps@founddir\ps@dir#1}
	\fi
   \else
	\closein1
	\def\@p@sfile{#1}
	\def\@p@sbbfile{#1}
   \fi
}
\def\@p@@sfile#1{\@p@@sfigure{#1}}
\def\@p@@sbbllx#1{
		\@bbllxtrue
		\dimen100=#1
		\edef\@p@sbbllx{\number\dimen100}
}
\def\@p@@sbblly#1{
		\@bbllytrue
		\dimen100=#1
		\edef\@p@sbblly{\number\dimen100}
}
\def\@p@@sbburx#1{
		\@bburxtrue
		\dimen100=#1
		\edef\@p@sbburx{\number\dimen100}
}
\def\@p@@sbbury#1{
		\@bburytrue
		\dimen100=#1
		\edef\@p@sbbury{\number\dimen100}
}
\def\@p@@sheight#1{
		\@heighttrue
		\dimen100=#1
   		\edef\@p@sheight{\number\dimen100}
}
\def\@p@@swidth#1{
		\@widthtrue
		\dimen100=#1
		\edef\@p@swidth{\number\dimen100}
}
\def\@p@@srheight#1{
		\@rheighttrue
		\dimen100=#1
		\edef\@p@srheight{\number\dimen100}
}
\def\@p@@srwidth#1{
		\@rwidthtrue
		\dimen100=#1
		\edef\@p@srwidth{\number\dimen100}
}
\def\@p@@sangle#1{
		\@angletrue
		\edef\@p@sangle{#1} 
}
\def\@p@@ssilent#1{ 
		\@verbosefalse
}
\def\@p@@sprolog#1{\@prologfiletrue\def\@prologfileval{#1}}
\def\@p@@spostlog#1{\@postlogfiletrue\def\@postlogfileval{#1}}
\def\@cs@name#1{\csname #1\endcsname}
\def\@setparms#1=#2,{\@cs@name{@p@@s#1}{#2}}
%
%
\def\ps@init@parms{
		\@bbllxfalse \@bbllyfalse
		\@bburxfalse \@bburyfalse
		\@heightfalse \@widthfalse
		\@rheightfalse \@rwidthfalse
		\def\@p@sbbllx{}\def\@p@sbblly{}
		\def\@p@sbburx{}\def\@p@sbbury{}
		\def\@p@sheight{}\def\@p@swidth{}
		\def\@p@srheight{}\def\@p@srwidth{}
		\def\@p@sangle{0}
		\def\@p@sfile{} \def\@p@sbbfile{}
		\def\@p@scost{10}
		\def\@sc{}
		\@prologfilefalse
		\@postlogfilefalse
		\@clipfalse
		\if@noisy
			\@verbosetrue
		\else
			\@verbosefalse
		\fi
}
%
%
\def\parse@ps@parms#1{
	 	\@psdo\@psfiga:=#1\do
		   {\expandafter\@setparms\@psfiga,}}
%
%
\newif\ifno@bb
\def\bb@missing{
	\if@verbose{
		\ps@typeout{psfig: searching \@p@sbbfile \space  for bounding box}
	}\fi
	\no@bbtrue
	\epsf@getbb{\@p@sbbfile}
        \ifno@bb \else \bb@cull\epsf@llx\epsf@lly\epsf@urx\epsf@ury\fi
}	
\def\bb@cull#1#2#3#4{
	\dimen100=#1 bp\edef\@p@sbbllx{\number\dimen100}
	\dimen100=#2 bp\edef\@p@sbblly{\number\dimen100}
	\dimen100=#3 bp\edef\@p@sbburx{\number\dimen100}
	\dimen100=#4 bp\edef\@p@sbbury{\number\dimen100}
	\no@bbfalse
}
\newdimen\p@intvaluex
\newdimen\p@intvaluey
\def\rotate@#1#2{{\dimen0=#1 sp\dimen1=#2 sp
		  \global\p@intvaluex=\cosine\dimen0
		  \dimen3=\sine\dimen1
		  \global\advance\p@intvaluex by -\dimen3
		  \global\p@intvaluey=\sine\dimen0
		  \dimen3=\cosine\dimen1
		  \global\advance\p@intvaluey by \dimen3
		  }}
\def\compute@bb{
		\no@bbfalse
		\if@bbllx \else \no@bbtrue \fi
		\if@bblly \else \no@bbtrue \fi
		\if@bburx \else \no@bbtrue \fi
		\if@bbury \else \no@bbtrue \fi
		\ifno@bb \bb@missing \fi
		\ifno@bb \ps@typeout{FATAL ERROR: no bb supplied or found}
			\no-bb-error
		\fi
		%
%
		\count203=\@p@sbburx
		\count204=\@p@sbbury
		\advance\count203 by -\@p@sbbllx
		\advance\count204 by -\@p@sbblly
		\edef\ps@bbw{\number\count203}
		\edef\ps@bbh{\number\count204}
		\if@angle 
			\Sine{\@p@sangle}\Cosine{\@p@sangle}
	        	{\dimen100=\maxdimen\xdef\r@p@sbbllx{\number\dimen100}
					    \xdef\r@p@sbblly{\number\dimen100}
			                    \xdef\r@p@sbburx{-\number\dimen100}
					    \xdef\r@p@sbbury{-\number\dimen100}}
%
                        \def\minmaxtest{
			   \ifnum\number\p@intvaluex<\r@p@sbbllx
			      \xdef\r@p@sbbllx{\number\p@intvaluex}\fi
			   \ifnum\number\p@intvaluex>\r@p@sbburx
			      \xdef\r@p@sbburx{\number\p@intvaluex}\fi
			   \ifnum\number\p@intvaluey<\r@p@sbblly
			      \xdef\r@p@sbblly{\number\p@intvaluey}\fi
			   \ifnum\number\p@intvaluey>\r@p@sbbury
			      \xdef\r@p@sbbury{\number\p@intvaluey}\fi
			   }
			\rotate@{\@p@sbbllx}{\@p@sbblly}
			\minmaxtest
			\rotate@{\@p@sbbllx}{\@p@sbbury}
			\minmaxtest
			\rotate@{\@p@sbburx}{\@p@sbblly}
			\minmaxtest
			\rotate@{\@p@sbburx}{\@p@sbbury}
			\minmaxtest
			\edef\@p@sbbllx{\r@p@sbbllx}\edef\@p@sbblly{\r@p@sbblly}
			\edef\@p@sbburx{\r@p@sbburx}\edef\@p@sbbury{\r@p@sbbury}
		\fi
		\count203=\@p@sbburx
		\count204=\@p@sbbury
		\advance\count203 by -\@p@sbbllx
		\advance\count204 by -\@p@sbblly
		\edef\@bbw{\number\count203}
		\edef\@bbh{\number\count204}
}
%
%
\def\in@hundreds#1#2#3{\count240=#2 \count241=#3
		     \count100=\count240	
		     \divide\count100 by \count241
		     \count101=\count100
		     \multiply\count101 by \count241
		     \advance\count240 by -\count101
		     \multiply\count240 by 10
		     \count101=\count240	
		     \divide\count101 by \count241
		     \count102=\count101
		     \multiply\count102 by \count241
		     \advance\count240 by -\count102
		     \multiply\count240 by 10
		     \count102=\count240	
		     \divide\count102 by \count241
		     \count200=#1\count205=0
		     \count201=\count200
			\multiply\count201 by \count100
		 	\advance\count205 by \count201
		     \count201=\count200
			\divide\count201 by 10
			\multiply\count201 by \count101
			\advance\count205 by \count201
		     \count201=\count200
			\divide\count201 by 100
			\multiply\count201 by \count102
			\advance\count205 by \count201
		     \edef\@result{\number\count205}
}
\def\compute@wfromh{
		\in@hundreds{\@p@sheight}{\@bbw}{\@bbh}
		\edef\@p@swidth{\@result}
}
\def\compute@hfromw{
	        \in@hundreds{\@p@swidth}{\@bbh}{\@bbw}
		\edef\@p@sheight{\@result}
}
\def\compute@handw{
		\if@height 
			\if@width
			\else
				\compute@wfromh
			\fi
		\else 
			\if@width
				\compute@hfromw
			\else
				\edef\@p@sheight{\@bbh}
				\edef\@p@swidth{\@bbw}
			\fi
		\fi
}
\def\compute@resv{
		\if@rheight \else \edef\@p@srheight{\@p@sheight} \fi
		\if@rwidth \else \edef\@p@srwidth{\@p@swidth} \fi
}
%
\def\compute@sizes{
	\compute@bb
	\if@scalefirst\if@angle
	\if@width
	   \in@hundreds{\@p@swidth}{\@bbw}{\ps@bbw}
	   \edef\@p@swidth{\@result}
	\fi
	\if@height
	   \in@hundreds{\@p@sheight}{\@bbh}{\ps@bbh}
	   \edef\@p@sheight{\@result}
	\fi
	\fi\fi
	\compute@handw
	\compute@resv}
\def\OzTeXSpecials{
	\special{empty.ps /@isp {true} def}
	\special{empty.ps \@p@swidth \space \@p@sheight \space
			\@p@sbbllx \space \@p@sbblly \space
			\@p@sbburx \space \@p@sbbury \space
			startTexFig \space }
	\if@clip{
		\if@verbose{
			\ps@typeout{(clip)}
		}\fi
		\special{empty.ps doclip \space }
	}\fi
	\if@angle{
		\if@verbose{
			\ps@typeout{(rotate)}
		}\fi
		\special {empty.ps \@p@sangle \space rotate \space} 
	}\fi
	\if@prologfile
	    \special{\@prologfileval \space } \fi
	\if@decmpr{
		\if@verbose{
			\ps@typeout{psfig: Compression not available
			in OzTeX version \space }
		}\fi
	}\else{
		\if@verbose{
			\ps@typeout{psfig: including \@p@sfile \space }
		}\fi
		\special{epsf=\ps@predir\@p@sfile \space }
	}\fi
	\if@postlogfile
	    \special{\@postlogfileval \space } \fi
	\special{empty.ps /@isp {false} def}
}
\def\DvipsSpecials{
	\special{ps::[begin] 	\@p@swidth \space \@p@sheight \space
			\@p@sbbllx \space \@p@sbblly \space
			\@p@sbburx \space \@p@sbbury \space
			startTexFig \space }
	\if@clip{
		\if@verbose{
			\ps@typeout{(clip)}
		}\fi
		\special{ps:: doclip \space }
	}\fi
	\if@angle
		\if@verbose{
			\ps@typeout{(clip)}
		}\fi
		\special {ps:: \@p@sangle \space rotate \space} 
	\fi
	\if@prologfile
	    \special{ps: plotfile \@prologfileval \space } \fi
	\if@decmpr{
		\if@verbose{
			\ps@typeout{psfig: including \@p@sfile.Z \space }
		}\fi
		\special{ps: plotfile "`zcat \@p@sfile.Z" \space }
	}\else{
		\if@verbose{
			\ps@typeout{psfig: including \@p@sfile \space }
		}\fi
		\special{ps: plotfile \@p@sfile \space }
	}\fi
	\if@postlogfile
	    \special{ps: plotfile \@postlogfileval \space } \fi
	\special{ps::[end] endTexFig \space }
}
%
%
\def\psfig#1{\vbox {
	%
	\ps@init@parms
	\parse@ps@parms{#1}
	\compute@sizes
	\ifnum\@p@scost<\@psdraft{
		\PsfigSpecials 
		\vbox to \@p@srheight sp{
			\hbox to \@p@srwidth sp{
				\hss
			}
		\vss
		}
	}\else{
		\if@draftbox{		
			\hbox{\fbox{\vbox to \@p@srheight sp{
			\vss
			\hbox to \@p@srwidth sp{ \hss 
			 \hss }
			\vss
			}}}
		}\else{
			\vbox to \@p@srheight sp{
			\vss
			\hbox to \@p@srwidth sp{\hss}
			\vss
			}
		}\fi

	}\fi
}}
\psfigRestoreAt
\setDriver
\let\@=\LaTeXAtSign

\title{HST Observations of 10 Two-Image
       Gravitational Lenses\footnote{
             Based on Observations made with the NASA/ESA
             Hubble Space Telescope, obtained at the Space Telescope
             Science Institute, which is operated by AURA, Inc., 
             under NASA contract NAS 5-26555.} }

\author{ J. Leh\'ar\altaffilmark{2},
	 E. E. Falco\altaffilmark{2}, 
	 C. S. Kochanek\altaffilmark{2}, 
         B. A. McLeod\altaffilmark{2}, 
	 J. A. Mu\~noz\altaffilmark{2},
	 C. D. Impey\altaffilmark{3}, 
	 H.-W. Rix\altaffilmark{4},
         C. R. Keeton\altaffilmark{3},
	 and C. Y. Peng\altaffilmark{3}}

\altaffiltext{2}{Harvard-Smithsonian Center for Astrophysics, 
	         60 Garden St., Cambridge, MA 02138}
\altaffiltext{3}{Steward Observatory, University of Arizona, Tucson, AZ 85721}
\altaffiltext{4}{Max-Planck-Institut f\"{u}r Astronomie, Keonigstuhl 17, 
                                Heidelberg, D-69117, Germany}

\begin{abstract}
We report on a program to obtain HST observations of galaxy-mass
gravitational lens systems at optical and infrared wavelengths.  Here
we discuss the properties of 10 two-image gravitational lens systems
(Q~0142--100=UM~673, B~0218+357, SBS~0909+532, BRI~0952--0115,
LBQS~1009--0252, Q~1017--207=J03.13, B~1030+074, HE~1104--1805,
Q~1208+1011, and PKS~1830--211). We grouped these 10 systems because
they have limited lens model constraints and often show poor contrast
between the images and the lens galaxy.  Of the 10 lens galaxies, 7
are probably early-type galaxies, 2 are probably late-type galaxies
(B~0218+357 and PKS~1830--211), and one was not detected
(Q~1208+1011). We detect the host galaxies of the $z_s=4.50$ lensed
quasar in BRI~0952--0115, the $z_s=2.32$ lensed quasar in
HE~1104--1805, and the unlensed $z=1.63$ quasar near LBQS~1009--0252.
We fit a set of four standard lens models to each lens that had
sufficient constraints to compare isothermal dark matter and constant
mass-to-light lens models, and to explore the effects of local tidal
shears.
\end{abstract}


\section{Introduction}

Gravitational lenses have become extraordinarily useful tools for
cosmology, galactic structure and galactic evolution (for example, see
Schneider, Ehlers \& Falco 1992; Blandford \&~Narayan~1992).  With
approximately 60 known lenses produced by galaxies 
(see~Keeton \& Kochanek 1996 for a recent but now 
quite incomplete summary\footnote{
    See http://cfa-www.harvard.edu/castles for a more complete summary.}), 
small number statistics are
rapidly becoming a problem of the past.  Preliminary studies have
shown that lens systems have mass-to-light ratios ($M/L$) that are
50\%--100\% higher than found in constant $M/L$ stellar dynamical
models locally, consistent with the expectation for early-type
galaxies embedded in massive dark halos (Kochanek 1996; Keeton,
Kochanek \& Falco 1998, hereafter KKF; Jackson et~al.\ 1998).  The mass
and light distributions are generally aligned, but have different
quadrupole moments (KKF).  Where simple ellipsoidal models suggest a
mass distribution that is misaligned with the light, the cause seems
to be external tidal perturbations from nearby objects.  In fact,
Keeton, Kochanek \& Seljak (1997) found that all four-image lenses
required two axes for the angular structure of the potential
to obtain a good fit, where the two axes are presumably the major axis
of the primary lens and the axis of the local tidal gravity field. The lens
galaxies generally have the colors and shapes of early-type galaxies
rather than spiral galaxies (KKF), consistent with theoretical
expectations for their relative lensing cross-sections 
(e.g.\ Turner, Ostriker \& Gott 1984, Fukugita \& Turner 1991, 
Maoz \& Rix 1993, Kochanek 1994, 1996).  KKF also
measured the evolution in the mass-to-light ratios of the lens
galaxies with redshift, and found results consistent with passive
evolution models (e.g.\ Leitherer et~al.\ 1996) and fundamental
plane measurements in rich, high-redshift clusters (e.g.\ van Dokkum
et~al.\ 1998).  These early results are based on limited observational
samples.

The CASTLES project ({\bf C}fA/{\bf A}rizona--{\bf S}pace--{\bf
T}elescope--{\bf LE}ns--{\bf S}urvey) is a survey of known
galaxy-mass lens systems using the Hubble Space Telescope (HST) 
and a uniform set of filters.  Since the characteristic size of 
galaxy-scale lenses is $\Delta\theta\sim1\arcsec$, precision 
photometric studies of the lensing galaxies are only practical with HST. 
Before CASTLES, HST images existed for about one-half of the 
$\sim 30$ known
galaxy-mass lens systems, but the diverse range of filters and
exposure times that were used severely limited the usefulness of these
observations (see~KKF).  The goals of CASTLES are: (1)~accurate
astrometric measurements to refine lens models, particularly for
systems where a time delay may provide a direct measurement of $H_0$;
(2)~photometric redshift estimates for all the lens galaxies in the
sample, to sharpen lensing constraints on the cosmological constant
$\Lambda$ and galaxy evolution; (3)~direct estimates of the $M/L$
ratio of lens galaxies up to $z$\,$\sim$\,1, to derive new
constraints on their mean formation epoch and star formation history;
(4)~a comparison of the dark matter and stellar light distributions in
the lens galaxies, and thus new constraints on the structure and shapes 
of galaxies; (5)~measurements of the properties of the interstellar
medium in distant galaxies, using differential extinction between the
lensed images; (6)~identification of as yet
undetected lens galaxies in known multiple-image systems, which would
confirm lensing and enlarge the sample for cosmological studies;
(7)~understanding of the environments of lens galaxies and their
role in the lensing phenomenon; and (8)~estimates of the photometric
properties of the host galaxies of lensed quasars.

The survey consists of all $\sim60$ small-separation ($\Delta\theta <10\arcsec$)
gravitational lenses.  The lenses were found as a produce of optical quasar
surveys, radio lens surveys and serendipity (see Kochanek 1993).  In all
cases, there is a dominant lens galaxy which may be a member of a group
or small cluster.  The heterogeneity of the overall sample is important for
some questions (e.g. the separation distribution), but relatively unimportant
for others (e.g. the evolution of the lens galaxies).  
We observed our targets in the near
infrared through the F160W (H-band) filter with the NICMOS camera
NIC2.  In cases where the lens was expected to be very bright, we
split the exposures to also include an F205W (K-band) exposure.  The
infrared observations are complemented by new or archival WFPC2
imaging in the F814W (I-band), F675W (R-band) or F555W (V-band) to
obtain uniform VIH or RIH multi-color photometry of the systems.
When available, we have also analyzed archival NIC1 images.  

In this paper, we restrict our attention to 10 two-image lens systems
(Table~1).  Detailed investigations of two-image lensed systems 
are limited by the small number of observational constraints. 
In many cases the contrast is too high, or the image separation 
is too small, to permit precise photometric modeling of the lens
(B~0218+357, SBS~0909+532,
LBQS~1009--0252, Q~1017--207, Q~1208+1011 and PKS~1830--211).  
The CASTLES observations of four-image lenses will be presented 
in a later paper (McLeod et~al.\ 2000), and some individual targets
with more constraints have been published
(e.g.\ PG~1115+080, Impey et~al.\ 1998; MG~1131+0456 Kochanek et~al. 2000a).
Surveys of the physical properties of the lenses are considered elsewhere,
such as extinction (Falco et~al.\ 2000), 
the fundamental plane (Kochanek et~al.\ 2000b), 
and the properties of lensed quasar host galaxies (Rix et~al.\ 1999, 2000).

We start by outlining our general procedures for carrying out, reducing and
interpreting the observations.  In \S2 we describe the data reduction,
in \S3 the decomposition of the images into a photometric
model, and in \S4 we describe the interpretation of the data using 
gravitational lens models.  In \S5 we discuss the properties of the 
10 systems analyzed in this paper, and in \S6 we summarize the results.
An Appendix presents the data reduction procedures we have developed
for NICMOS data.

\section{Observations and Reduction}

Table~1 lists the available NICMOS and WFPC2 observations of the 
10 lenses we consider.
We acquired near-infrared imaging with the NICMOS/NIC2 camera,
observing the targets for one-half, one or two orbits using the
H-band (F160W) filter.  Whenever one-half of an orbit was sufficient,
the remainder was used to obtain a K-band (F205W) image.  For each 
target we obtained four dithered exposures to minimize the effects 
of detector defects and edges, and cosmic ray persistence following 
South Atlantic Anomaly (SAA) passages.    
Initially, we used sub-arcsecond dithers, but experience 
with the first data sets showed that larger dithers ($\sim4\arcsec$) 
allow for better background subtraction.  
The early HST pipeline image reductions were inadequate for 
performing accurate image modeling and photometry;
therefore, we developed our own custom package ``NICRED'',
based on C-programs and IRAF\footnote{
    IRAF (Image Reduction and Analysis Facility) is distributed by the
    National Optical Astronomy Observatories, which are operated by the
    Association of Universities for Research in Astronomy, Inc., under
    contract with the National Science Foundation.}
routines (McLeod 1997).
As it is applicable to many other NICMOS data sets,
we describe our reduction in the Appendix. 
The archival NICMOS/NIC1 images 
were reduced using the same methods, but with significantly poorer
results because they were not dithered and there were not enough
data to construct sky darks.

We reduced new or archival WFPC2 images of the systems in a manner 
similar to the last steps of NICRED.  We ignored archival WFPC1 images of
the lenses because the extended point spread function (PSF) skirts of the
bright quasars generally obscured the properties of the lens galaxy. 
We first produced shifted subimages for each exposure,
as in NICRED, using a 3-sigma ``ccdclip'' rejection algorithm
(or 3-sigma ``crreject'' in cases where
there were fewer than 4 exposures on a target).  
This rejection removed nearly all the cosmic rays from the combined
images.  The combined images were then averaged together with 
exposure-time weighting.  Many hot pixels and chip defects
remain in the final images created from undithered archival 
observations.  

We use Vega-referenced magnitudes for the HST filters used in the 
observations.  The NICMOS calibration was determined by fitting PSF 
models to observations of the standard star P330E, and then referencing to
ground-based photometry (Persson et~al.\ 1998).  The transformations
from ground-based filters to the NICMOS filters were done
synthetically.  The resulting zero-point magnitudes (for 1~count/s
in an infinite aperture) 
are 21.88, 21.80, and 22.47, respectively for
the F205W, F160W, and F110W filters.  The WFPC2 calibrations are from
Holtzman et~al.\ (1995) after changing to a gain of 7 and correcting
from a finite aperture to an infinite aperture ($\sim0.1$\,mag),
as needed for our
fitting procedures.  The zero-point magnitudes we applied are: 21.69
for F814W; 22.08 for F675W; and 22.57 for F555W.  We refer to the
F555W, F675W, F814W, F110W, F160W, and F205W bands as V, R, I, J, H
and K respectively.  We estimate the uncertainty in the zero points to be
$\sim0.05$ mag.

We used the SExtractor package (Bertin \& Arnouts 1996) to catalog the
objects found in the NICMOS and WFPC2 fields. We present the
catalogs for all our targets in Table~2. The objects were
morphologically classified as galaxies or stellar by SExtractor,
followed by visual confirmation.  In ground-based images, SExtractor
classification accuracy depends on seeing and image depth, 
but is typically $\sim90\%$ accurate for objects more than 1
magnitude brighter than the detection limit.  The total object
magnitudes were estimated using Kron-type 
(Bertin \& Arnouts 1996) automatic apertures. 
We tested the galaxy photometry by adding $\sim100$ synthetic galaxies
to WFPC2 and NICMOS images, and measuring them using our procedure.
SExtractor accurately determined the magnitudes for objects
with Gaussian profiles, but systematically underestimated the flux 
for de~Vaucouleurs and exponential profile galaxies. 
The effect varied with the size of the galaxy and the PSF,
and for a typical de~Vaucouleurs $R_e$ of $0\farcs3$,
the shift was $\approx0.1\pm0.2$ and $\approx0.5\pm0.2$ magnitudes
for the WFPC2 and NICMOS chips, respectively.
The quoted photometric uncertainties are a quadrature sum of the 
scatter in this systematic shift
and the internal SExtractor uncertainties.  Galaxy colors
were determined within fixed circular apertures ($0\farcs3$,
$0\farcs56$, $1\farcs0$, or $1\farcs7$), with results given for the
smallest aperture whose diameter exceeds $\sqrt{2}$ times the mean
object major diameter in the deepest optical exposure. We 
used synthetic PSF models (see~\S3) to determine that PSF color terms
are $\lesssim 0.02$ mag, sufficiently small to be ignored for these
galaxy colors.  We estimated our detection limits by randomly adding 
$0\farcs3$ FWHM Gaussians to each field and determining the magnitude
at which 50\% are detected by SExtractor.  
We first cataloged all objects within
$20\arcsec$ of the lens based on the deepest optical image (galaxies
and stars labeled G\# and S\#, respectively, in order of decreasing
brightness).  We then added any galaxies detected only in the H-band
images (labeled H\#) and more distant galaxies which could tidally
influence the lens system (labeled T\#, see~\S4).

\section{Image Decomposition}

The angular extent of a lens galaxy is usually comparable to
the angular separation of the images that it produces, so the
data consists of overlapping images of objects at different
redshifts.  This problem is compounded by the fact that the 
angular separations of components are also comparable to the 
size of the HST point spread function. A successful
analysis therefore requires a rigorous 
simultaneous image decomposition into source and lens 
components.

Both the WFPC2 and NICMOS point-spread functions (PSF) consist of a
sharp central core (FWHM $\sim$0\farcs1), the first Airy ring, and an
outer skirt of low intensity ($\leq$ 1\% of the peak intensity)
with a complex spatial variation in azimuth.  For models of
the WFPC2 and NIC1 data we used TinyTim v4.4 (Krist \& Hook 1997) model
PSFs taking into account the location of the object and, for the NIC1
data, the position of the Pupil Alignment Mechanism.  For models of the
NIC2 data we used a set of 13 stellar images
(McLeod, Rieke \& Storrie-Lombardi 1999). 
The empirical PSF images were
reduced exactly like the images of the lenses.  When we fitted the 
PSFs to isolated stellar sources, the peak residuals were less 
than 2\% of the peak intensity of the source and the 
RMS residuals
integrated over the PSF were less than 1\% of the total flux. 
The systematic residuals are due to the time variability of the PSF
as NICMOS lost coolant, as well as to 
the limited temporal and spatial sampling of the available PSFs.
We fitted each lensed system using all the empirical PSFs in our library,
and selected as a reference that model whose PSF yielded the best fit
to the observed data. 
Which PSF provided the best fit did not correlate with observation time
or position on the sky. 

We used parameterized models to separate lens and source components:
in most cases, the source images were assumed to be unresolved and the lens
galaxies were described by ellipsoidal exponential disk 
or de~Vaucouleurs models.  Thus a typical model for a two-image
lens system is specified by 12 parameters: the position and flux of
each point source, and the position, flux, major axis scale length,
axis ratio, and position angle (PA) of the lens galaxy.  Where necessary,
we included additional parameters to model the background level and an
empirical Gaussian blur to improve the PSF match.  The model, convolved
with the PSF, was fitted to the image by adjusting its parameters 
to minimize the sum of the squared residuals, using
Powell's algorithm (Press et~al.\ 1988).  The residuals
are dominated by systematic deviations created by the PSF models, rather than
statistical uncertainties.  Therefore, we weighted the pixels uniformly in
computing the $\chi^2$ during the fit optimization.  Changes in the 
weighting had little effect on the results. Parameter
uncertainties were determined by combining three terms in quadrature:
(1) a statistical error estimated from the parameter range found from
separate fits to the individual exposures;
(2) a PSF modeling error term from the range of parameters found
from using different PSF models; and (3) a modeling
term found from using two independent image fitting programs.
The lens galaxy structure was determined from the filter with
the highest signal-to-noise for 
the lens galaxy (usually the NIC2 H image, but the WFPC2 I
image for B~1030+074) and the galaxy structure was then held fixed for the
fits to the other filters.  
Tests on several of the brightest early-type lenses in our sample showed
no significant structural parameter differences between independent fits
to the various filters.  We did find some structural color-dependence for 
late-type lenses, but both late-type lenses in this paper are too faint
for this effect to be measurable. 
When we could not estimate galaxy scale
lengths or axis ratios from our data, we assumed modest, fixed scale
lengths and a round galaxy.  The uncertainties in the galaxy colors
were estimated with the photometric structure of the galaxy held
fixed.  The results of the photometric fits are presented in Table~3.

For the NIC2 images we fit the model components on the two-times
oversampled grid generated by NICRED, while for the NIC1 and WFPC2
data we fit the data on the original pixels. For the WFPC2 data we
used a two-times oversampled PSF model. We assumed pixel scales of
0\farcs04554 (Holtzman et~al.\ 1995) for the PC1 images. We adopt
pixel scales of 0\farcs076030 and 0\farcs075344 for the x and y axes
of NIC2, and 0\farcs043230 and 0\farcs043056 
for NIC1\footnote{
    http://www.stsci.edu/instruments/nicmos/nicmos\_doc\_platescale.html}.
We verified the astrometric accuracy of our fits through cross checks
between the different detectors and against VLBI positions for the
images in B~0218+357
($\hbox{difference}\pm\hbox{uncertainty}=2.0\pm1.0$~mas), B~1030+074
($6\pm3$~mas) and PKS~1830-211 ($1\pm4$~mas).  We impose a
conservative minimum positional rms uncertainty of 3~mas for all
objects.

\section{Lens Modeling}

After the images are decomposed into components, we can interpret the
image geometry with a lens model.  The most likely contributors
to gravitational lensing are the principal lensing galaxy (which can
be described by a central position, the mass within the Einstein ring
radius, and the lens ellipticity and orientation), the tidal
effects of nearby galaxies, which can usually be approximated by an
external shear and its orientation, and the tidal effects produced
by large scale structure along the ray path.

\subsection{The Main Lens Galaxy}

Previous studies of lenses (e.g.\ Kochanek 1995), stellar dynamics
(e.g.\ Rix et~al.\ 1997) and X-ray studies (e.g.\ Fabbiano 1989) all
suggest that the isothermal radial mass distributions ($\rho \propto
1/r^2$) are realistic representations of the total mass. Other lens
models that have traditionally been used (e.g.\ ellipsoidal modified
Hubble models, Plummer models, or the more general power law surface
density $\Sigma\propto[R^2+S^2]^{-\alpha}$) are known to be unphysical
-- we know that the central regions of galaxies are cuspy, and do not
have homogeneous cores (e.g.\ Faber et~al.\ 1997). Therefore the
centers of lens models should be described by a density cusp $\rho
\propto r^{-a}$ with $1 \lesssim a \lesssim 2$ for $r \ll 1$.  For the
isothermal model $a=2$, for the de~Vaucouleurs model $a \simeq 1.25$,
and $a=1$ for the Hernquist (1990) or NFW models (Navarro, Frenk \& White
1996). A radial scale for these models is specified by the ``break
radius" between the inner cusp and a steeper outer profile.  In dark
matter models the break radius is sufficiently large compared to the
size of the lens to be ignored (see~Navarro, Frenk \& White 1996),
while in constant M/L models it is comparable to the effective radius.

We consider three standard lens models.  The singular isothermal ellipsoid
(SIE), the de~Vaucouleurs model, and the exponential disk. 
The surface density of the SIE in units of the critical surface density
for lensing is 
\begin{equation}
  \kappa(x,y) = {b \over 2}\, \left[ 
    {2 q^2 \over 1+q^2}\left(x^2 + y^2/q^2\right) \right]^{-1/2} ,
\end{equation}
where  $q$ is the axis ratio, and $b$ would be the 
critical radius for a singular isothermal sphere ($q\equiv1$).
In cases where we have a good  photometric model of the lens galaxy
we also consider constant mass-to-light ratio (M/L) models
where the lensing mass distribution matches the light distribution
of the observed galaxies, for most cases a de~Vaucouleurs profile.
The scaled surface density of the de~Vaucouleurs model is defined by
\begin{equation}
  \kappa(x,y) = {b \over 2 {\cal N} r_e}\,
  \exp\left[-k \left( {x^2+y^2/q^2 \over r_e^2} \right)^{1/8} \right],
\end{equation}
where $r_e$ is the major-axis effective radius, $k=7.67$, and 
${\cal N} = \int_0^\infty v\,e^{-k v^{1/4}}\,dv$.  
To model spiral galaxies, we used an exponential disk model, with
 \begin{equation}
   \kappa(x,y) = {b \over q}\,\exp\left(-{\sqrt{x^2+y^2/q^2} \over
      R_d}\right)\,,
 \end{equation}
where  $R_d$ is the scale length and $b$ sets the mass scale.
The axis ratio is $q=|\sin i|$ for a thin disk at inclination angle $i$
($i=0$ for edge-on).  
Although none of the late-type galaxies in this paper have sufficient
signal to determine $q$, we will use $q\neq1$ models elsewhere.
A lens model consisting of an isolated 
exponential disk usually produces a third or central
image that is never observed in real lenses, making it an 
unacceptable model if used by itself (Keeton \& Kochanek 1998).
A third image can be suppressed by adding a more singular 
bulge or halo component.

\subsection{External Shear From Nearby Galaxies}

At some level, all gravitational lenses are perturbed by 
the tidal gravity produced by nearby galaxies or other mass
concentrations.  Keeton, Kochanek \& Seljak (1997) found that
every four-image system they considered required two quadrupole axes
to obtain a good fit, independent of the choice for the radial mass
distribution.  The two axes are presumably the axis of the mean mass
distribution of the primary lens galaxy and the axis of the mean tidal
gravity field, although misaligned dark-matter halos are also
possible.  Thus, realistic models for any lens must include an
external shear field.  

We used the SExtractor catalogs (see~\S2) to estimate the local tidal
fields produced by nearby galaxies. We assumed that each nearby galaxy
has an SIS mass distribution, the same M/L as the lens 
(after adjusting for the SExtractor magnitude offset for de~Vaucouleurs
galaxies, see~\S2) and lies at the lens redshift.  While the neighbors can be 
at different redshifts, affecting the estimated shears, we cannot 
make reliable photometric redshift and type estimates for the neighboring 
galaxies with the limited data available.
For a primary lens with a critical radius of $b_l$,
luminosity $L_l$, and a neighbor at distance $r_n$ with luminosity
$L_n$, the Faber-Jackson relationship combined with the SIS lens model
predicts a shear and convergence of
$\gamma_T=\kappa_T=(1/2)(b_l/r_n)(L_n/L_l)^{1/2}$, if the neighbor's
mass distribution extends to 
the lens system\footnote{
    We define the sign of the shear such that for a perturbing galaxy
    G$_T$, the PA of its shear matches the PA of a vector extending from
    the lens galaxy to G$_T$ (modulo 180\arcdeg).
    If a lens is modeled as an isothermal sphere with an external shear,
    the PA of the shear defined in this way will match the major axis PA
    of an equivalent isothermal ellipsoid model. }.
The $20\arcsec$ cutoff radius
for our standard catalog encompasses halo sizes of
$\sim100{}h_{60}^{-1}$~kpc for $z_l \gtrsim 0.3$.  However, if the dark
halos of galaxies extend to $300 h_{60}^{-1}$~kpc (see, e.g., Zaritsky
1994), then galaxies up to $\sim1\arcmin$ from the lens could
have significant tidal effects.  For this reason our field catalogs
include distant, luminous galaxies whose estimated shears exceed 1\%
(the T\# entries).  The halo truncation radius $a$ is crucial for
determining the influence of distant galaxies on the primary lens -- a
halo with critical radius $b$ and a truncation radius $a$ produces
shear $\gamma_T = b/2r$ for radii $r\ll a$ and $b a/r^2$ for radii
$r\gg a$ (see~Brainerd, Blandford \&~Smail 1996).  
Differences in the galaxy types
affect the shear estimates because the mass-to-light ratios of late
type galaxies are roughly three times lower than those of early-type
galaxies, and the shear estimates are proportional to $\gamma_T
\propto (M/L)^{1/2}$.

The total shear $\gamma_T$ and convergence $\kappa_T$ (see Table~4)
are found by taking the tensor sum of the individual contributions.  The
convergence terms simply add, while the shear terms can cancel, so the
totals for the two terms will differ.  In the absence of detailed
information on the redshifts and mass-to-light ratios of the
neighbors, these estimates should be viewed as a qualitative picture
of the strength and orientation of the tidal perturbations for each
lens.   For comparison we also include the shear $\gamma_{M/L}$ expected 
for galaxies without dark matter by computing the shear with the halos 
truncated at the lens critical radius ($a=b$).  Without dark matter,
neighboring galaxies produce negligible tidal perturbations.

\subsection{Large Scale Structure Shear}

To the extent that all the neighboring galaxies are clustered at the
lens redshift, our local tidal shear estimate is only a partial
accounting of the sources of tidal perturbations, because potential
fluctuations from large scale structure along the ray path also
contribute to the lens optics (Bar-Kana 1996).  Although their full
effects are more complicated than a simple external shear (see~Bar-Kana
1996), their overall strength can be characterized by an effective
shear $\gamma_{LSS}$.  In general, the perturbations from other
galaxies at the lens redshift are more important (see~Keeton et~al.\
1997), but not by a large amount.  Unfortunately, if $\gamma_T \sim
\gamma_{LSS}$ we should see little correlation between the locations
of nearby galaxies and the orientations of the external shear
perturbations needed to model the lens.  On the other hand, Keeton et~al.\
(1997) noted that most of the cosmic or large scale structure
shear was generated by the non-linear parts of the power spectrum
which correspond to regions that collapse and produce halos filled
with galaxies.  This means that galaxies with small projected
separations but different redshifts from the lens galaxy may be a
reasonable tracer of the strength and orientation of $\gamma_{LSS}$.
For each lens we used the semi-analytic results in Keeton et~al.\ (1997)
to estimate the LSS shear expected for each lens (see Table~4).

\subsection{The Standard Lens Models}

Based on these physical considerations, we selected a set of four
standard models to fit to all lens systems: (1) a dark matter model
(the SIE model), (2) a model based on the photometric fits (the
constant M/L model), (3) the dark matter model in an external shear
field (the SIE$+\gamma$ model), and (4) the photometric model in an
external shear field (the M/L$+\gamma$ model).  The lens models are
fit to each lens system using the astrometric and structural
properties from Table~3, augmented by more accurate VLBI data for the
component separations when available.  We used either the
extinction-corrected optical flux ratios from Falco et~al.\ (2000) or
the observed radio flux ratios.  The accuracy of flux ratio
measurements is dominated by systematic problems (e.g., time variability
or microlensing) rather than measurement precision, so we
assumed a conservative minimum flux uncertainty of 5\% per component.
Time delays were computed in a flat $\Omega_0=1$ cosmology, using the
lens redshift estimates from Kochanek et~al.\ (2000b) where they had not
been measured spectroscopically.  In the M/L models we fixed the scale
length, axis ratio and position angle of the lens to the values found
in the best fit photometric model\footnote{
    Allowing the photometric parameters to vary independently, 
    constrained with their standard errors from Table~3, 
    does not lead to significantly reduced model $\chi^2$ values. 
    However, the resulting parameters can be grossly inconsistent
    with the observed properties of the galaxy because of strong
    correlations between the photometric parameters.}.

For two-image lenses we usually have a total of five constraints 
on the lens model: 
the positions of the two images relative to the lens and the flux ratio of
the images.  Where we lack complete information on
the lens properties we fit only the SIE model.  Parameter
uncertainties were estimated from the range over which
$\Delta\chi^2<1$.  For underconstrained models we surveyed the
solution space and present allowed parameter ranges defined by the
region with $\Delta\chi^2 < 1$.  The SIE+$\gamma$ model is highly
underconstrained for a two-image lens, so we fixed the orientation
of the SIE to that of the light, limited its ellipticity to be no 
larger than that of the light, and limited the external shear to be
no more than twice the total environmental estimate 
($\gamma_T$ or $\gamma_{LSS}$ from Table~4).  
The M/L+$\gamma$ models required no additional constraints
since the shape and orientation of the galaxy are fixed.

\section{Results}

Images of the ten two-image lenses are shown in Figure~1, where 
we show the original image and the residual image found after 
modeling and subtracting the bright, lensed point sources.
Figure~2 shows detailed images of the lensed 
quasar host galaxies found in BRI~0952--0115 and HE~1104--1805.  
Table~1 summarizes the data available for each system,
and the results of the analyses described in 
\S2, \S3 and \S4 are presented in Tables~2--5.  
Table~2 presents the SExtractor catalogs for each surrounding field.
Table~3 presents the source and lens components for each target. 
Table~4 gives the tidal shear estimates for each field, and 
Table~5 presents the lens models for each system.  
Figures~3 and 4 show the colors and magnitudes of the lens galaxies
as compared to their neighbors 
and the predictions of spectrophotometric models. 

Most of the lens galaxies for which we can accurately measure
the colors have properties consistent with those of passively evolving
early-type galaxies (KKF, Kochanek et~al.\ 2000b). 
To interpret colors,  we have used the solar
metallicity, GISSEL96 version of the 
Bruzual \& Charlot (1993, 2000) spectrophotometric models.
We assumed that galaxies begin forming stars at $z_f=3$
($H_0=65\kms$\,Mpc$^{-1}$, $\Omega_0=0.3$, $\lambda_0=0.7$) with a
Salpeter IMF for the early-type galaxies and a Scalo IMF for the
late-type galaxies.  For the early-type galaxies we use a ``burst''
model, in which star formation occurs at a constant rate for 1~Gyr and then
ceases, and an ``E/S0'' model with an exponentially decaying star
formation rate and a 1~Gyr e-folding time scale.  For the late-type
galaxies we used the star formation rate models for Sa, Sb and Sc
galaxies from Guiderdoni \& Rocca-Volmerange (1988).  

The lens galaxies are generally redder and brighter than their neighbors 
(see~Figure~3), and have colors consistent with early-type galaxies 
(see~Figures~3 and~4).  In fact, most of the lenses for which we can
measure the photometric properties lie on the passively evolving fundamental
plane of early-type galaxies (see~Kochanek et~al.\ 2000b), and we can
use this property to accurately estimate the unknown lens redshifts 
(average error $\langle \Delta z \rangle = -0.01\pm 0.09$).  
Where the lens is undetected or
the photometric models are too poor to test whether the lens lies on
the fundamental plane, the faintness of the lens favors an early-type
lens over a late-type lens because the higher $M/L$ of early-type
lenses makes them fainter for a given image separation (i.e. mass).
In the rest B band the late-type, spiral lenses would be $\sim 1.7$ mag
brighter for the same separation, based on the Faber-Jackson and
Tully-Fisher relations (see Fukugita \& Turner 1991, Kochanek 1996).
For example, the spiral lens galaxy of the smallest separation lens,
B~0218+357 at $z_l=0.68$, is brighter than many of the
early-type lenses found in the wider separation systems.  Only
B~0218+357 and PKS~1830--211 show evidence that they are
late-type galaxies. They are also contain 
molecular gas (e.g.\ Combes \& Wiklind 1997; Carilli et~al.\ 1998) and
they show large differential and total extinctions of the lensed
images (see~Falco et~al.\ 2000).

We now discuss particular characteristics of the individual lenses.
In particular, we compare the goodness of fit for simple constant 
$M/L$ models to those for isothermal dark matter lens models. 
We then add an external tidal field to the models and examine the improvement
in the fit and the relationship of the modeled external tide to our estimates
from the lens environments.  For all model classes we consider the
predicted time delays, and their uncertainty ranges, for use in
determinations of the Hubble constant.  The level of detail possible for each
object depends strongly on the data, geometry and complexity of the individual
systems.

\subsection{Q~0142--100 = UM~673}

The $z_s=2.72$ quasar Q~0142--100 (UM~673) was discovered 
by MacAlpine \& Feldman (1982), and found to be a $\Delta\theta=2\farcs2$
lens by  Surdej et~al.\ (1987, 1988). Surdej et~al.\ (1988) also
identified the lens galaxy with a $z_l=0.49$
Ca~II absorption system with $R \simeq 19$ mag. 
A later spectrum (Sargent \& Rauch 1998, private communication)
showed three Mg~{\sc ii} absorption systems located at
$z_a=0.43$, 0.49, and 0.56 (the strongest feature).
Based on the WFPC2 V and R images, KKF found that 
the lens galaxy is well fit by a de~Vaucouleurs profile, and its colors
match those of a passively evolving early-type galaxy 
at $z_l\approx 0.5$ (see Figures~3 and~4).  

A constant mass-to-light ratio lens model, with both the profile and
ellipticity matched to the photometric models, provides a poor fit to the
image positions and magnifications, with $\chi^2\sim80$.  As also found 
by KKF, the SIE model must be significantly misaligned relative to the 
light to obtain a good fit -- either the dark matter halo is itself 
misaligned, or external objects are exerting a strong tidal shear.
The predicted time delays differ considerably between the two models, 
so a delay measurement and an estimate of $H_0$ could be used to 
discriminate between them.
Either model can be aligned with the lens and have an acceptable
$\chi^2$ with the addition of an external shear of $\gamma\sim 0.07$,
but the strength and orientation of the shear show no correlation 
with shear estimates for the nearby galaxies.  Since the cosmic
shear estimate is comparable to that for the nearby galaxies, we
should not expect such a correlation.  

\subsection{B~0218+357}

B~0218+357 (Patnaik et~al.\ 1993) consists of two images of a compact,
flat spectrum radio core, offset from an Einstein ring image of the
associated radio jet.  The source redshift is $z_s=0.96$ (Lawrence
1996). The lens redshift of $z_l=0.685$ (Browne et~al.\ 1993) was
confirmed by the detection of H{\sc i} (Carilli \& Rupen 1993), CO and other
molecules (Wiklind \& Combes 1995, Gerin et~al.\ 1997, Combes \&
Wiklind 1997) in the lens galaxy.  Grundahl \& Hjorth (1995) detected
the lens galaxy in the optical, but could not characterize its
properties.  B~0218+357 is of particular interest because it has a
measured time delay of $10.5\pm0.4$\, days (Biggs et~al.\ 1999).

We simultaneously fit the H image with 2 point sources and an
exponential disk galaxy.  The estimated B--A quasar separation of
($0\farcs307$, $0\farcs126$) matches the separation of the B1--A1 VLBI
peaks ($0\farcs3092$, $0\farcs1274$) to an accuracy of 2 mas.  We
detect the lens galaxy in all 3 bands, and find that it is closer to
the bright optical/faint radio core B, as expected from the radio flux
ratio of the images.  We quote our final results in Table~3 for a
circular galaxy, because our fits could not distinguish the effects of
a non-zero ellipticity.  The lens galaxy is better fit by an
exponential disk than by a de~Vaucouleurs lens, but the lens colors
are not consistent with a single spectrophotometric model -- the V--I
color is consistent with an $L\sim L_*$ spiral galaxy at the observed
redshift, while the I--H color is more consistent with a less luminous
elliptical.  The discrepancies could be explained either by a red
bulge dominating the IR luminosity, or contamination of the colors due
to the systematic errors in fitting this highly blended system.  The
NIC2 image has a galaxy position of (0\farcs178, 0\farcs046) relative
to A; the 25\,mas positional uncertainty is dominated by the
variations in the position found using different PSF models.

As a check of our NIC2 results, we also reduced and fitted the
archival NICMOS/NIC1 image of the B~0218+357 system (Xanthopoulos,
Jackson, \& Browne 2000).  The undithered NIC1 image has the advantage
of significantly smaller pixels and the disadvantages of a higher
background noise level and poorer flattening.  While the NIC1
and NIC2 quasar separations ($0\farcs305$, $0\farcs126$) agree within 
2 mas, the NIC2 measurement is closer to the VLBI separation. The NIC1
lens position, $(0\farcs184, 0\farcs092)$ relative to A, is 46 mas
away from the NIC2 value (see Figure~5).  Both the NIC1 and NIC2
images yielded markedly worse residuals when we used the lens galaxy
position from one image to fit the other.  We are forced to conclude
that the existing data are unable to determine the position of G
accurately.  We adopt the mean of the NIC1 and NIC2 positions for the
lens position, and one-half their difference as its uncertainty ($\sim
30$ mas; see Table~3 and Figure~5).

We used both NICMOS and VLBI constraints to fit lens models.  The A/B
VLBI images of B~0218+357 each consist of 2 sub-images called A1/A2
and B1/B2, with a separation within each pair of $\sim 1$ mas.  The
subimage separation vectors are roughly equal and radial, suggesting
that successful lens models require nearly flat rotation curves to
avoid large differences in the radial magnification at the two images.
Following Patnaik, Porcas \& Browne (1995) we identified the VLBI
components A1 and B1 with the A and B quasar images. Note, however,
that an error in identification would have negligible effects on the
models because of the very small pair separations.  We modeled
B~0218+357 using the A1--B1 and A2--B2 VLBI image pairs and
magnifications (Patnaik et~al.\ 1995) combined with our position for
the lens galaxy relative to the A1--B1 pair.  We assumed that the VLBI
positions of the components were accurate to 0.1~mas and used the
Patnaik et~al.\ (1995) flux density uncertainties.

We fit only the SIE model to the system.  When the lens position is
constrained by the NICMOS results (model SIE in Table 5), we obtain an
extraordinarily low estimate for the Hubble constant,
$H_0=(20\pm20)(\Delta t/10.5\hbox{days})$, given the Biggs
et~al.\ (1999) time delay.  The low delay is entirely a consequence of
the lens position, as we show in Figure~5.  Using no constraints 
Biggs et~al.\ (1999) derived a lens position of
($0\farcs252\pm0\farcs012$, $0\farcs115\pm0\farcs005$), but they
appear to have underestimated the full extent of the degenerate region.
When we remove the lens position constraints, keeping all other
constraints the same, the $\chi^2$ contours have a
degenerate region parallel to the A--B quasar separation.  
The estimated Hubble constant rises monotonically as the lens approaches
image~B.  An alternative constraint on the lens center comes from the
centroid of the radio ring (at
$0\farcs27\pm0\farcs01$, $0\farcs11\pm0\farcs01$; Patnaik et~al.\
1993), although the systematic errors in such a centroid are of order
$2\epsilon_\phi b \gtrsim 0\farcs03$ where $\epsilon_\phi$ is the
ellipticity of the lens potential and $b$ is the critical radius.  In
model SIE$'$ we constrain the lens center to the location of the ring
centroid.  The fit has $\chi^2=1.5$ and gives a larger Hubble
constant, $H_0=76\pm7$, but it requires a large ellipticity for the
lens ($\epsilon=1-b/a=0.4\pm0.1$).  The luminous galaxy is not this flat
(see Figure 1), and the tidal shear is negligible.  Thus, it is likely
that both the NICMOS and ring centroid estimates of the lens position
are systematically incorrect.  It is important to obtain an accurate
direct measurement of the lens position.

Other changes to the lens models appear relatively unimportant.  We
experimented with changing the lens potential by using a pseudo-Jaffe
model and varying the break radius $a$\footnote{
    The pseudo-Jaffe model is the difference of two SIE models (eqn.~1)
    with scale lengths of $0$ and $a$ respectively.  It
    corresponds to a three-dimensional density distribution with
    $\rho \propto 1/[r^2(r^2+a^2)]$ similar to the Jaffe (1983) model
    with $\rho \propto 1/[r^2(r+a)^2]$.  When the break radius $a$ is
    large compared to the ring radius the model acts like an SIE,
    but by reducing $a$ we model a more compact, finite mass
    distribution (see Keeton \& Kochanek 1998).}.
When the break radius becomes smaller than the
ring radius, the $\chi^2$ of the fit begins to rise along with the
Hubble constant estimate.  But the rise in $H_0$ is insignificant
unless the $\chi^2$ of the fit is so large that the model must be
rejected.

Of all ten lenses, B~0218+357 is the most isolated in the sense that
the estimated tidal shear at the location of the lens is only 
$\gamma_T \sim 1\%$. This is in marked contrast to the other systems
with measured time delays, Q~0957+561 (Schild \& Thomson 1995;
Kundi\'c et~al.\ 1997), 
PG~1115+080 (Schechter et~al.\ 1997), and B~1608+656 (Fassnacht 1997)
and PKS~1830--211 (Lovell et~al.\ 1998)
which are all embedded in small groups or clusters
of galaxies.  
The morphology-density relationship (e.g.\ Dressler 1980) suggests that
late-type galaxies are more likely to be isolated; 
and of the six lenses with time delays, 
only B~0218+357 and PKS~1830--211 are late-type galaxies.  
The low source redshift means that the
large scale structure shear of $\gamma_{LSS}\simeq 0.02$ is also small.
   
\subsection{SBS~0909+532}

SBS~0909+532 was discovered by Kochanek et~al.\ (1997) as two images
of a $z_s=1.377$ quasar separated by $1\farcs11$. These redshifts were
confirmed by Oscoz et~al.\ (1997).  With the NICMOS observations we
have discovered a large, bright galaxy between the quasar images,
confirming that SBS~0909+532 is a gravitational lens.  The lens
redshift is now measured to be $z_l=0.83$ (Lubin et~al.\ 2000).  A
significant residual corresponding to the core of the lens is also
found in the WFPC2 I~band image after subtracting the two quasars, but
the lens is undetected in the WFPC2 V~band image. The lens galaxy has
a large effective radius, with a correspondingly low surface
brightness.  The colors of the lens are poorly measured but consistent
with those of an early-type galaxy at the observed redshift.  The lens
is in a typical tidal environment, with $\gamma_T \simeq 0.05$ and
$\gamma_{LSS}\simeq0.03$.

\subsection{BRI~0952--0115}

BRI~0952--0115 was discovered by McMahon, Irwin \& Hazard (1992) as a
pair of $z_s=4.5$ quasars separated by $0\farcs9$.  From WFPC2 imaging and 
photometry, KKF found that the lens galaxy was an R $\simeq22$ mag, 
flattened early-type galaxy.

The lens appears to be a typical early-type lens galaxy, with a
fundamental plane redshift estimate of $z_{FP}=0.41\pm0.05$ (Kochanek
et~al.\ 2000b).  A constant M/L lens model fails to fit the lens
constraints ($\chi^2=109$) but a modest external shear in a
M/L$+\gamma$ model yields a good fit. We expect no correlations of the
shear with the local environment because the local estimate of
$\gamma_T \simeq 0.03$ is significantly smaller than $\gamma_{LSS}
\simeq 0.05$.  Moreover, galaxy G3, which dominates the local shear
estimate, is likely to be a late-type galaxy (it is unusually blue in
R--H), so we have probably overestimated its tidal
effects.  While we cannot differentiate between SIE and M/L models based
on the image constraints, the time delays depend strongly on the
amount of dark matter in the lens.
 
After subtracting the quasars and the lens, the residual $R$-band
image reveals a prominent arc, extending asymmetrically from quasar
image A (see Figure~2).  Its shape matches the shape predicted for the
lensed image of an extended source centered on the quasar, suggesting
that it is part of an Einstein ring image formed from the $z_s=4.5$
quasar host. The H-band image hints at similar but significantly
fainter residuals.  The limiting magnitudes of the observations in
$\simeq 0\farcs2$ square boxes (4 PC pixels or 5 subsampled NICMOS
pixels) are approximately 24.9 and 22.8 mag arcsec$^{-2}$ in R and H
respectively.  The brightest part of the ring, which is well separated
from the quasar, has a mean surface brightness of 22.4 mag
arcsec$^{-2}$ in R.  All of our spectrophotometric models for
formation redshifts of $z_f\approx5-10$ predict R--H colors in the
range 1.0 to 1.5 in the absence of extinction.  Therefore, for a
stellar population we would expect the H surface brightness of the arc
to be approximately 20.4 mag arcsec$^{-2}$, which should easily have
been detected in our H~band image.  The R filter is almost exactly
centered on the Ly$\alpha$ line of the quasar, so the extended
emission could be dominated by line emission similar to the Ly$\alpha$
emission regions seen near MG~2016+112 (Schneider et~al.\ 1986).  Such
emission regions are rare (a modest amount of dust will quench the
Ly$\alpha$ emission) and surveys for high redshift galaxies based on
detecting high equivalent width Ly$\alpha$ emission have been mostly
unsuccessful (e.g.\ Thompson, Djorgovski \& Trauger 1995). Thus, the
origin of the lensed host light remains unclear.
  
\subsection{LBQS~1009--0252}

LBQS~1009--0252 (Hewett et~al.\ 1994, Surdej et al. 1993) consists of two $z_s=2.74$
quasars, A~and~B, separated by $1\farcs53$.  A third quasar, C, at a
redshift of $z_C=1.62$ lies only $4\farcs6$ to the Northwest of the lens.
No lens galaxy had been detected prior to our observations, although a 
strong Mg~{\sc ii} absorption line system at $z=0.869$ has been proposed 
for the lens redshift.

We clearly detect the lens galaxy near quasar~B in the H and I-band images. 
A de~Vaucouleurs profile fits the lens with an effective radius of
$\approx0\farcs2$, but there was insufficient flux to determine its
ellipticity.  The galaxy is not detected in the shorter V-band observation.
The fundamental plane lens redshift estimate, 
$z_{FP}\approx0.8\pm0.1$ (Kochanek et~al.\ 2000b),
agrees well with the Mg~{\sc ii} absorption redshift. 
In the H-band image we find a residual component G$'$ near quasar~A
and an extended residual, C$'$, around quasar~C.  G$'$ might be the 
lensed quasar host or a fainter companion to the lens galaxy, and
C$'$ is presumably the host galaxy to quasar C.  G$'$ and 
C$'$ are not detected in the optical images. 

We estimate that the quasar host galaxy C$'$ can produce
an effective shear of $\simeq 0.08$ at the lens after correcting for
the redshift differences.  This is comparable to the estimated shear
generated by the other galaxies, but in an orthogonal direction. 
Given the uncertainties in our estimate, either C$'$ or the other
neighbors could dominate the external shear.  
While we fit only an SIE lens model, the
derived major axis of the SIE model is consistent with an 
external shear dominated by C$'$.  Thus, C$'$ may be an extreme example
of the ``large scale structure'' shear contribution being associated
with a detectable object that we would include in our estimate of
the local shear environment in the absence of redshift information. 
The surprisingly blue I--H color for G3 is not due to SExtractor errors,
it is simply much fainter in the H-band image than its neighbor G4.

\subsection{Q~1017--207 = J03.13 = CTQ~286}

Q~1017--207 (Claeskens, Surdej \& Remy 1996) was confirmed by 
Surdej et~al.\ (1997) to be a pair of $z_s=2.545$ quasars separated 
by $0\farcs85$.  Remy, Claeskens, \& Surdej (1998) obtained WFPC2 V 
and I-band images, in which they found no sign of a lens galaxy.

The lens galaxy is clearly detected in our NICMOS H-band image.
Because the unfavorable contrast 
precludes a well-constrained fit to the lens galaxy, we fit a round 
de~Vaucouleurs model with scale length of $0\farcs3$.
Using the astrometric and structural properties from the H image we can
estimate the luminosity of the galaxy in the I band, 
because the core of the galaxy is seen in the I image at 
the same location as in the H band after subtracting
the quasar images.  The V-band image yielded
only an upper limit.  The I--H color of $2.56\pm0.48$
mag is consistent with an early type galaxy at
$z_{FP}=0.78\pm0.07$ (Kochanek et~al.\ 2000b). 
We fit only the SIE model, finding good fits with a moderately
elliptical galaxy having an East-West orientation for the major axis.
However, the external shear estimates (including the expected
cosmological shear) for the field suffice to allow significant freedom
to the true orientation of the lens.

\subsection{B~1030+074}

B~1030+074 was discovered by Xanthopoulos et~al.\ (1998) as a pair of
flat-spectrum radio sources, A and~B, separated by $1\farcs56$.  Both
radio sources had optical quasar counterparts in WFPC2 V and I images.
They noted that the lens consists of two components G and G$'$.
This is confirmed by their later NIC1 image.  We include this lens in
our discussion because its general morphology and HST observations
match our CASTLES sub-sample, and because the VLBI data provide
a useful absolute astrometry check on our procedures.

We measured the B--A separation between the two quasars 
on the WFPC2 I image to be 
($0\farcs928\pm0\farcs003$, $-1\farcs257\pm0\farcs003$),
in agreement with the VLBI offset of 
($0\farcs934\pm0\farcs001$, $-1\farcs258\pm0\farcs001$) 
from Xanthopoulos et~al.\ (1998) 
to within 6~mas, for a joint $\chi^2/N_{dof}=3.6/2$\,.  
We could not obtain better results from the NIC1 data because
the combination of chip defects, chip edges, cosmic ray persistence and the
lack of dithering leads to a very poor image.
We find that G is well fit by a de~Vaucouleurs profile,
and that it has the color, flux, and scale-length expected for 
a red, early-type galaxy at redshift 0.60\,. 
Although the color of G$'$ is nearly the same as that of G, it is
better fit by an exponential disk profile.
The two-component photometric model fits the image well, and we 
conclude that G and G$'$ are distinct galaxies.  

For the lens models, we used as position constraints our NICMOS galaxy 
position and the Xanthopoulos et~al.\ (1998) 
VLBI image radio positions (with an assumed error of 2\,mas).
We used an image flux ratio of $A/B=15\pm6$ based on the 
average radio flux ratio, and used an error equal to twice
the dispersion in the flux ratio. 
We used two-component lens mass models to represent B~1030+074.  
To reduce the number of free parameters in the dark matter model, 
we allowed only galaxy G to be an elliptical SIE, while 
G$'$ was modeled as a circular SIS with a critical radius scaled
to that of G by the square root of their luminosities.  Without an
external shear, the orientation of the SIE models is inconsistent
with that of the light. The constant M/L lens model, however,
which included an elliptical de~Vaucouleurs lens for G and a circular
exponential lens for G$'$ with masses scaled to their relative 
brightnesses, provided a good fit to the lens constraints,
with $\chi^2/N_{dof}=0.7/2$.  This is the only case in our sample 
where a constant M/L model can fit the constraints in the
absence of an external perturbation.

B~1030+074 is one of the two systems where the local shear
$\gamma_T\simeq0.11$ estimated from the nearby galaxies is
significantly larger than the large scale structure contribution
$\gamma_{LSS}\simeq 0.03$, and we might hope to find a strong
correlation between the shear required by the lens model and the
predictions from the environment.  
Unfortunately, the galaxy dominating the local shear estimate, G1, 
is too blue (V--I=0.9~mag) to be a early-type galaxy at the lens redshift
(V--I$\simeq$1.7~mag), which means that we have overestimated its
shear contribution, and cannot make firm statements about the environment.

\subsection{HE~1104--1805}

HE~1104--1805 was discovered by Wisotzki et~al.\ (1993, 1995) as two
$z_s=2.319$ quasars separated by $3\farcs19$.  Courbin, Lidman \&
Magain (1998) detected the lens galaxy in ground-based observations,
with estimated magnitudes of J$=19.0\pm0.2$ and K$=17.1\pm0.2$. 
Remy et~al.\ (1998) detected the lens galaxy in their WFPC2 I images but
not their V images, and suggested that it was an early-type galaxy at
$z_l=1.32$, where the quasar spectrum has a metal absorption feature.
Since all quasars show an average of one metal line absorption system
per unit redshift down to a rest equivalent width of around 1\AA$\,$
(Steidel 1990), independent of whether or not they are lensed,
there is no convincing reason to adopt the redshift advocated by Remy
et~al.\ (1998)\footnote{
    See the discussion in Siemiginowska et~al.\ (1998) on using metal
    line absorption features to estimate lens redshifts.}.

The lens galaxy's colors are consistent with a high-redshift early-type
galaxy, and we estimate its redshift is 
$z_{FP}=0.77\pm0.07$ based on the fundamental
plane (Kochanek et~al.\ 2000b).  HE~1104--1805 is very unusual in that 
the lens is near the bright image, rather than the faint one.  Simple 
models can match this configuration only for a narrow range of parameters
which imply a large misalignment between the light and simple ellipsoidal
lens models.  The mass can be aligned with the light only if the shear
field is twice as strong as estimated in Table~4.  The image separation 
of $3\farcs2$ is also much larger than that of a typical lens
(the median is $1\farcs5$), suggesting that the separation is boosted by the
presence of a group or cluster, although none is obviously present
in the WFPC2 fields.  The narrow parameter range is a consequence
of the peculiar flux ratio, which may be distorted by either intrinsic 
source variability or microlensing (see Wisotzki et~al.\ 1995).  Although we
can find less extreme lens models if we drop the flux ratio as a 
constraint, the required change in the flux ratio (a factor of 4)
is inconsistent with the structure of the host galaxy (see below), 
where the A image is clearly more magnified than the B image.

In the residuals of the H image we clearly see two arc images of the
quasar host galaxy.  We fitted a circular exponential host galaxy
model, lensed by the SIE+$\gamma$ model, and derived an intrinsic host
galaxy magnitude of H=21.1, corresponding to $\sim 0.3 L_*$ for a
burst population evolution model.  Fitting a de~Vaucouleurs model
gives H=20.5 mag, with slightly larger residuals.  The corrections to the
magnitudes of the quasar nuclei are $<0.1\%$ as a result of the
additional host component.  By scaling and adding the H image of the
arcs to the I and V images, we estimate that the arc has I--H$ >3$ mag
and V--H$ >2.5$ mag.  This is significantly redder than all but the
burst spectrophotometric models. The E/S0 and Sabc models all have
I--H$\simeq 2.4\pm0.4$ mag for formation redshifts of 
$3 \lesssim z_f\lesssim 15$, 
while the burst models have I--H$\simeq 3.5$ to $5.0$ mag.
Thus the host galaxy either has little ongoing star formation or is
dusty.

\subsection{Q~1208+1011}

Q~1208+1011 was discovered by Maoz et~al.\ (1992) and Magain et~al.\ 
(1992) as two images of a $z_s=3.8$ quasar separated by
$0\farcs45$.  The indistinguishable redshifts were confirmed by
Bahcall et~al.\ (1992a). WFPC1 images (Bahcall et~al.\ 1992b) yielded
no evidence for a lens galaxy.  Siemiginowska et~al.\ (1998) estimated
that an early-type lens galaxy would have $20 < H < 21$ mag for the
most plausible lens redshift range, while a late-type lens would be
0.5-1.0 mag brighter.  The most plausible redshift range includes 
a $z_a=1.13$ Mg~{\sc ii} absorption system, and we use this redshift
to compute lens model time delays. 

We failed to detect the lens galaxy in either the NICMOS or the WFPC2
images.  Our detection limit for an $r_e=0\farcs3$ de~Vaucouleurs galaxy
located exactly between the quasar images is approximately 20 H mag, 
still consistent with the system being a normal gravitational lens where
the lens galaxy is masked by the bright quasars.  It is purely a problem of
contrast, as the detection limit in the field away from the quasars is
closer to 22 H mag for the same galaxy.  The detection limits for the
lens galaxy at V and I are significantly worse.  Without even a position
for the lens we cannot explore models more complicated than a simple,
circular SIS model.  

\subsection{PKS~1830--211}

PKS~1830--211 was identified as a lens by Subrahmanyan et~al.\ (1990),
although the details became clear only with the higher resolution
radio maps of Jauncey et~al.\ (1991).  It is an Einstein ring
containing two bright images of the radio core with a
separation of $0\farcs971\pm0\farcs002$ (Jones et~al.\ 1993).  
The source is strongly variable and a time delay of
$26\pm5$~days has been measured (Lovell et~al.\ 1998). 
Molecular absorption has been detected at $z_l=0.886$ 
(Wiklind \& Combes 1996, Gerin et~al.\ 1997, Carilli et~al.\ 1998)
largely in front of the B (Southwest) image (Frye et~al.\ 1999). 
Lovell et~al.\ (1996) also found a $z_a=0.19$ H{\sc i} absorption feature.  
X-ray spectra also show a large
amount of absorption due to dust and gas (Mathur \& Nair 1997).
Using infrared observations, 
the source components have been resolved
(Courbin et~al.\ 1998; Frye et~al.\ 1999),
and the quasar redshift of $z_s=2.507$ has been measured 
(Lidman et~al.\ 1999). 

We detect the bright quasar and the lens galaxy in I, H and K, while
the faint quasar is only detected unambiguously in the K-band.  The
lens galaxy is best fit by an exponential disk which is
sufficiently round to make estimates of the PA unreliable.  The
astrometry for the lens galaxy is complicated by the presence of a
feature, labeled P in Table~3, between the quasars and near the center 
of the lens galaxy.
It appears to be point-like, but its colors are 
redder than those of the nearby M star, and similar to those of the
lens galaxy G (see Table~3). In the region we used for our fits
there are 10 stars in an area corresponding to 36 Einstein ring areas,
leading to an expectation of $\sim 0.3$ stars per ring area.  
Thus, the feature could be either the bulge of the lens galaxy or a
superposed, red Galactic star. 
We decided to assume that the feature is a superposed Galactic star,
and to adjust the position of the lens and the feature separately.

The positional uncertainty for P is $< 6$ mas. The corresponding uncertainty for 
G is much larger, $\sim 80$ mas, because of its low surface brightness and
the proximity of A and P.  The K-band image yields
positions for the lensed compact images A and B with positional
uncertainties $< 3$ mas, as expected for bright point-like sources.
Courbin et~al.\ (1998) noted that their position for the SE component
is likely to be contaminated by the lens galaxy. 
Indeed, while their location for the brighter NE component
is consistent with our determination, their position for 
the SW component is shifted towards G, disagreeing with ours at
the 3--4$\sigma$ level for their $0\farcs05$ standard errors.  
Our estimated separation of $0\farcs973\pm0\farcs004$ is
an excellent match to the mean VLBI separation of
$0\farcs974\pm0\farcs002$ (Jones et~al.\ 1993).  The galaxy G is
close to the lens model position of Nair, Narasimha \& Rao (1993),
while the feature P is close to the lens model position of
Kochanek \& Narayan (1992).  Figure~5 shows the locations of the
various components as well as the lens positions in the two models.

For the lens models, we used the NICMOS lens position, combined with the more 
precise VLBI image positions and flux ratios.  We do not use any constraints 
from the radio ring.  Given the Lovell et~al.\
(1998) time delay and an SIE model for G, we find a Hubble constant of 
$H_0= (73\pm35)(\Delta t/26\hbox{days})$~km~s$^{-1}$~Mpc$^{-1}$.  The 
uncertainties in $H_0$ from the lens model are dominated by the 
uncertainties in the lens position (see Figure~5).  The system can also
be well fit using two SIS lenses (model SIS+SIS in Table 5), one located at the position of G
and the other at the position of the close neighbor G2 (see~below),
and in these models the values of $H_0$ are reduced by approximately
15\%.  

We identified several other galaxies in the field, in particular the
bright galaxy G2, only $2\farcs5$ to the South of the lens system.
The lens galaxy, G, and the nearby galaxies G1 and G3 have similar 
colors (I--H$=2.6$ to $2.8$ mag), while the very close neighbor
G2 is significantly bluer (I--H$=1.8$ mag).  Although the lens is better
fit by an exponential profile, its I--H color is closer to that of
an early-type galaxy than a late-type galaxy at $z_l=0.9$.  
However, the high estimates (Courbin et~al.\ 1998; Falco et~al.\ 2000)
for the differential ($\Delta E_{B-V} \approx 3.0$ mag) 
and total extinction for the blue image ($E_{B-V} \approx 0.6$ mag) mean
that the lens galaxy colors are significantly affected by extinction.   
The molecular absorption is almost certainly associated with the primary
lens galaxy, and the similar colors of G1 and G3 suggest that they lie
at the same redshift.  

The relatively blue galaxy G2 is the only candidate for producing the
H{\sc i} absorption at $z_a=0.19$ reported by Lovell et~al.\ (1996).  If it
is a low redshift galaxy, it is not an important contributor to the
lens model.  An $L_*$ galaxy at $z=0.2$ has $H_*=$15--16 mag (early
to late type) compared to $H_*=$18--20 mag at $z=0.9$.  If G2 is at
the lens redshift, it is an $\sim L_*$ galaxy producing strong tidal
perturbations ($\gamma_T\simeq0.08$), whereas if it is at the H{\sc i}
absorption redshift of $0.19$ it is a $\sim 0.01L_*$ dwarf galaxy
contributing a negligible tidal shear ($\gamma_T\simeq0.004$).  
A model consisting of two SIS lenses representing galaxies G and
G2 fits the data well (model SIS+SIS in Table 5), 
and the critical radius ratio of $b_{G2}/b_{G1}=0.7\pm0.4$
implies a luminosity difference of $0.5\pm0.5$~mag which is consistent with
the H-band magnitude difference of $0.6\pm0.6$ mag. The uncertainties
in the luminosity ratio are dominated by the uncertainties in the galaxy 
scale lengths.  G2 also has the well structured appearance of a massive 
galaxy and an H-band surface brightness similar to that of G, G1 and G3. We
conclude that G2 lies at the redshift of G.

\section{Conclusions}

Because of the small angular scales of gravitational lenses, HST is
essential for accurate surface photometry of gravitational lens galaxies.
Furthermore, NICMOS allows us to observe lens galaxies where they emit
most of their energy, rather than beyond the sharp spectral breaks that
make them difficult to study in the optical.  Here we outlined our survey
procedures and present ten double lenses where a combination of limited 
constraints and extreme contrast between the lens and the source restricts
our analysis to relatively simple models.  For these 10 targets, we have made 
the first detections of a lens galaxy in 3 cases 
(SBS~0909+532, LBQS~1009--0252 and Q~1017--207), and  
we were unable to detect the lens in only one case, Q~1208+1011.  
Given our estimated detection limits on the Q~1208+1011 lens galaxy,
our failure to find the lens galaxy is still consistent with
the lens hypothesis.  We 
obtained quantitative surface photometry on 6 systems, Q~0142--100,
BRI~0952--0115, LBQS~1009--0252, Q~1017--207, B~1030+074, HE~1104--1805,
and PKS~1830--211, and we clearly detected the host galaxy of the lensed quasar
in BRI~0952--0115 and HE~1104--1805.

Most of the lens galaxies appear to be early-type galaxies based on
their colors and luminosities, as previously noted by KKF and
predicted in most theoretical models (Fukugita \& Turner 1991,
Kochanek 1994, 1996, Maoz \& Rix 1993).  Usually where
we lack accurate photometry on the lens, an early-type galaxy is
favored because the lower $M/L$ of late-type galaxies should make
them brighter than we actually observe.  For example, the spiral lens
galaxy in B~0218+357 is quite luminous for a lens despite its being the
smallest separation lens known.  We have found (KKF; Kochanek et~al.\ 2000b) 
that most of the doubles for which we obtained quantitative
photometry (Q~0142--100, BRI~0952--0115, LBQS~1009--0252, B~1030+074
and HE~1104--1805) lie on the fundamental plane (Djorgovski \& 
Davis 1987; Dressler et~al.\ 1987).  The two
clear late-type lenses, B~0218+357 and PKS~1830--211, also show large
differential extinctions (see Falco et~al.\ 2000) and contain molecular gas
(e.g., Combes \& Wiklind 1997; Carilli et~al.\ 1998)

We fit a set of four standard lens models designed to explore the
issues of dark matter and the lens environment.  We used either a
constant M/L lens model matched to our photometric fits or a singular
isothermal ellipsoid.  We fit each model in isolation and then with an
external shear to represent perturbations to the model from nearby
galaxies or potential perturbations along the ray path.  Without
external tidal perturbations, the constant M/L models usually could
not fit the lens constraints, and the dark matter models could only
fit the lens constraints if misaligned relative to the luminosity.
With the addition of a modest external shear, either model could fit
all the lenses because the two-image lenses provide so few constraints
on the models.  In general, the constant M/L and dark matter models
predict very different time delays for the two-image lenses, and are
hence distinguishable.

We cataloged the galaxies found within $\sim 100h^{-1}$~kpc of the
lenses and estimated the local tidal environment based on the
luminosities of the neighboring galaxies.  If the galaxies have
extended dark matter halos, they produce significant shears at the
lens galaxy of $\gamma_T \sim 0.05$, while if they have a constant M/L
they produce a negligible perturbation of $\gamma_T < 0.01$.  The dark
matter estimates for the tidal shear are comparable to the shears
necessary to find good lens models, but there is no clear correlation
between the shear from the lens model and the estimate from the local
environment.  Part of the problem is that the shear perturbations
created by large scale structure along the ray path (see Bar-Kana
1996, Keeton et~al.\ 1997) are of comparable strength to those from
nearby galaxies.  The poorly constrained two-image lenses are ill
suited for quantitative studies of their tidal environments.

Two of these 10 lenses, B~0218+357 (Biggs et~al.\ 1999) and PKS~1830--211
(Lovell et~al.\ 1998), have measured time delays.  Unfortunately, the
short exposures necessary for a survey project cannot measure the
position of the lens galaxy in these two systems with sufficient
precision to accurately determine the Hubble constant --  we can measure
the positions with an accuracy of 0\farcs03--0\farcs07, but we need
the positions to an accuracy $<0\farcs01$.  For PKS~1830--211 only
longer observations are necessary -- the data obtained here
demonstrate that the uncertainties would be easily resolved given a
higher signal-to-noise ratio.  For B~0218+357 the solution is less
clear, because we are limited by the small component separations
rather than the signal-to-noise.  Better image sampling (currently
obtainable only in the ultraviolet for HST) combined with concurrent PSF
measurements might resolve the problem.

\acknowledgements
Acknowledgements: 
We would like to thank K.\ McLeod for her library of empirical PSF data.
We also thank N.\ Jackson and the CLASS collaboration for 
discussions of their NIC1 data.
Support for the CASTLES project was provided by NASA through grant
numbers GO-7495 and GO-7887 from the Space Telescope Science
Institute which is operated by the Association of Universities for
Research in Astronomy, Inc.\ under NASA contract NAS5-26555. 
This research was supported in part by
the Smithsonian Institution. CSK is also supported by NASA grant NAG5-4062. 

\section{Appendix: NICMOS Image Reduction}

We developed the NICRED reduction package\footnote{
    the package may be downloaded from http://cfa-www.harvard.edu/castles}
for NICMOS data sets
which are acquired in the MULTIACCUM mode (MacKenty et~al.\ 1997),
where the detector is read out non-destructively at intervals
throughout the exposure. By recording all readouts, pixels that
become saturated or are struck by cosmic rays during the exposure can
still be used.  The ability to avoid saturation is an enormous 
advantage for lens systems because of the extreme surface brightness 
contrast between the quasar images and the lens or host galaxies.

The first step in NICRED is to subtract the dark current and
bias frame by frame from each MULTIACCUM subexposure.
The library darks from the HST archive left unacceptably large
dark-frame structures in the reduced images.  Therefore we
constructed ``sky dark'' frames from our own images, which contained 
mainly empty sky: we masked pixels with significant detected flux in 
all our raw images, and 
median-combined the masked images to produce the sky dark image.
A change in the flight software on August 21, 1997 altered the dark 
frame structure and forced us to construct two separate sky-darks.
Among the data taken before August 1997 only two
images (Q~0142--100 and Q~1634+267) had sufficient empty sky to build
the sky-dark.  Here we first combined the masked raw images separately
for each target, and then kept only the lowest of the two target
signals in each pixel.

Following dark subtraction, we fit, for each pixel, a linear function to 
the run of counts as a function of exposure time at readout.  We
allow for the possibility of a cosmic ray hit between readouts by
fitting two line segments, with the same slope, before and after 
each readout.  Thus for a sequence with $N$ readouts, we considered
$N$ trial intervals where a cosmic ray might have hit.  For each
trial, we derive the slope and two intercepts.  If the difference
between the two intercepts is large, compared with the residuals of
the fit, a cosmic ray hit is likely.  Out of the $N$ trials, we choose
the trial with the largest difference-to-residual ratio $r_{\rm diff}$
as that where the hit occurred.  If no trial has a statistically
significant $r_{\rm diff}$, no cosmic ray hit has occurred, and we
perform a traditional linear regression with a single intercept.
The slope derived for each pixel represents its count rate.

We cannot apply a simple flat-field correction, because the NICMOS
detectors suffer from the ``pedestal effect'' (Skinner 1997), a global
shift in the signal level that affects each readout.  In general, the
shift is different for each readout, but is the same for all quadrants 
of the NICMOS detector.  As a consequence, the final image reflects
the actual count rates plus some global constant.  When the flat 
field correction is made, the additive constant makes an incorrect
contribution to the flat field pattern, resulting in a poorly
flattened image.  Because our images are predominantly blank sky, we can
correct for this problem by masking out the objects in each image, and
assuming that the rest of the image is a constant, multiplied by the flat,
plus another constant for the pedestal effect.  We simultaneously determine the
two constants with a least-squares fit, and remove them with the
appropriate flat-field correction.

Each of the individual readouts is affected by the pedestal effect, 
so the counts in the individual pixel fits are not a linear function 
of exposure time, and the cosmic ray rejection algorithm can fail. 
We use an iterative scheme to correct for this non-linearity.
For each read-out, we scale the initial estimate of the count 
rates by the appropriate fractional exposure time at each read-out, 
and subtract them pixel by pixel from each readout. The pixel median 
of each resulting frame is then an estimate of the ``pedestal constant''
for that readout and can be subtracted. Finally, the entire slope-fitting 
and cosmic-ray rejection process is repeated, yielding a sky-subtracted, 
flat-fielded image for each pointing of the telescope.

Next, each of the four individual target images, which has been constructed 
from the multiple read-outs, is magnified by a factor of two using third-order 
spline interpolation. Then all four are registered, after the dither shifts 
are determined from a two-dimensional cross-correlation. To construct the
combined, final image, they are weighted inversely by the sky variance and
co-added. During the image
combination, bad pixels are removed using a mask file, and the IRAF
``avsigclip'' 3-sigma rejection algorithm.  We also produce separate
combinations for each exposure where the other exposures have been
relative weights a factor of 1000 smaller.
This results in shifted, masked
and clipped images for each subexposure, which we used to
estimate measurement errors. 

The analysis of NIC1 data (B~0218+357 and B~1030+074) is similar to that 
for the NIC2 data.  However, the failure to dither the observations 
and the limited use of the NIC1 observing mode led to significantly
poorer results than for the NIC2 data.  The lack of dithering, the 
use of two different exposure sequences and the limited use of the 
observing mode made it impossible to construct empirical sky-darks.  
The lack of dithering made it impossible to remove bad pixels or chip 
boundaries.  The most important problem was the enormous increase in
the background noise caused by cosmic ray persistence (dark current
variations) from SAA passages (Najita, Dickinson, \& Holfeltz 1998).  
The background noise in the NIC1 
observations was dominated by cosmic ray persistence rather than
the actual sky background.

 \vspace*{-0.2in}
 \begin{figure}
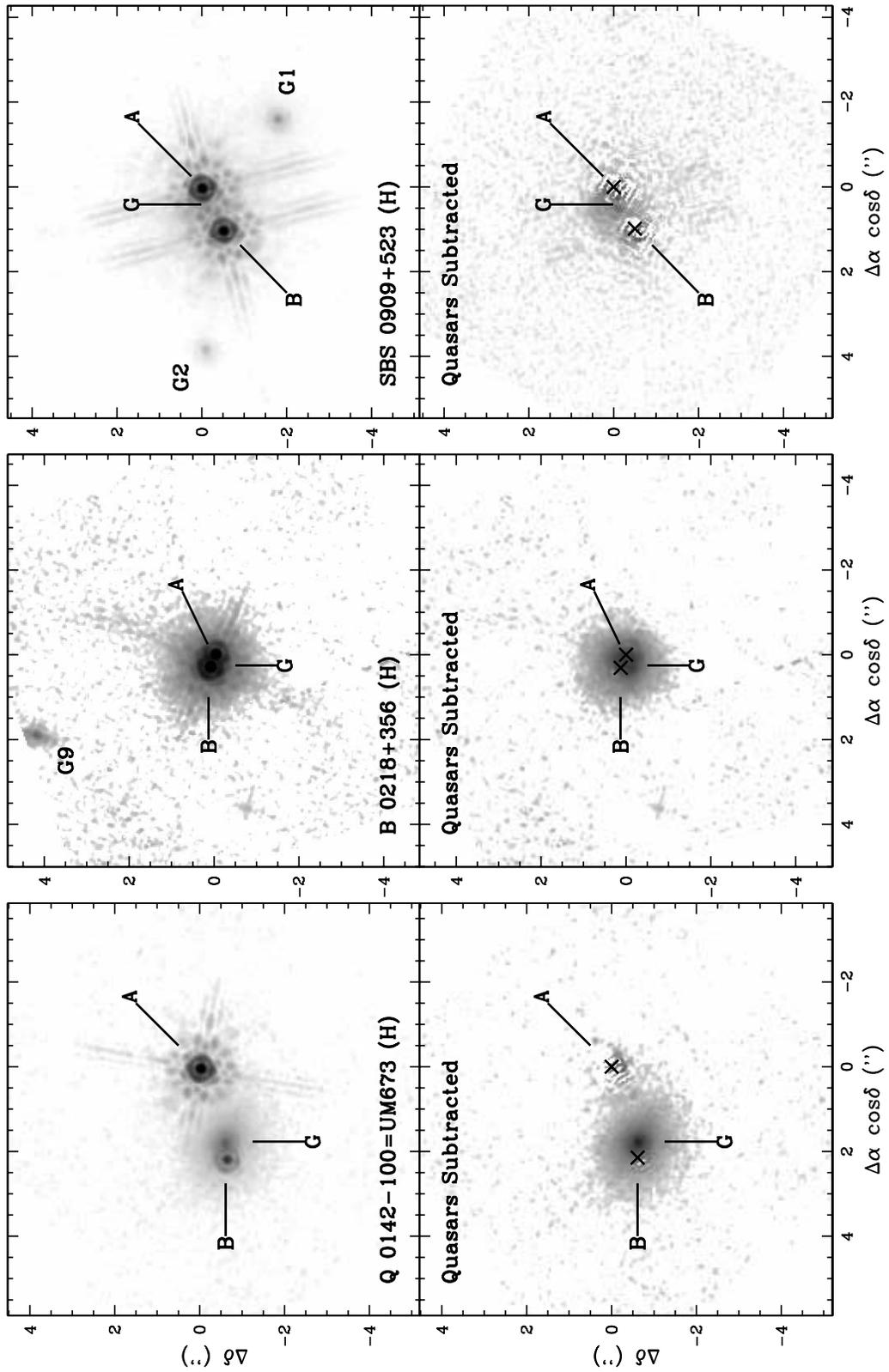

 \label{fig-obs1}
 \figurenum{1}
\vspace*{-1.7cm}
\centerline{
  \psfig{figure=fig1e.eps,width=3.0in,angle=90.}
  \hspace*{-1.50cm}
  \psfig{figure=fig1f.eps,width=3.0in,angle=90.}
  }
\vspace*{-0.7cm}
\centerline{
  \psfig{figure=fig1c.eps,width=3.0in,angle=90.}
  \hspace*{-1.50cm}
  \psfig{figure=fig1d.eps,width=3.0in,angle=90.}
  }
\vspace*{-0.7cm}
\centerline{
  \psfig{figure=fig1a.eps,width=3.0in,angle=90.}
  \hspace*{-1.50cm}
  \psfig{figure=fig1b.eps,width=3.0in,angle=90.}
  }
 \caption{
   HST images of the 10 lens systems, with the filter indicated.
   Top panels show the final image of the system,
   with components labeled as in Tables~3 and~5.
   Bottom panels show the residuals after subtracting the quasars,
   with crosses marking the location of the quasar images.
   }
 \end{figure}

 \begin{figure}
 \label{fig-obs1}
 \figurenum{1}
\vspace*{-1.7cm}
\centerline{
  \psfig{figure=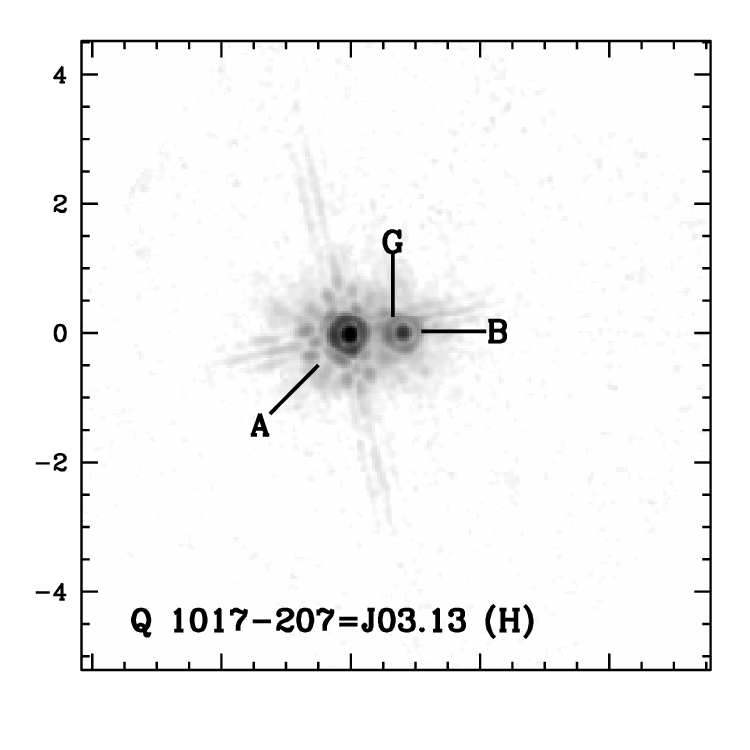,width=3.0in,angle=90.}
  \hspace*{-1.50cm}
  \psfig{figure=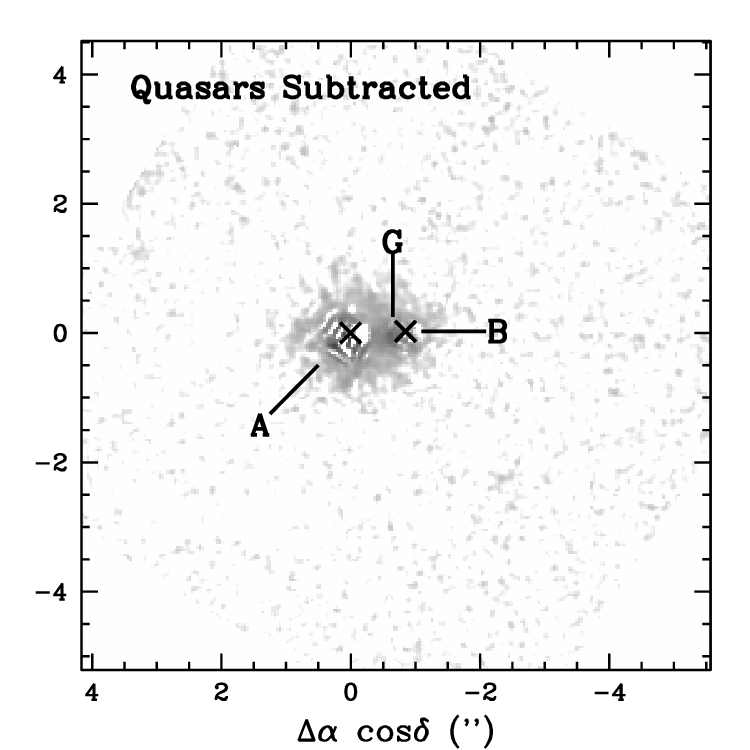,width=3.0in,angle=90.}
  }
\vspace*{-0.7cm}
\centerline{
  \psfig{figure=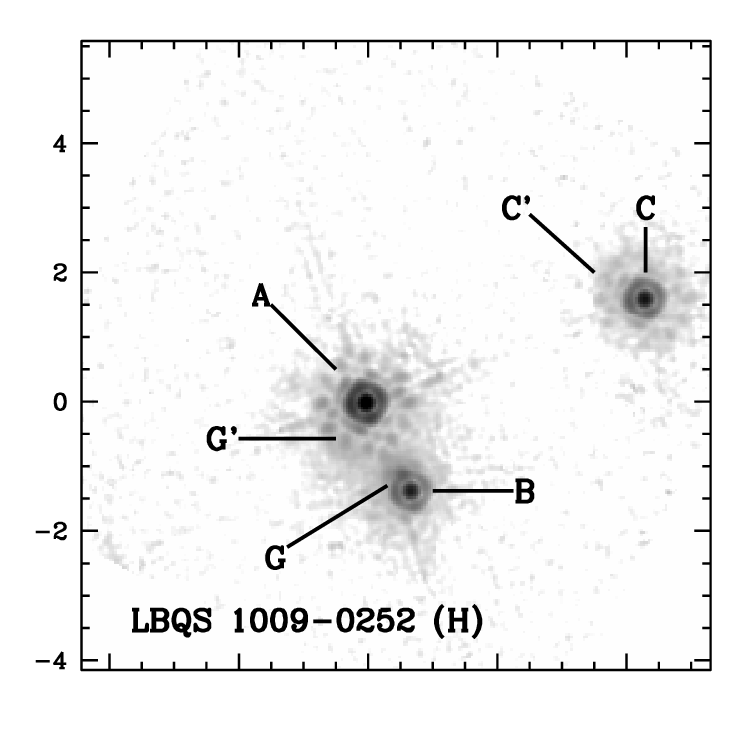,width=3.0in,angle=90.}
  \hspace*{-1.50cm}
  \psfig{figure=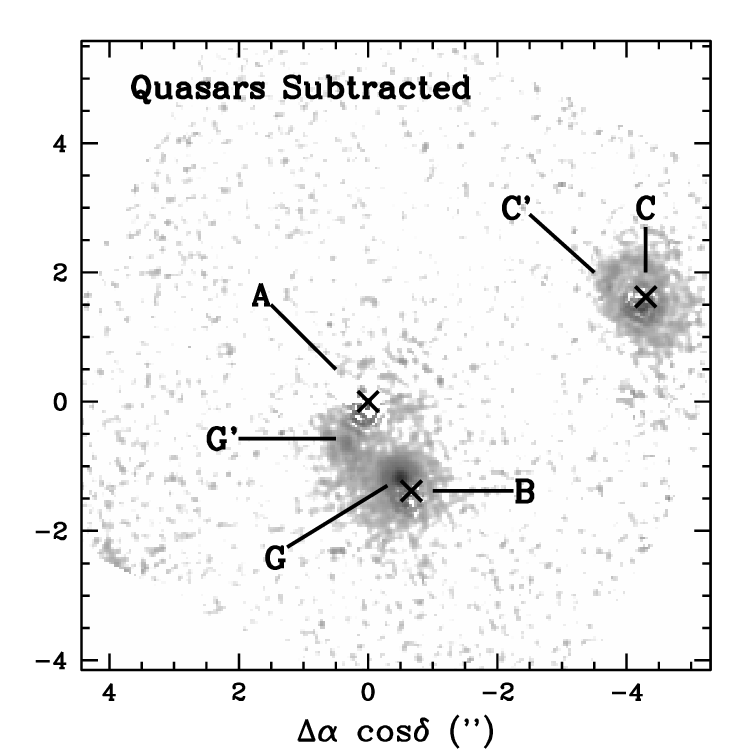,width=3.0in,angle=90.}
  }
\vspace*{-0.7cm}
\centerline{
  \psfig{figure=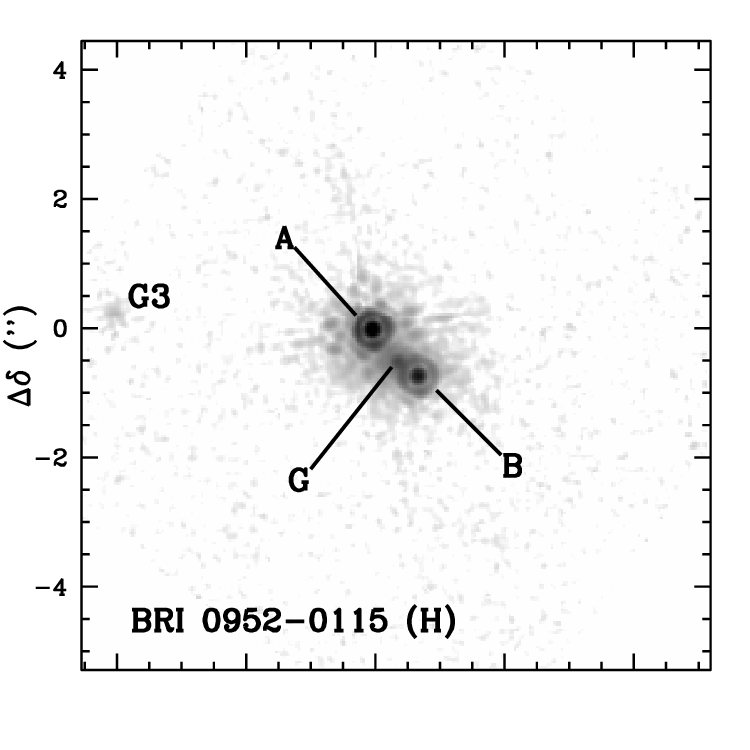,width=3.0in,angle=90.}
  \hspace*{-1.50cm}
  \psfig{figure=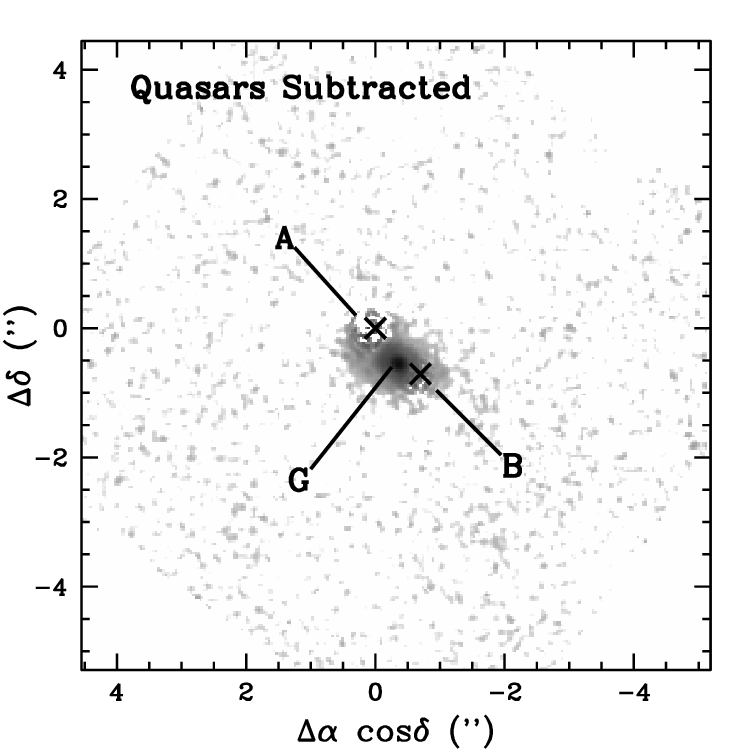,width=3.0in,angle=90.}
  }
 \caption{Continued}
 \end{figure}

 \begin{figure}
 \label{fig-obs1}
 \figurenum{1}
\vspace*{-1.7cm}
\centerline{
  \psfig{figure=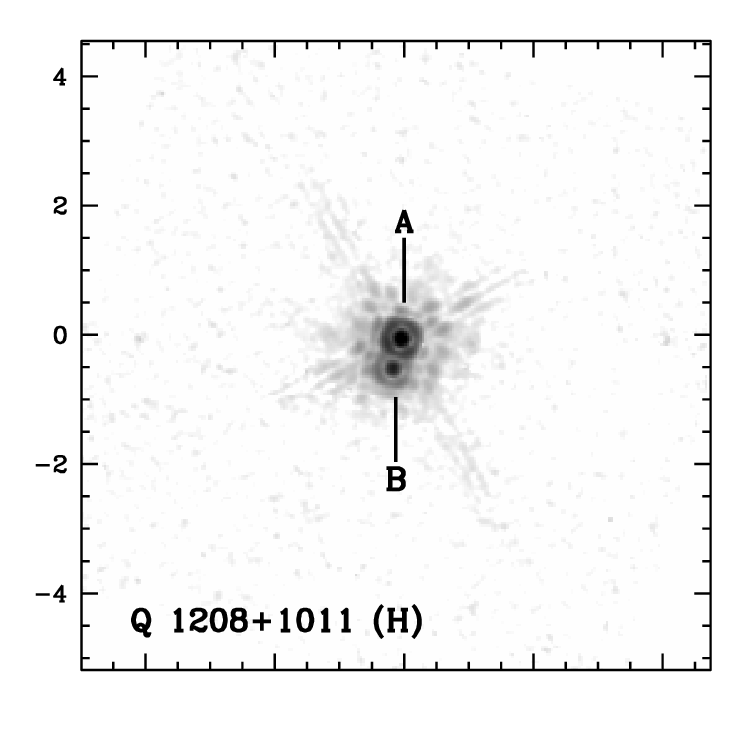,width=3.0in,angle=90.}
  \hspace*{-1.50cm}
  \psfig{figure=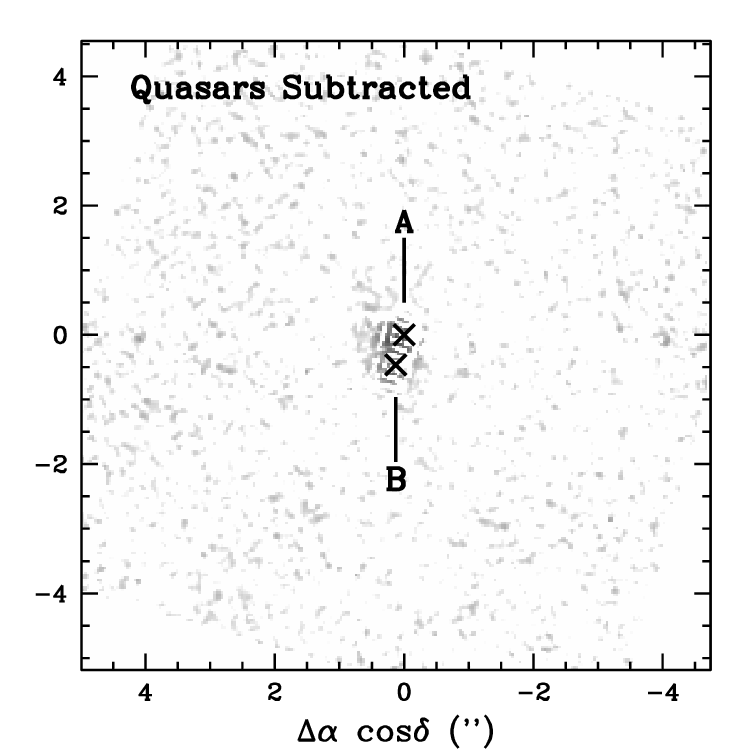,width=3.0in,angle=90.}
  }
\vspace*{-0.7cm}
\centerline{
  \psfig{figure=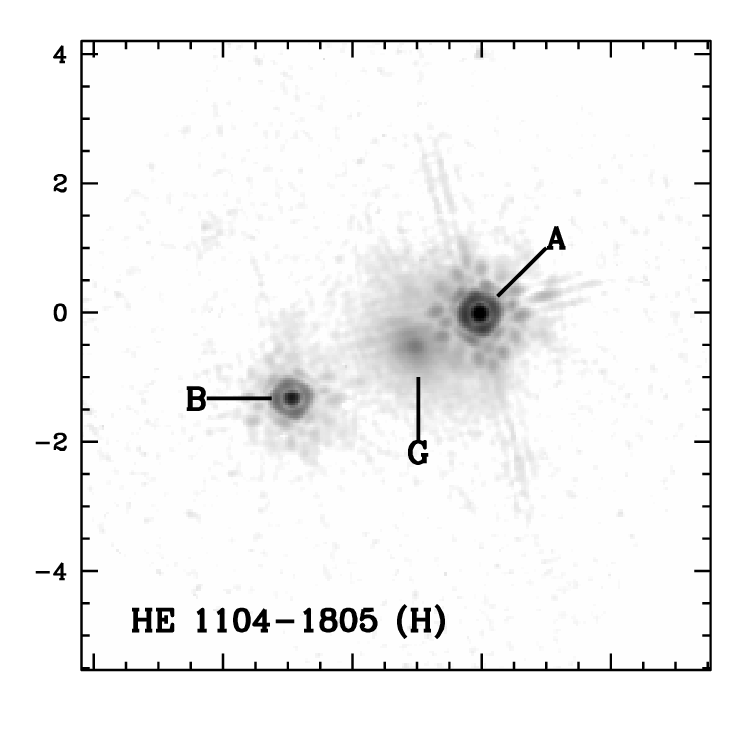,width=3.0in,angle=90.}
  \hspace*{-1.50cm}
  \psfig{figure=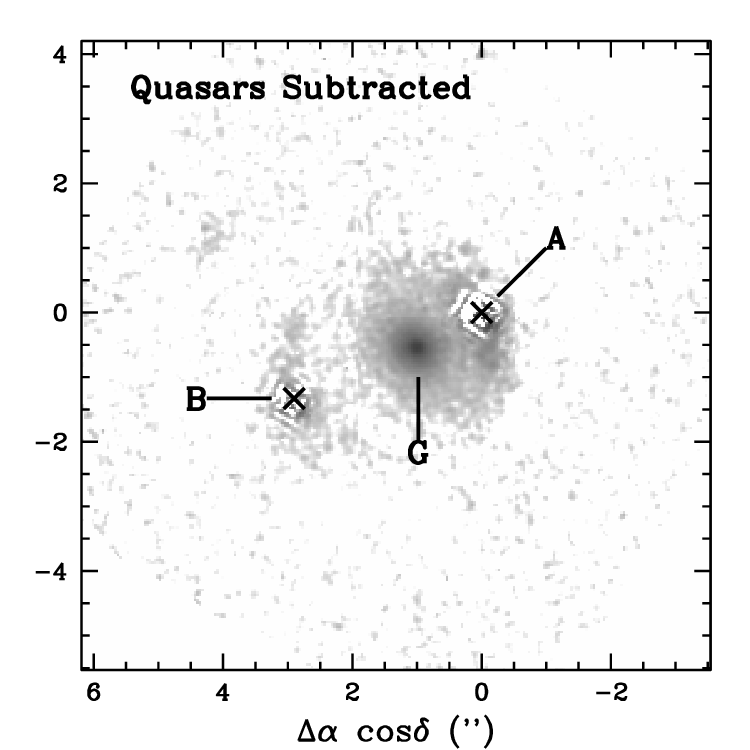,width=3.0in,angle=90.}
  }
\vspace*{-0.7cm}
\centerline{
  \psfig{figure=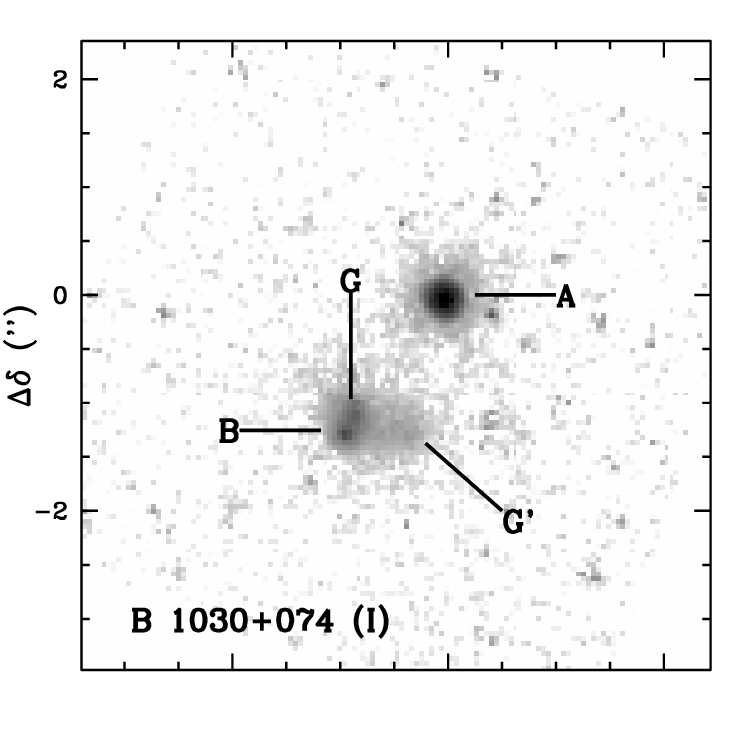,width=3.0in,angle=90.}
  \hspace*{-1.50cm}
  \psfig{figure=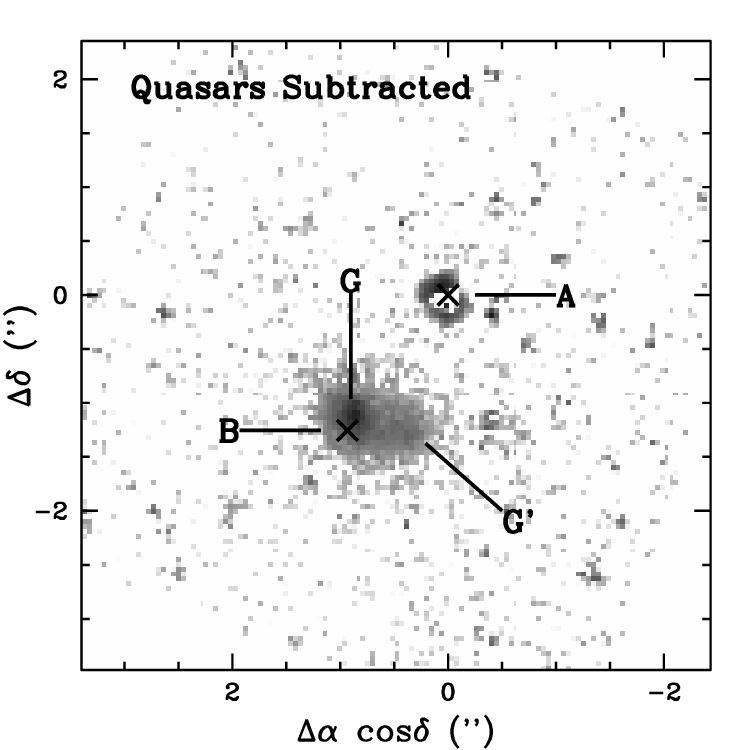,width=3.0in,angle=90.}
  }
 \caption{Continued}
 \end{figure}

 \begin{figure}
 \label{fig-obs1}
 \figurenum{1}
\vspace*{-1.7cm}
\centerline{
  \psfig{figure=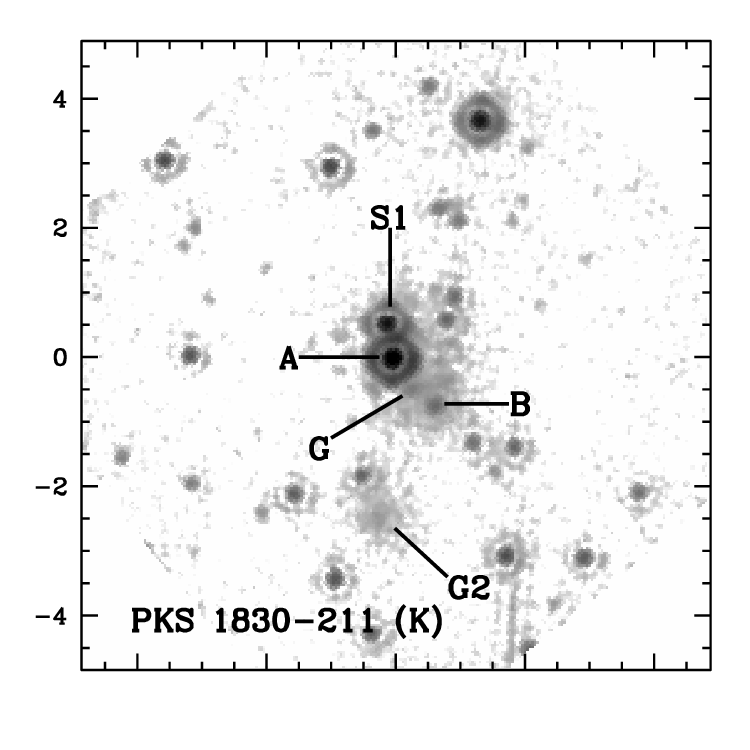,width=3.0in,angle=90.}
  \hspace*{-1.50cm}
  \psfig{figure=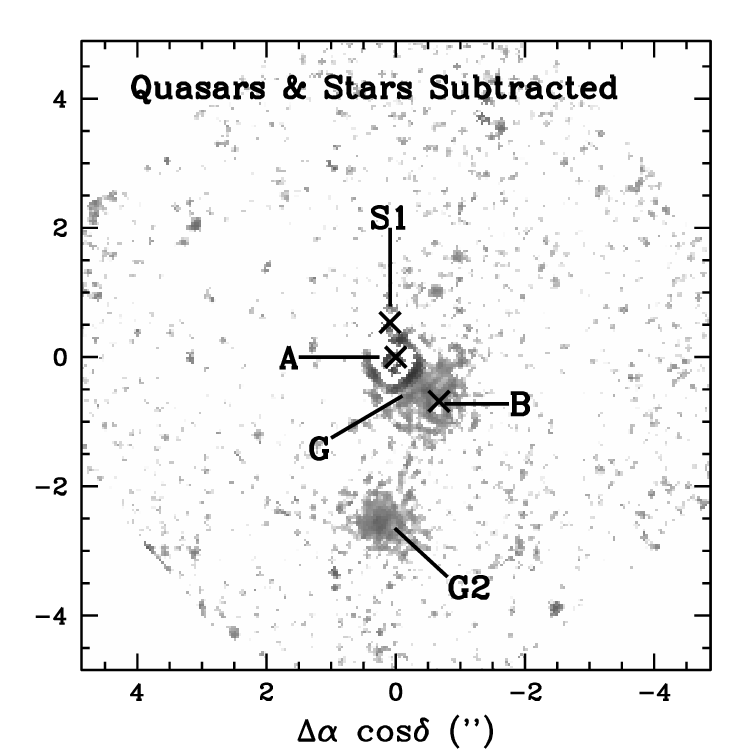,width=3.0in,angle=90.}
  }
\vspace*{-0.7cm}
\centerline{
  \psfig{figure=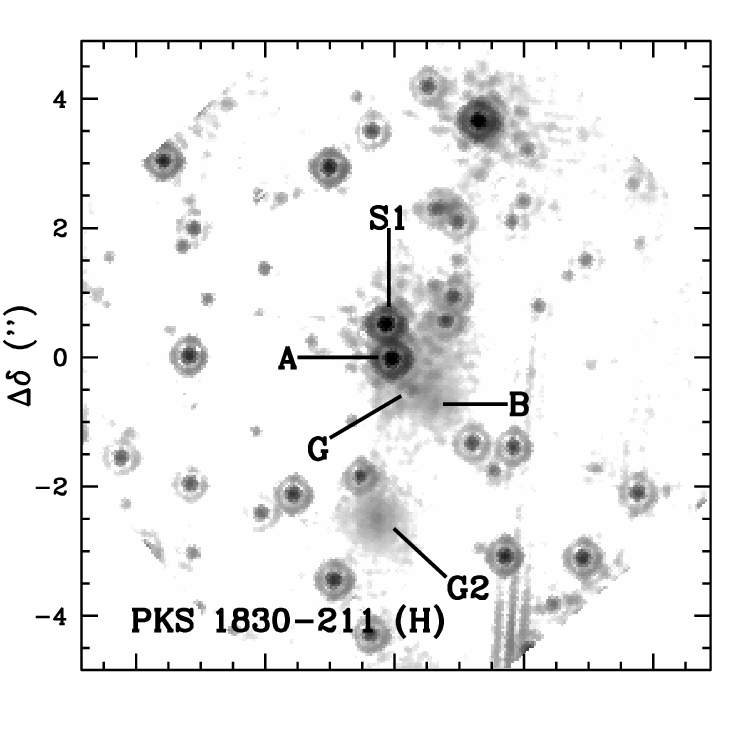,width=3.0in,angle=90.}
  \hspace*{-1.50cm}
  \psfig{figure=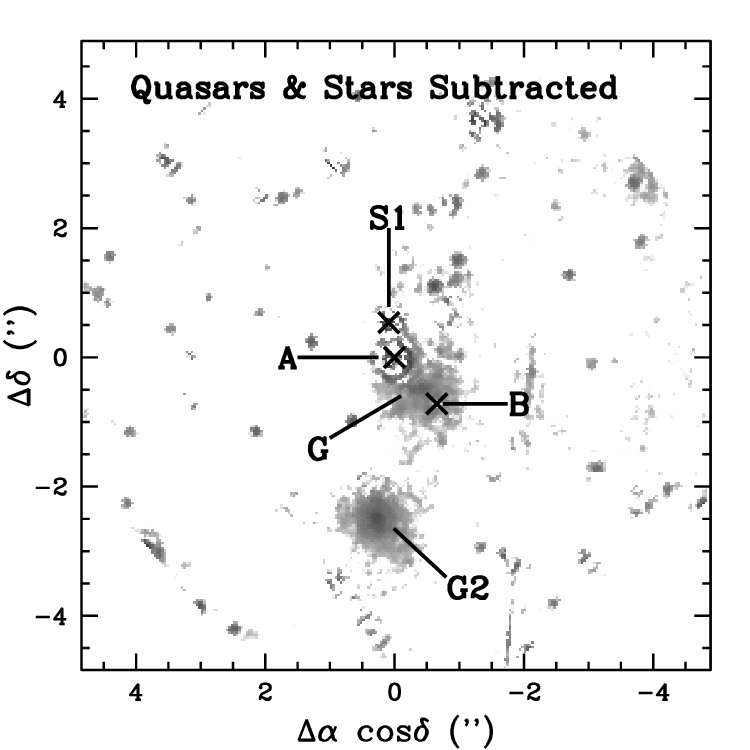,width=3.0in,angle=90.}
  }
 \caption{Continued}
 \end{figure}

 \begin{figure}
 \label{fig-host}
 \figurenum{2}
 \centerline{
   \psfig{figure=fig2b.eps,height=8cm}
   \psfig{figure=fig2d.eps,height=8cm}
   }
 \vspace*{-1.33cm}
 \centerline{
   \psfig{figure=fig2a.eps,height=8cm}
   \psfig{figure=fig2c.eps,height=8cm}
   }
\caption{The quasar host galaxies we detected in BRI~0952--0115 (left)
         and HE~1104--1805 (right).  The top
         panel shows the images after subtracting both the quasar
         images and the lens galaxy, and the bottom panel shows the same
         image after smoothing with a 2-pixel FWHM Gaussian filter. 
         }
 \end{figure}

\begin{figure}
\label{fig-neighbphot}
\figurenum{3}
 \vspace*{-2.4cm}
\centerline{
   \vspace*{ -1.5cm}
   \psfig{figure=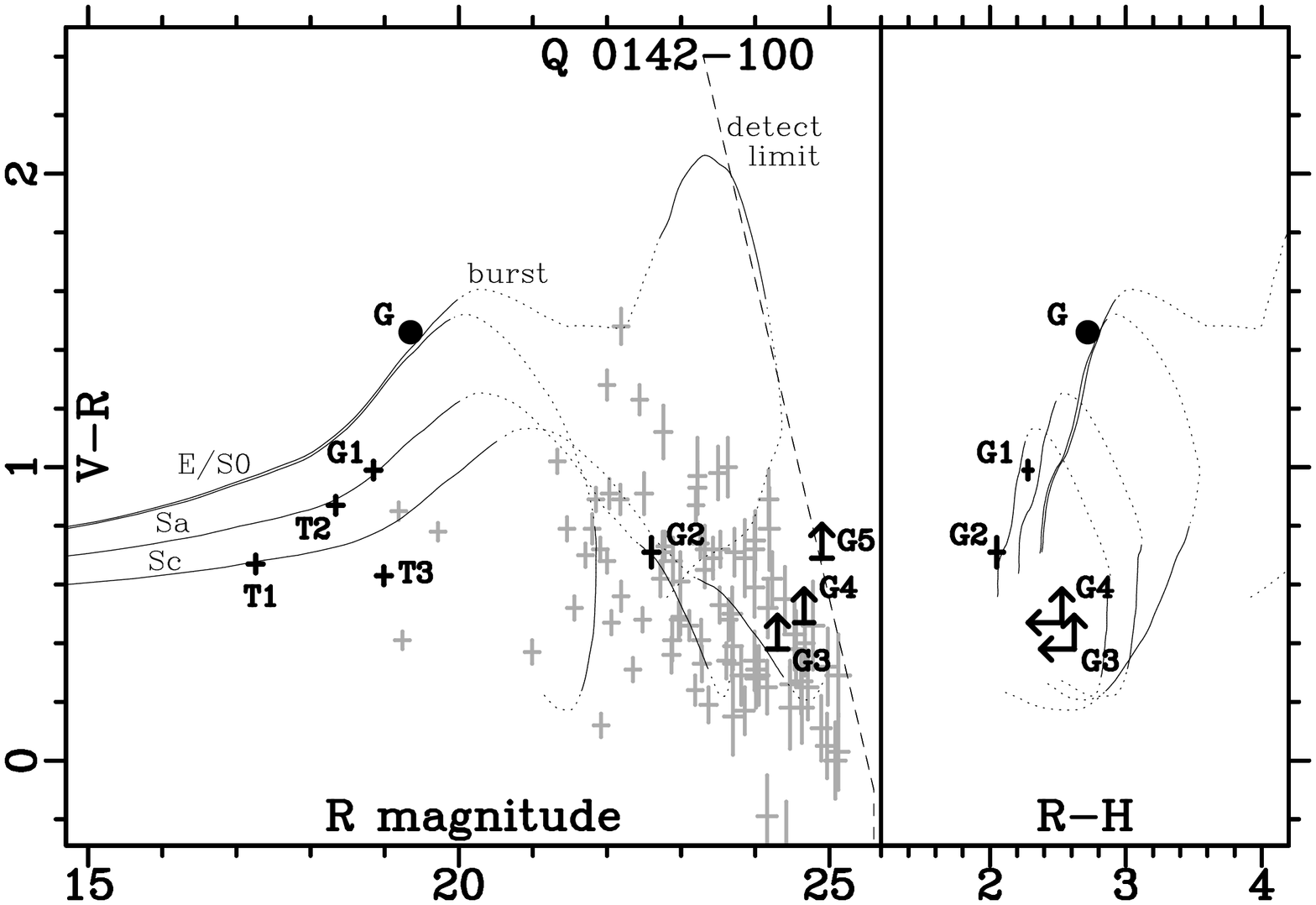,height=3.2in,angle=-90.}}
\centerline{
   \vspace*{ -1.5cm}
   \psfig{figure=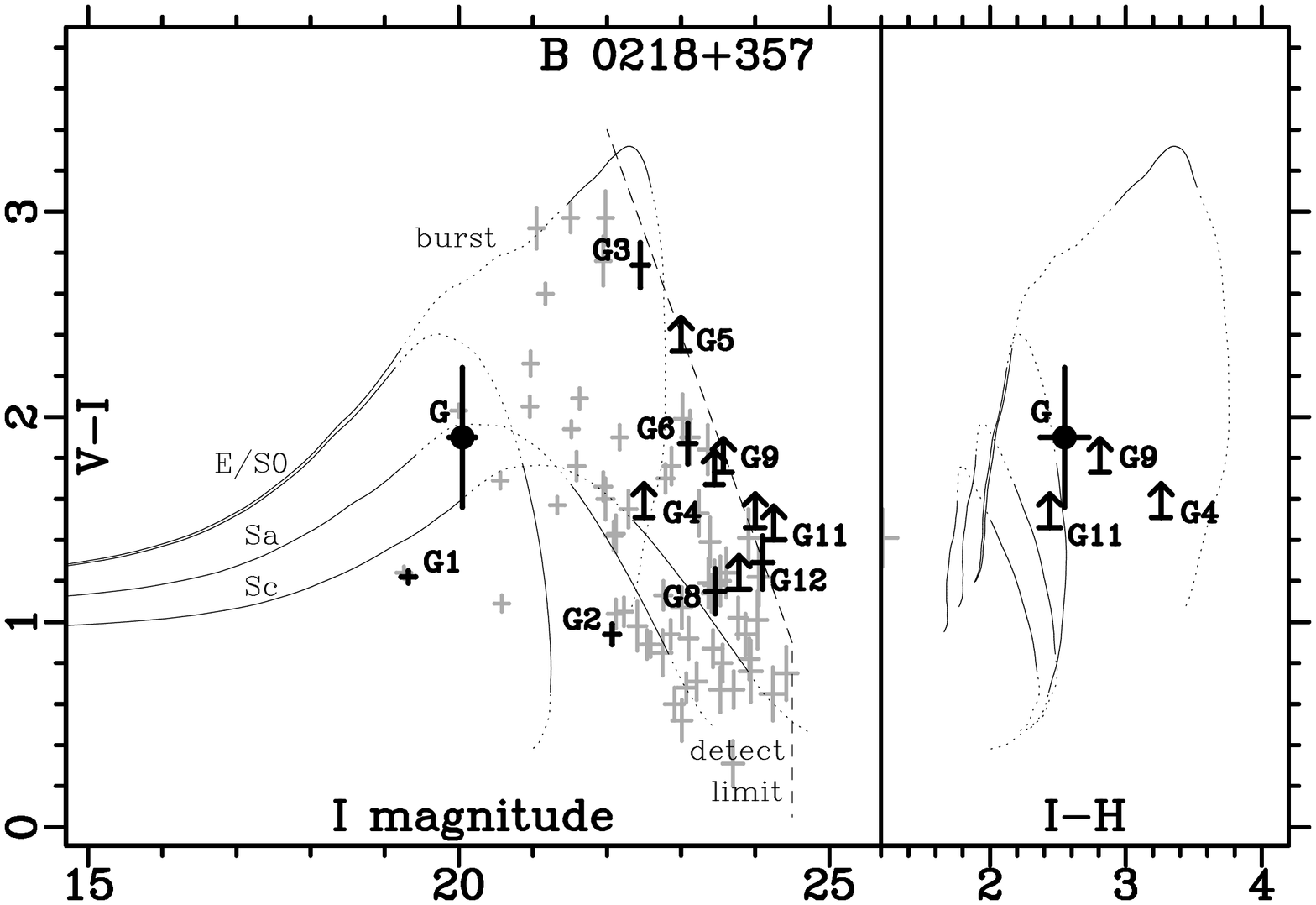,height=3.2in,angle=-90.}}
\centerline{
   \vspace*{ -1.5cm}
   \psfig{figure=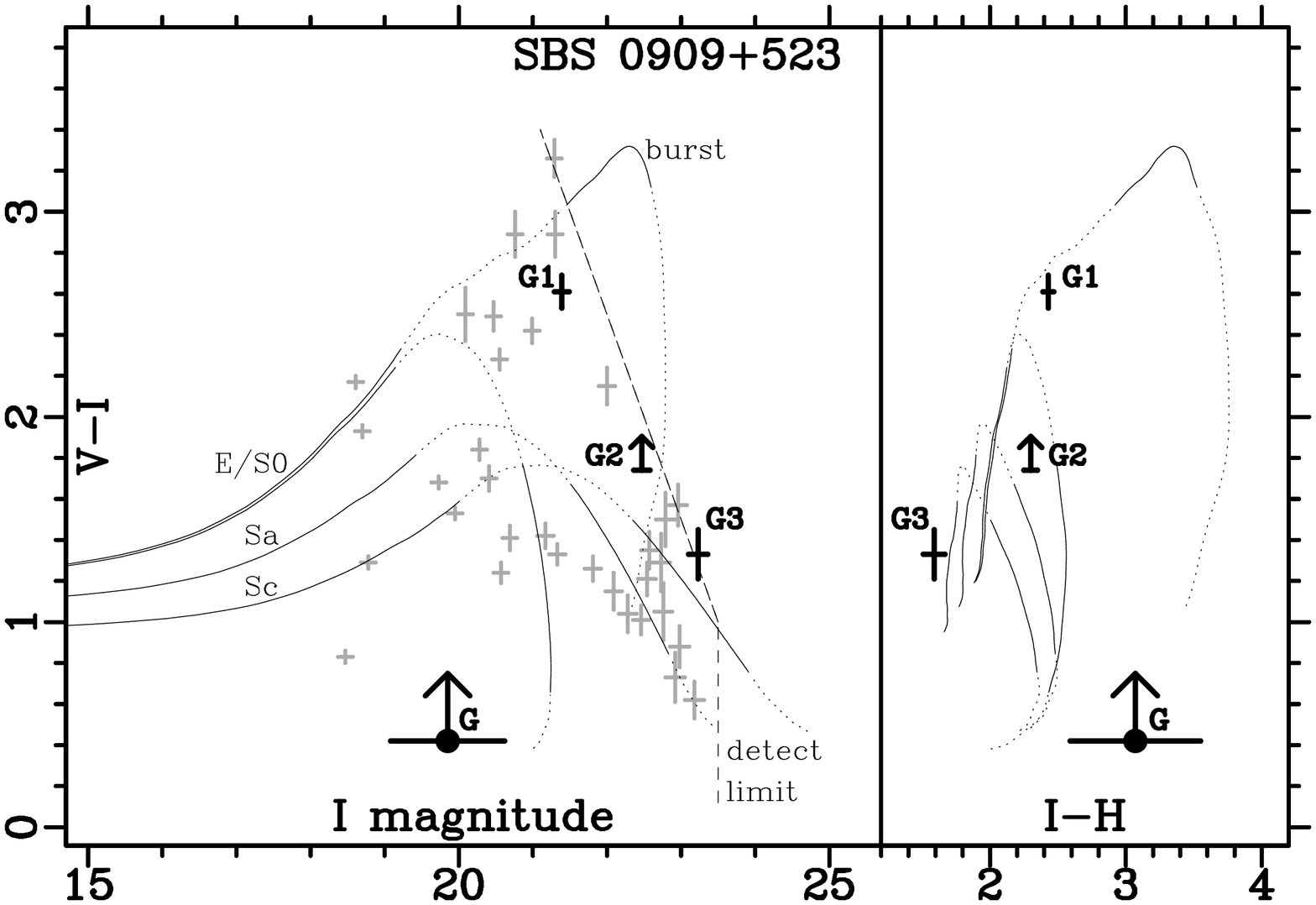,height=3.2in,angle=-90.}}
 \vspace*{0.3cm}
\caption{
   Colors and total magnitudes for the lens galaxies and their neighbors.
   Galaxies outside a radius of $20\arcsec$ are shown in gray. 
   SExtractor total magnitudes are offset-corrected (see~\S2). 
   The curves show photometric evolution models for $L_*$ galaxies,
   alternating between solid and dotted lines 
   at intervals of $\Delta z=0.5$\,. 
   Galaxies with $L<L_*$ should lie on the color-color curve
   which corresponds to their morphology type, 
   but will be shifted to the right in color-magnitude plots.
  }
\end{figure}

\begin{figure}
\label{fig-neighbphot}
\figurenum{3}
 \vspace*{-2.0cm}
\centerline{
   \vspace*{ -1.5cm}
   \psfig{figure=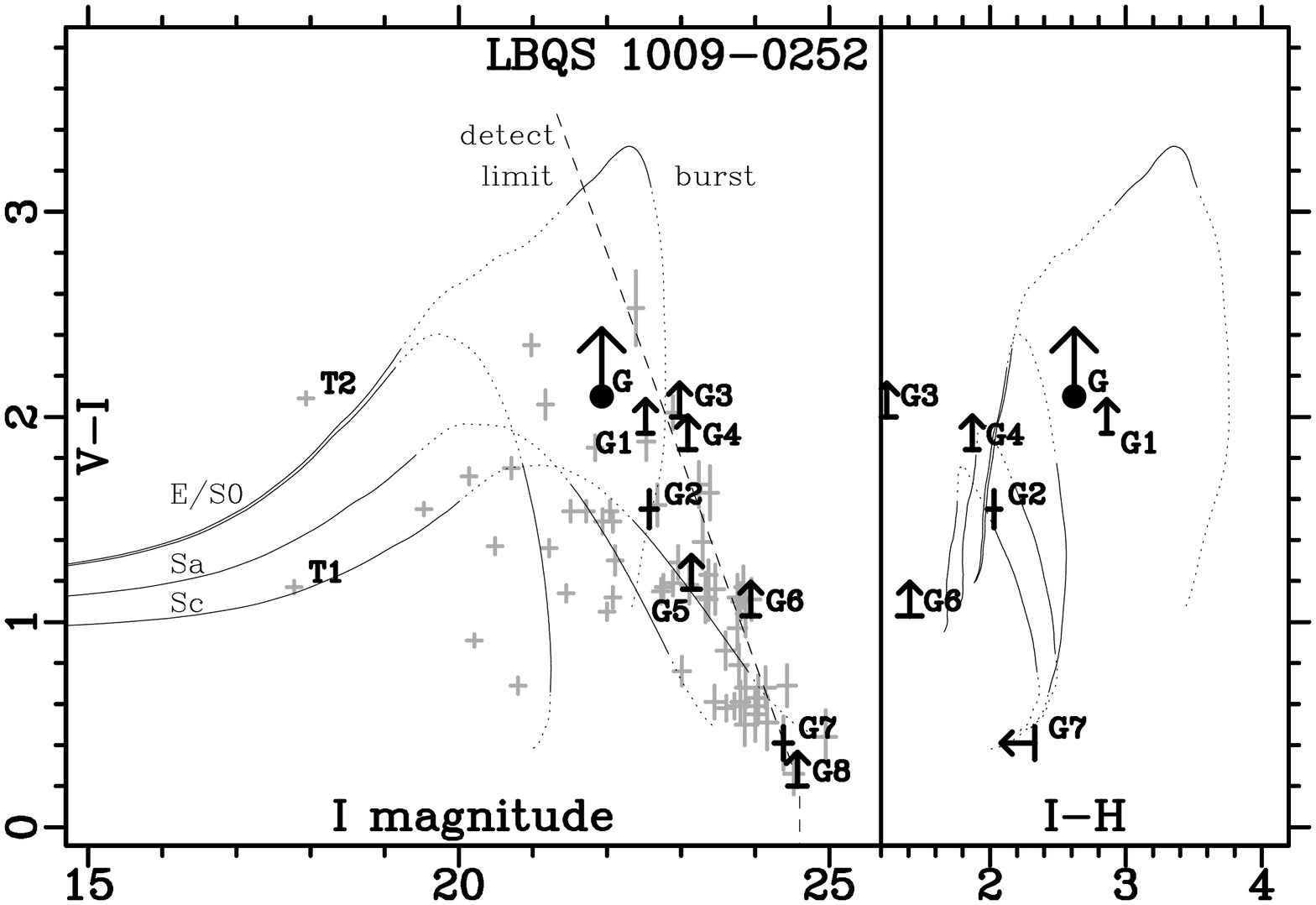,height=3.2in,angle=-90.}}
\centerline{
   \vspace*{ -1.5cm}
   \psfig{figure=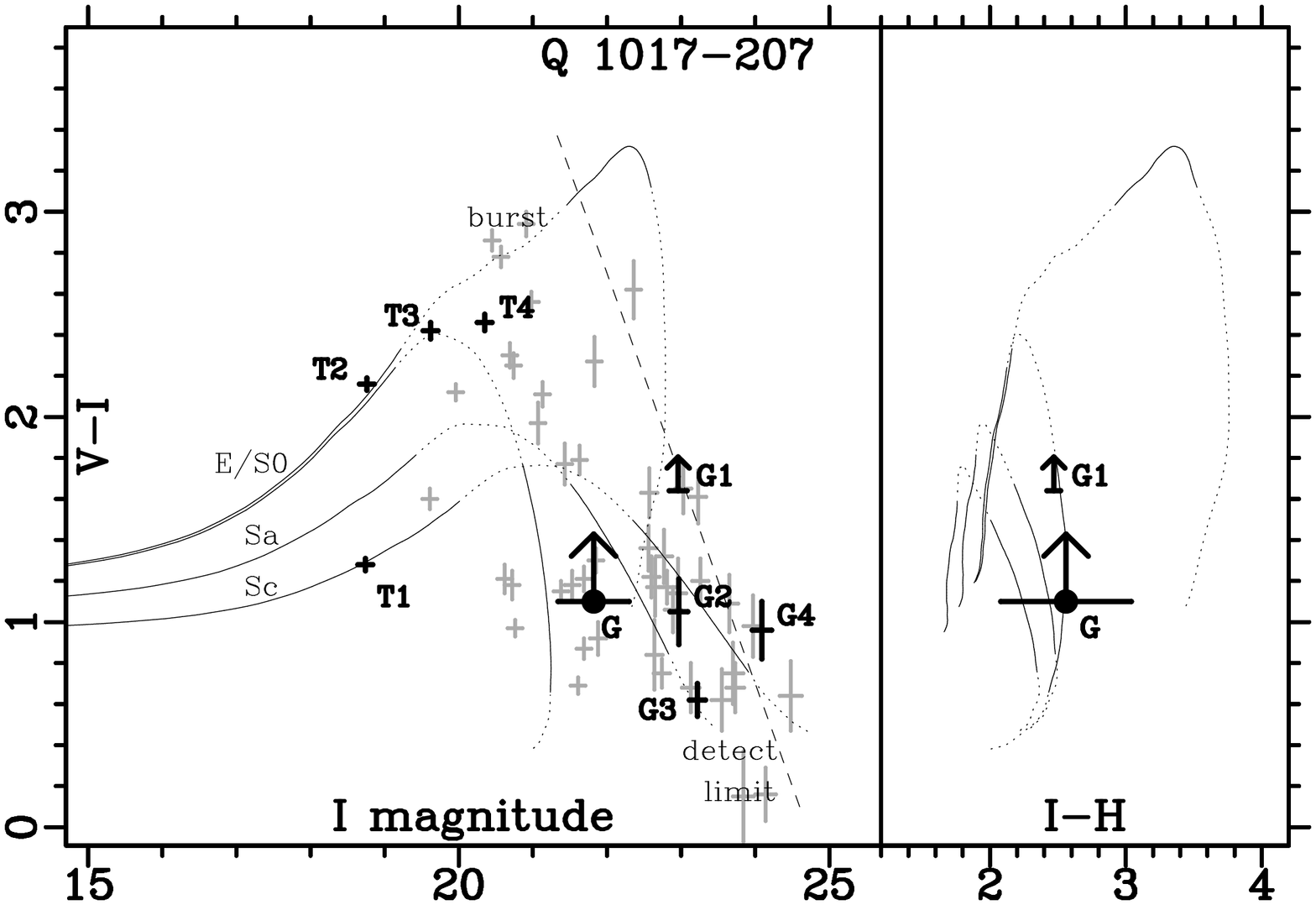,height=3.2in,angle=-90.}}
\centerline{
   \vspace*{ -1.5cm}
   \psfig{figure=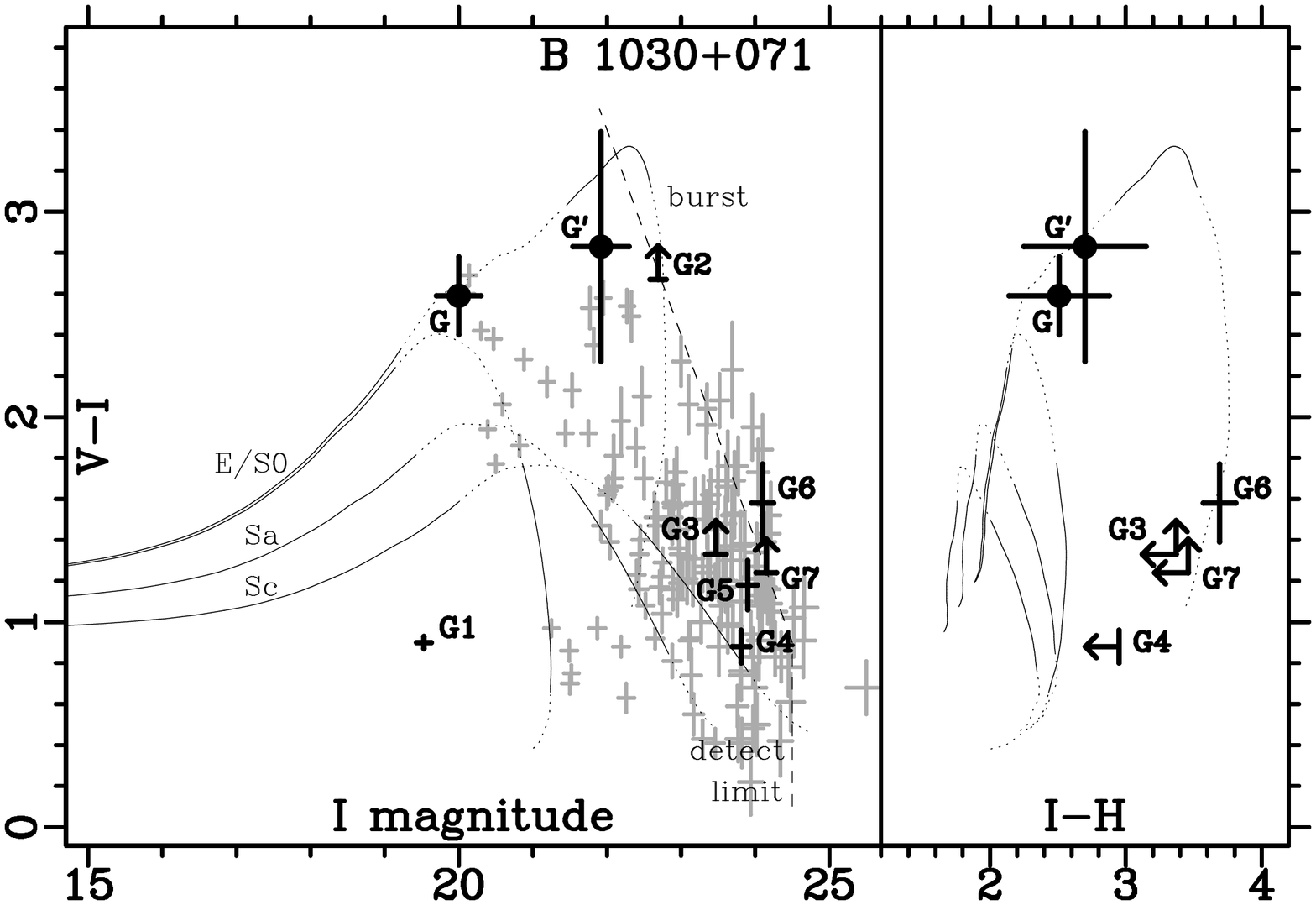,height=3.2in,angle=-90.}}
 \vspace*{0.3cm}
\caption{Continued}
\end{figure}

\begin{figure}
\label{fig-neighbphot}
\figurenum{3}
 \vspace*{-2.0cm}
\centerline{
   \vspace*{-1.5cm}
   \psfig{figure=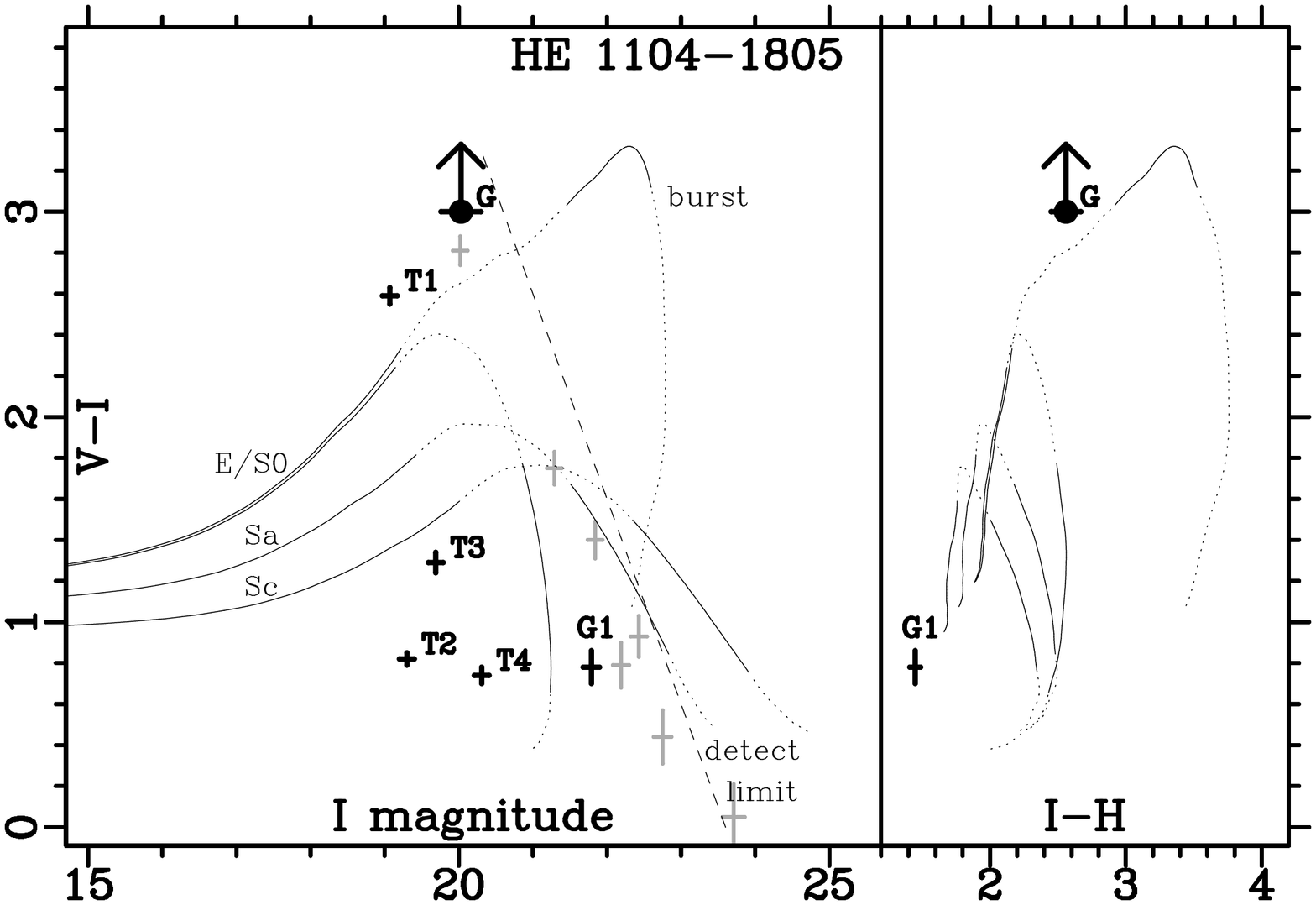,height=3.2in,angle=-90.}}
\centerline{
   \vspace*{ -1.5cm}
   \psfig{figure=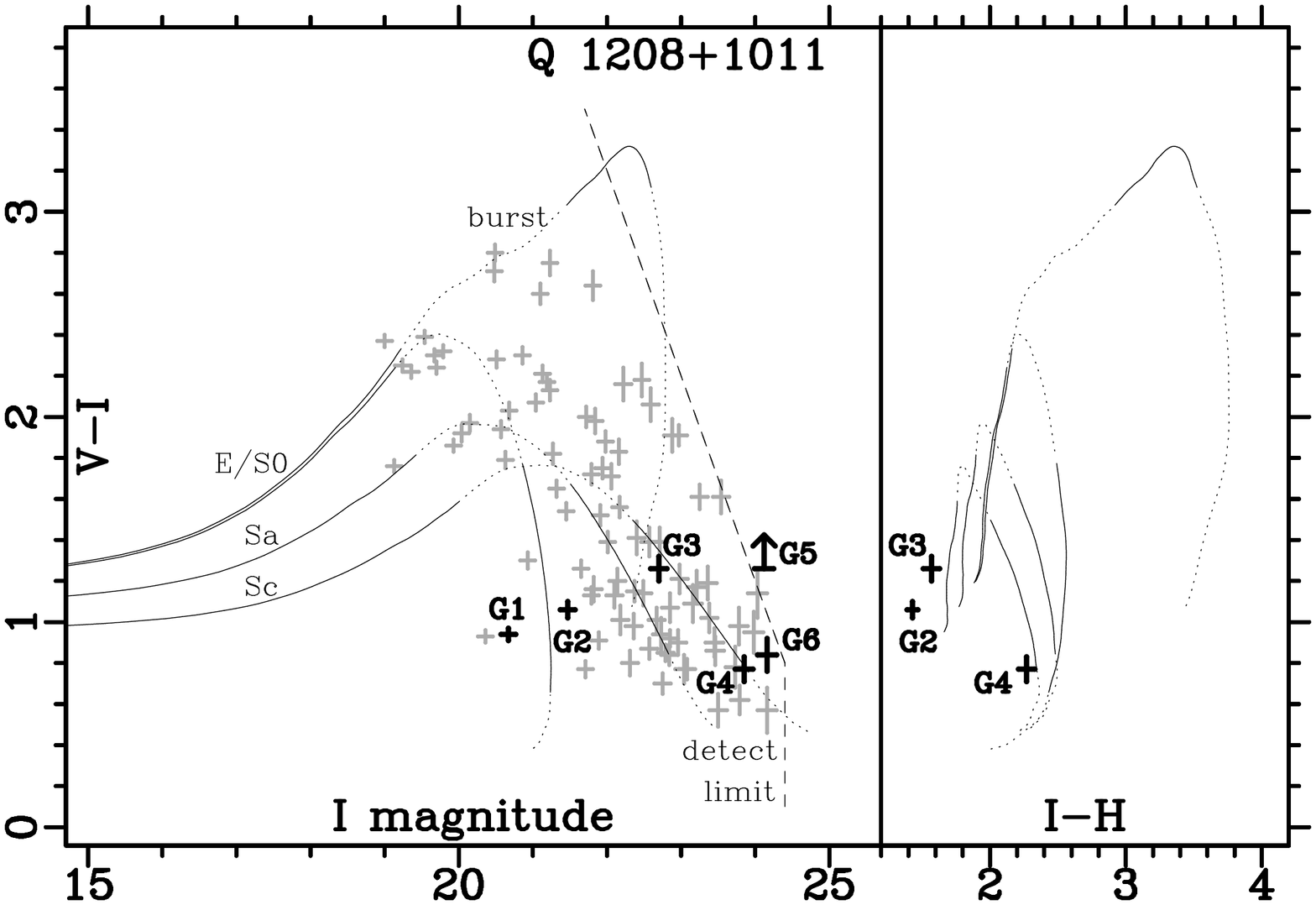,height=3.2in,angle=-90.}}
\centerline{
   \vspace*{ -1.5cm}
   \psfig{figure=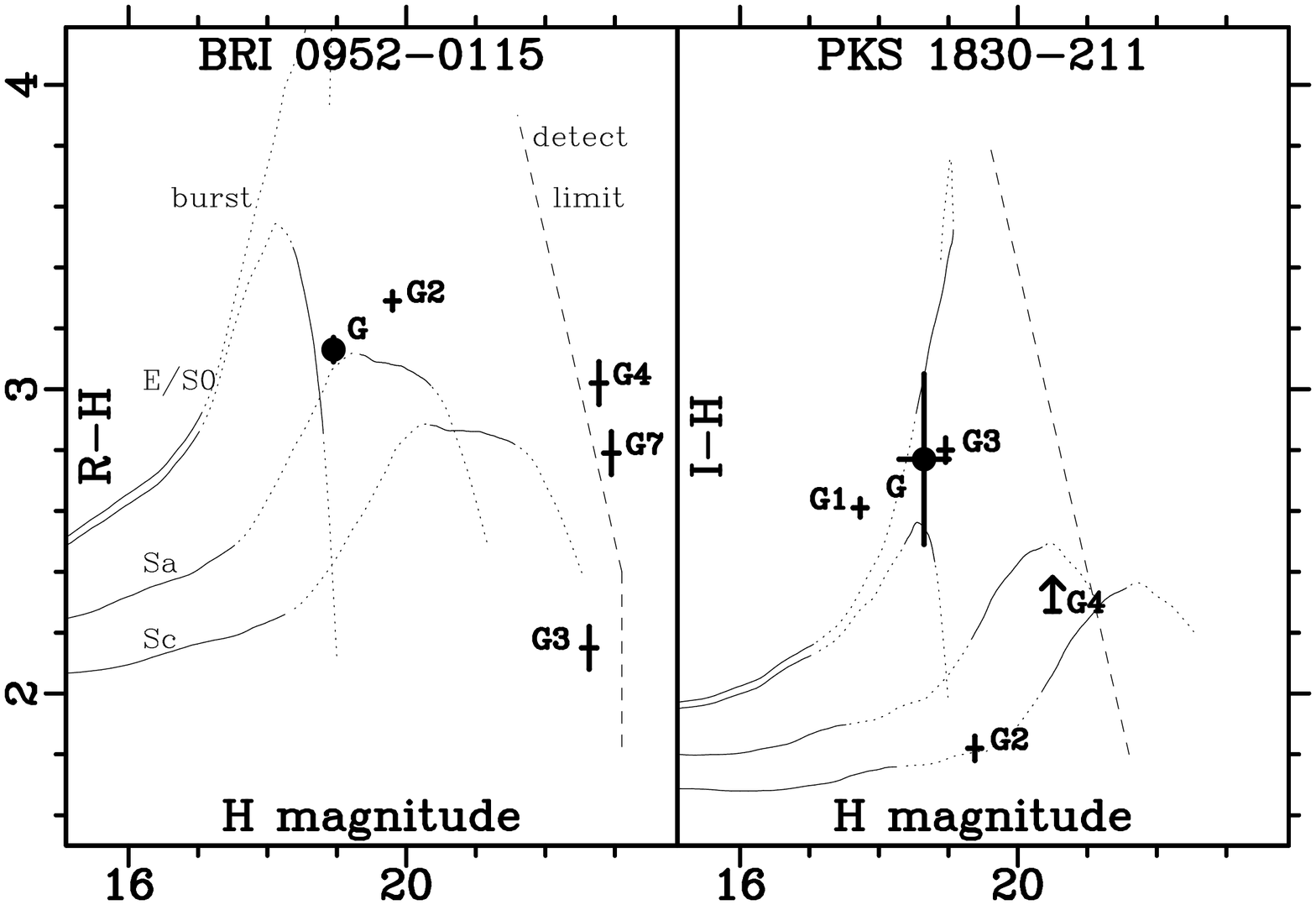,height=3.2in,angle=-90.}}
 \vspace*{0.3cm}
\caption{Continued}
\end{figure}

\begin{figure}
\label{fig-lenscol}
\figurenum{4}
\centerline{ \psfig{figure=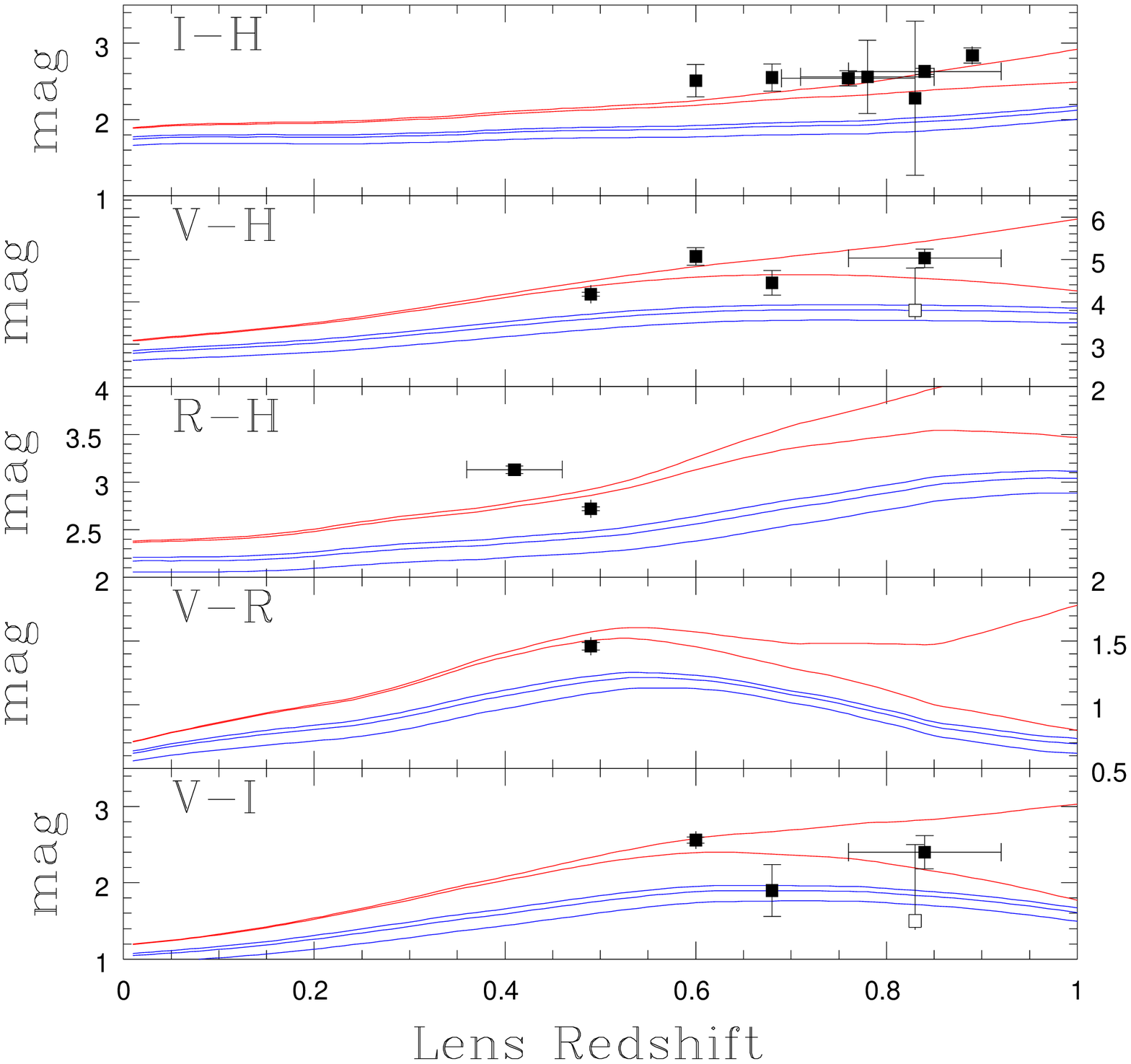,height=5in} }
\caption{ 
   Lens galaxy colors. The curves (from top to bottom) show the
   burst, E/S0, Sa, Sb, and Sc evolution models.  The solid points
   show the measured colors, and the open points show lower limits.
   The redshift error bars are the formal uncertainties in the 
   fundamental plane/photometric redshifts estimated by Kochanek
   et~al.\ (1999).
   }
\end{figure}

\begin{figure}
\figurenum{5}
\centerline{ \psfig{figure=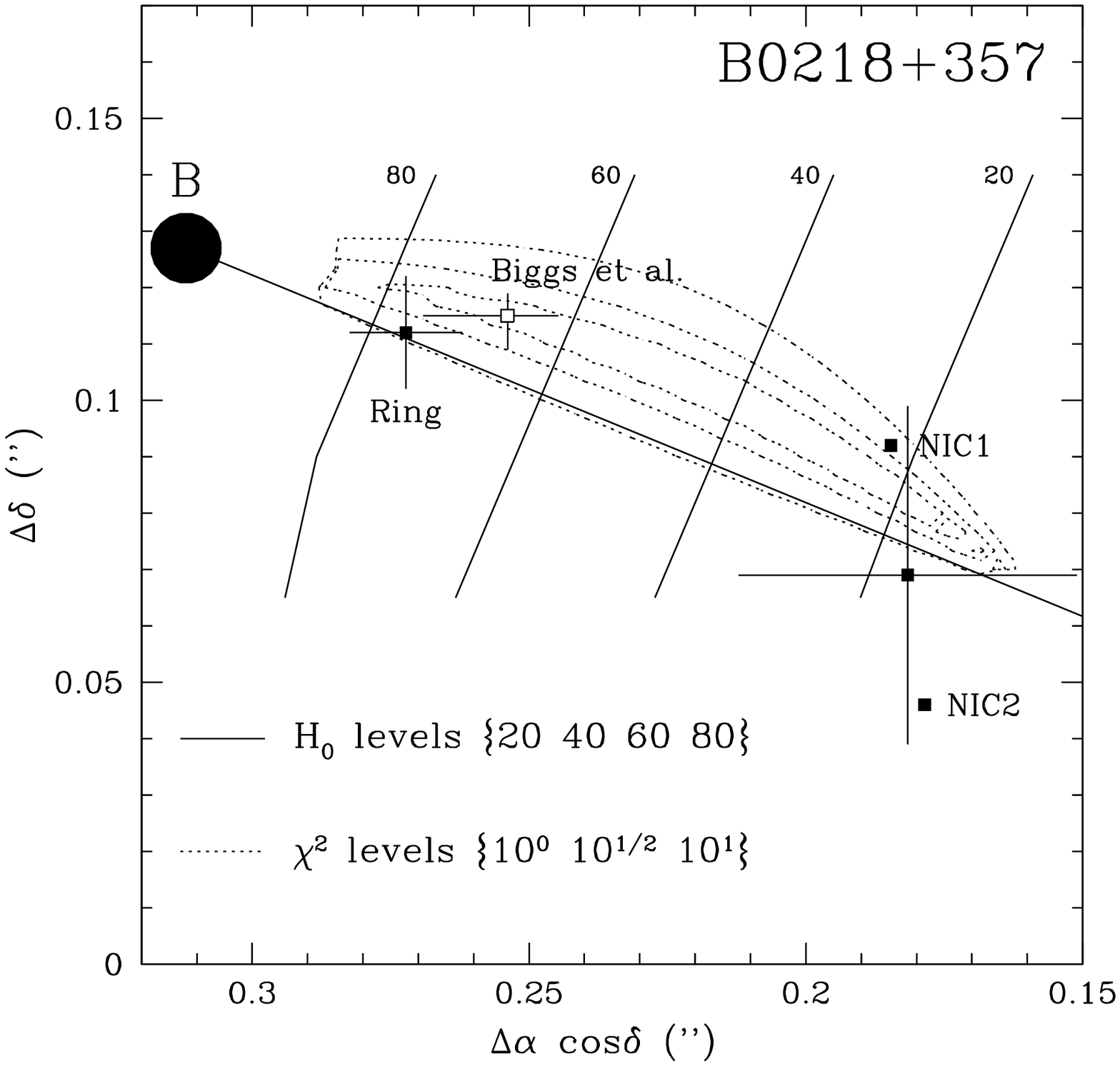,height=3.5in} 
             \psfig{figure=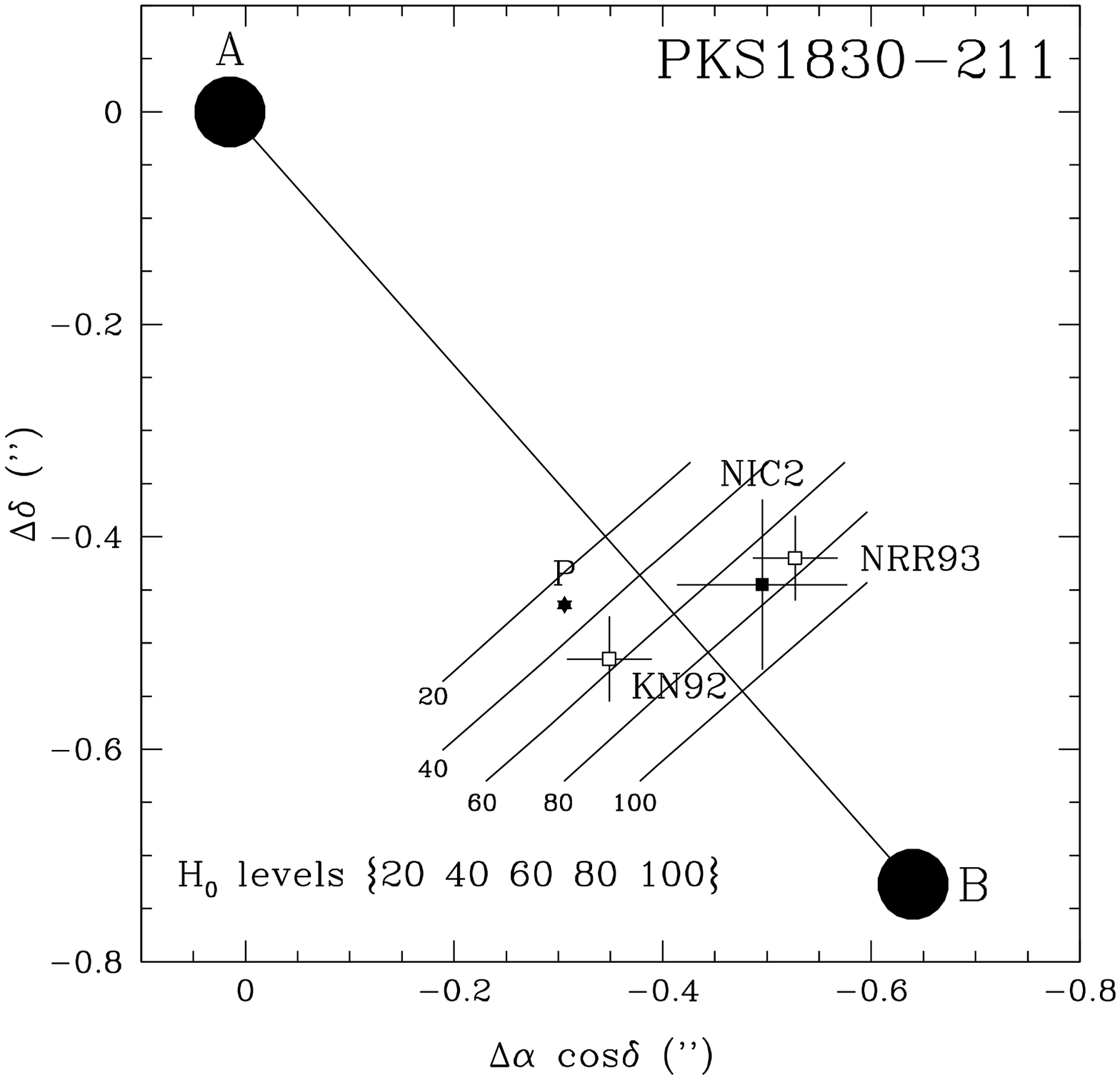,height=3.5in} }
\caption{The variation of $H_0$ with lens galaxy position
in B~0218+357 (left) and PKS~1830--211 (right).  
The solid contours show the variation in $H_0$ with the position 
of the lens given the measured time delays. 
For B~0218+357 the positions derived from the NICMOS observations
are the filled points labeled NIC1 and NIC2, and the point at their 
mean position with the large error bars is the lens position we used 
in our basic models.  The point labeled ``Ring'' marks the estimated 
center of the Einstein ring (Patnaik et~al.\ 1993).  
Using no direct constraints on the lens position,
Biggs et~al.\ (1999) found the position marked by the open point from their
lens models.  Using the same constraints, we find a much broader degeneracy
in the models for the lens position, illustrated with the dashed 
$\chi^2$ contours.  The filled points for PKS~1830--211 mark the positions 
of the lens galaxy (labeled NIC2) and the point-like source P.  The open
points give the lens positions found in the models of Kochanek \& Narayan
(1992, KN92) and Nair et~al.\ (1993, NRR93). }
\end{figure}

\pagestyle{plain}
\begin{deluxetable}{llrlccl}
\tablenum{1}
\label{tab-obslog}
\tablewidth{0pt}
\tablecolumns{7}
\tablecaption{Summary of Observations}
\scriptsize
\tablehead{ Target & Band=Camera/Filt &\multicolumn{1}{c}{Time} 
    &\multicolumn{1}{c}{N/Dith} &\multicolumn{1}{c}{Detect} &UT Date &Source }
\startdata
Q0142--100        & H=NIC2/F160W  & 2560\,s & $4/1\farcs1$ & 22.5 & 1997.08.15 & CASTLES \\
{}                & R=WFPC2/F675W & 2500\,s & $5/$---      & 25.6 & 1994.11.22 & KKF \\
\vspace*{0.06in}
{}                & V=WFPC2/F555W & 1200\,s & $3/$---      & 25.5 & 1994.11.22 & KKF \\
B0218+357         & H=NIC1/F160W  & 7866\,s & $5/$---      & 21.7 & 1997.08.07 & Xanthopoulos et al.\ 1999 \\
{}                & H=NIC2/F160W  & 2560\,s & $4/0\farcs6$ & 23.7 & 1997.08.19 & CASTLES \\
{}                & I=WFPC2/F814W & 1000\,s & $2/$---      & 24.5 & 1995.07.22 & Xanthopoulos et al.\ 1999 \\
\vspace*{0.06in}
{}                & V=WFPC2/F555W & 1000\,s & $2/$---      & 25.4 & 1995.07.22 & Xanthopoulos et al.\ 1999 \\
SBS0909+523       & H=NIC2/F160W  & 2816\,s & $4/7\farcs9$ & 22.9 & 1997.11.07 & CASTLES \\
{}                & I=WFPC2/F814W &  640\,s & $4/1\farcs1$ & 23.5 & 1999.03.18 & CASTLES \\
\vspace*{0.06in}
{}                & V=WFPC2/F555W &  640\,s & $4/1\farcs1$ & 24.5 & 1999.03.17 & CASTLES \\
BRI0952--0115     & H=NIC2/F160W  & 5120\,s & $8/8\farcs0$ & 23.1 & 1997.10.17 & CASTLES \\
\vspace*{0.06in}
{}                & R=WFPC2/F675W & 5400\,s & $3/$---      & 25.5 & 1994.10.22 & KKF \\
LBQS1009--0252    & H=NIC2/F160W  & 2560\,s & $4/7\farcs7$ & 23.6 & 1997.11.15 & CASTLES \\
{}                & I=WFPC2/F814W & 2600\,s & $2/$---      & 24.6 & 1999.01.01 & GO-6790, Surdej \\
\vspace*{0.06in}
{}                & V=WFPC2/F555W & 1600\,s & $4/0\farcs4$ & 24.8 & 1999.01.01 & GO-6790, Surdej \\
Q1017--207        & H=NIC2/F160W  & 2560\,s & $4/8\farcs0$ & 23.3 & 1997.11.14 & CASTLES \\
{}                & I=WFPC2/F814W & 2300\,s & $6/2\farcs3$ & 24.8 & 1995.11.28 & Remy et al.\ 1998 \\
\vspace*{0.06in}
{}                & V=WFPC2/F555W &  800\,s & $5/2\farcs3$ & 24.7 & 1995.11.28 & Remy et al.\ 1998 \\
B1030+071         & H=NIC1/F160W  & 2624\,s & $2/$---      & 20.7 & 1997.11.20 & GO-7255, Jackson \\
{}                & I=WFPC2/F814W & 1000\,s & $2/$---      & 24.5 & 1997.02.03 & Xanthopoulos et al.\ 1998 \\
\vspace*{0.06in}
{}                & V=WFPC2/F555W & 1000\,s & $2/$---      & 25.4 & 1997.02.03 & Xanthopoulos et al.\ 1998 \\
HE1104--1805      & H=NIC2/F160W  & 2560\,s & $4/7\farcs9$ & 22.9 & 1997.11.22 & CASTLES \\
{}                & I=WFPC2/F814W & 1000\,s & $6/2\farcs3$ & 23.8 & 1995.11.19 & Remy et al.\ 1998 \\
\vspace*{0.06in}
{}                & V=WFPC2/F555W &  200\,s & $2/$---      & 23.6 & 1995.11.19 & Remy et al.\ 1998 \\
Q1208+1011        & H=NIC2/F160W  & 2560\,s & $4/7\farcs1$ & 23.0 & 1997.08.15 & CASTLES \\
{}                & I=WFPC2/F814W & 2400\,s & $6/1\farcs0$ & 24.4 & 1999.06.04 & CASTLES \\
\vspace*{0.06in}
{}                & V=WFPC2/F555W & 2400\,s & $6/1\farcs0$ & 25.2 & 1999.06.04 & CASTLES \\
PKS1830--211      & K=NIC2/F205W  & 1408\,s & $8/5\farcs8$ & 21.3 & 1997.10.31 & CASTLES \\
{}                & H=NIC2/F160W  & 3072\,s & $8/5\farcs8$ & 22.1 & 1997.10.31 & CASTLES \\
{}                & I=WFPC2/F814W & 1600\,s & $4/1\farcs0$ & 23.2 & 1999.05.12 & CASTLES \\
\enddata

\tablecomments{
   ``N/Dith'' refers to the number of exposures and largest dither separation.
   The detection limit is the magnitude at which 50\% of randomly
   added $0\farcs3$ FWHM Gaussians are found by SExtractor (see~\S2). 
   }
\end{deluxetable}

\def\heada#1#2#3#4{
      \hline
      #1 & ID & \multicolumn{1}{c}{RA($''$)} & \multicolumn{1}{c}{DEC($''$)}
           &\multicolumn{1}{c}{#2} & \multicolumn{1}{c}{#3} & \multicolumn{1}{c}{#4} 
           & \multicolumn{1}{c}{$\gamma_T$} & \multicolumn{1}{c}{PA ($^\circ$)} &Comments \\
      \hline }
\def\headc#1#2#3#4{
       \hline\hline
       \headb{#1}{#2}{#3}{#4}
       \tablebreak
       }
\def\headb#1#2#3#4{\tablehead{
                      \colhead{#1} &
                      \colhead{ID} &
                      \colhead{RA($''$)} &
                      \colhead{DEC($''$)} &
                      \colhead{#2} &
                      \colhead{#3} &
                      \colhead{#4} &
                      \colhead{$\gamma_T$} &
                      \colhead{PA ($^\circ$)} &
                      \colhead{Comments}
                      } }
\begin{deluxetable}{lcrrccccrl}
\tablenum{2}
\label{tab-nearby}
\scriptsize
\tablecaption{Nearby Objects}
\tablewidth{0pt}
\headb{Q0142--100}{R (mag)}{V--R (mag)}{R--H (mag)}
\startdata
 &G1   &    2.9 & --10.1 & $18.95\pm0.10$ &  $0.99\pm0.03$ &  $2.28\pm0.03$ &  0.079 &  --7 & \\
 &G2   &    5.8 &  --2.5 & $22.70\pm0.10$ &  $0.71\pm0.05$ &  $2.05\pm0.05$ &  0.030 & --64 & \\
 &S1   & --16.7 &  --9.8 & $23.98\pm0.11$ &  $0.17\pm0.06$ &                &        &      & \\
 &G3   &    4.5 &  --1.5 & $24.40\pm0.14$ &     $>0.38$    &     $<2.62$    &  0.021 & --72 & \\
 &G4   &  --1.3 &    9.5 & $24.76\pm0.13$ &     $>0.47$    &     $<2.53$    &  0.005 & --17 & \\
 &G5   &    7.6 &   11.0 & $25.00\pm0.14$ &     $>0.69$    &                &  0.004 &   27 & \\
 &T1   & --75.0 & --13.2 & $17.36\pm0.10$ &  $0.67\pm0.03$ &                &  0.020 &   81 & \\
 &T2   & --53.1 &   19.9 & $18.44\pm0.10$ &  $0.87\pm0.03$ &                &  0.016 & --70 & \\
 &T3   & --30.1 & --58.5 & $19.09\pm0.10$ &  $0.63\pm0.03$ &                &  0.011 &   29 & \\
\heada{B0218+357}{I (mag)}{V--I (mag)}{I--H (mag)}
 &G1   & --11.7 & --12.9 & $19.42\pm0.10$ &  $1.22\pm0.03$ &                &  0.005 &   42 & \\
 &G2   &   17.5 &  --8.9 & $22.17\pm0.10$ &  $0.94\pm0.05$ &                &  0.001 & --62 & \\
 &G3   &    2.9 &   15.8 & $22.55\pm0.11$ &  $2.74\pm0.11$ &                &  0.001 &   10 & \\
 &G4   &  --7.3 &  --5.1 & $22.60\pm0.12$ &     $>1.51$    &  $3.26\pm0.07$ &  0.002 &   55 & \\
 &G5   & --10.2 &  --1.9 & $23.10\pm0.12$ &     $>2.32$    &                &  0.002 &   79 & \\
 &G6   &    3.8 &    8.8 & $23.19\pm0.11$ &  $1.87\pm0.10$ &                &  0.002 &   23 & \\
 &G7   & --10.3 & --11.2 & $23.55\pm0.12$ &     $>1.67$    &                &  0.001 &   43 & \\
 &G8   &   15.2 &    0.9 & $23.56\pm0.13$ &  $1.15\pm0.11$ &                &  0.001 &   87 & \\
 &G9   &    2.0 &    4.1 & $23.67\pm0.12$ &     $>1.73$    &  $2.81\pm0.07$ &  0.003 &   24 & \\
 &G10  &  --8.5 & --12.4 & $23.88\pm0.15$ &     $>1.16$    &                &  0.001 &   35 & \\
 &G11  &    6.5 &    2.4 & $24.10\pm0.13$ &     $>1.46$    &  $2.44\pm0.08$ &  0.002 &   70 & \\
 &G12  & --13.1 &    7.6 & $24.20\pm0.14$ &  $1.29\pm0.13$ &                &  0.001 & --60 & \\
 &G13  &    2.5 &   14.2 & $24.35\pm0.15$ &     $>1.40$    &                &  0.001 &    9 & \\
 &H1   &  --4.1 &  --6.4 &                &                &     $>1.69$    &  0.001 &   33 & H=$22.13\pm0.12$ \\
 &H2   &    6.6 &  --0.5 &                &                &     $>0.91$    &  0.001 & --85 & H=$23.59\pm0.16$ \\
\heada{SBS0909+523}{I (mag)}{V--I (mag)}{I--H (mag)}
 &S1   &   13.7 &    7.0 & $17.18\pm0.10$ &  $1.86\pm0.03$ &                &        &      & \\
 &G1   &  --1.7 &  --1.7 & $21.49\pm0.10$ &  $2.61\pm0.08$ &  $2.43\pm0.04$ &  0.037 &   51 & \\
 &G2   &    3.7 &  --0.2 & $22.57\pm0.11$ &     $>1.74$    &  $2.30\pm0.05$ &  0.019 & --87 & \\
 &S2   &    4.3 &  --4.0 & $23.09\pm0.11$ &     $>1.63$    &  $2.67\pm0.05$ &        &      & \\
 &G3   &  --7.0 &  --5.3 & $23.33\pm0.13$ &  $1.33\pm0.12$ &  $1.59\pm0.08$ &  0.005 &   54 & \\
 &H1   &  --7.3 &  --5.5 &                &                &     $>1.59$    &  0.003 &   55 & H=$21.54\pm0.11$ \\
 &H2   &    4.4 &    3.9 &                &                &     $>1.69$    &  0.004 &   46 & H=$21.80\pm0.10$ \\
 &H3   &   11.4 &  --3.7 &                &                &     $>1.21$    &  0.002 & --71 & H=$21.83\pm0.11$ \\
 &H4   &    5.3 &    9.4 &                &                &     $>1.67$    &  0.002 &   28 & H=$21.84\pm0.11$ \\
 &H5   &    7.5 &    0.3 &                &                &     $>1.02$    &  0.003 &   87 & H=$21.98\pm0.11$ \\
\hline
\enddata
\end{deluxetable}
\begin{deluxetable}{lcrrccccrl}
\tablenum{2}
\scriptsize
\tablecaption{Continued}
\tablewidth{0pt}
\headb{BRI0952--0115}{R (mag)}{R--H (mag)}{\hphantom{R--H (mag)}}
\startdata
 &G1   &   15.6 &   10.5 & $21.80\pm0.10$ &                &  &  0.016 &   55 & \\
 &G2   &    6.9 &   12.6 & $22.96\pm0.10$ &  $3.29\pm0.03$ &  &  0.012 &   29 & \\
 &G3   &    4.0 &    0.2 & $24.19\pm0.11$ &  $1.71\pm0.06$ &  &  0.023 &   81 & \\
 &G4   &  --0.1 & --12.0 & $24.59\pm0.12$ &  $3.02\pm0.07$ &  &  0.007 &  --2 & \\
 &G5   &   15.5 &  --3.9 & $24.73\pm0.12$ &                &  &  0.005 & --78 & \\
 &G6   &    4.8 &  --9.5 & $24.84\pm0.15$ &                &  &  0.007 & --30 & \\
 &G7   &   10.9 &  --4.9 & $24.91\pm0.12$ &  $2.79\pm0.07$ &  &  0.006 & --69 & \\
 &T1   &   38.0 & --26.3 & $19.38\pm0.10$ &                &  &  0.020 & --56 & \\
 &T2   & --12.2 & --31.3 & $20.70\pm0.10$ &                &  &  0.016 &   21 & \\
 &T3   &   26.9 &    9.4 & $21.40\pm0.10$ &                &  &  0.013 &   70 & \\
\heada{LBQS1009--0252}{H (mag)}{V--I (mag)}{I--H (mag)}
 &S1=C &  --4.3 &    1.7 & $18.69\pm0.10$ &  $0.90\pm0.03$ &  $0.85\pm0.03$ &$\sim0.08$ & --53 & Shear from C$'$\\
 &S2   &  --0.9 &  --7.7 & $22.44\pm0.10$ &  $1.60\pm0.05$ &  $1.10\pm0.04$ &        &      & \\
 &G1   &  --6.8 &  --7.4 & $22.62\pm0.10$ &     $>1.92$    &  $2.86\pm0.04$ &  0.034 &   44 & \\
 &G2   &    6.0 &    8.5 & $22.67\pm0.11$ &  $1.55\pm0.09$ &  $2.03\pm0.05$ &  0.025 &   34 & \\
 &G3   &  --4.2 &    7.5 & $23.08\pm0.11$ &     $>2.00$    &  $1.24\pm0.07$ &  0.026 & --23 & \\
 &G4   &  --5.3 &   11.4 & $23.19\pm0.11$ &     $>1.84$    &  $1.87\pm0.05$ &  0.017 & --21 & \\
 &G5   &   10.3 & --15.9 & $23.24\pm0.12$ &     $>1.16$    &                &  0.012 & --36 & \\
 &G6   &  --4.3 &  --5.9 & $24.04\pm0.12$ &     $>1.03$    &  $1.41\pm0.09$ &  0.025 &   38 & \\
 &G7   &  --5.0 & --12.3 & $24.48\pm0.12$ &  $0.41\pm0.08$ &     $<2.33$    &  0.011 &   22 & \\
 &G8   &    8.4 & --10.7 & $24.67\pm0.13$ &     $>0.20$    &                &  0.009 & --43 & \\
 &H1   &    3.9 &  --2.4 &                &                &     $>1.18$    &  0.023 & --73 & H=$22.60\pm0.12$ \\
 &T1   &   26.0 &   24.4 & $17.88\pm0.10$ &  $1.17\pm0.03$ &                &  0.072 &   46 & \\
 &T2   &  101.8 & --64.4 & $18.04\pm0.10$ &  $2.09\pm0.03$ &                &  0.020 & --58 & \\
\heada{Q1017--207}{I (mag)}{V--I (mag)}{I--H (mag)}
 &S1   &    9.3 &    1.7 & $19.67\pm0.10$ &  $1.59\pm0.03$ &  $1.92\pm0.03$ &        &      & \\
 &S2   &    5.5 &   14.8 & $22.90\pm0.10$ &     $>1.89$    &    $<-0.49$    &        &      & \\
 &G1   &    3.7 &    2.6 & $23.06\pm0.12$ &     $>1.64$    &  $2.47\pm0.05$ &  0.026 &   59 & \\
 &G2   &    5.7 & --13.5 & $23.07\pm0.12$ &  $1.05\pm0.16$ &     $<0.73$    &  0.009 & --25 & \\
 &G3   &    6.7 & --14.2 & $23.32\pm0.11$ &  $0.62\pm0.08$ &                &  0.007 & --27 & \\
 &G4   &    2.6 &   14.9 & $24.19\pm0.13$ &  $0.96\pm0.14$ &                &  0.005 &   12 & \\
 &H1   &    2.4 &  --7.9 &                &                &     $>1.42$    &  0.005 & --21 & H=$22.51\pm0.13$ \\
 &T1   &  --5.0 & --43.1 & $18.84\pm0.10$ &  $1.28\pm0.03$ &                &  0.021 &    6 & \\
 &T2   &  --9.1 & --52.6 & $18.86\pm0.10$ &  $2.16\pm0.03$ &                &  0.017 &    9 & \\
 &T3   &    4.0 & --37.2 & $19.72\pm0.10$ &  $2.42\pm0.04$ &                &  0.017 &  --7 & \\
 &T4   &   18.8 & --16.1 & $20.45\pm0.10$ &  $2.46\pm0.04$ &                &  0.018 & --50 & \\
\hline
\enddata
\end{deluxetable}
\begin{deluxetable}{lcrrccccrl}
\tablenum{2}
\label{tab-nearby}
\scriptsize
\tablecaption{Continued}
\tablewidth{0pt}
\headb{B1030+071}{I (mag)}{V--I (mag)}{I--H (mag)}
\startdata
 &G1   &  --8.9 &    7.4 & $19.63\pm0.10$ &  $0.90\pm0.03$ &                &  0.044 & --49 & \\
 &G2   & --10.5 &  --8.9 & $22.79\pm0.11$ &     $>2.67$    &                &  0.010 &   55 & \\
 &G3   &  --1.1 &    2.4 & $23.57\pm0.14$ &     $>1.33$    &     $<3.37$    &  0.023 & --30 & \\
 &G4   &  --2.5 &    3.4 & $23.91\pm0.12$ &  $0.88\pm0.08$ &     $<2.95$    &  0.014 & --37 & \\
 &G5   &    0.3 &    5.6 & $24.00\pm0.13$ &  $1.18\pm0.12$ &                &  0.011 &  --5 & \\
 &G6   &    2.7 &  --4.9 & $24.20\pm0.15$ &  $1.58\pm0.19$ &  $3.69\pm0.12$ &  0.016 & --26 & \\
 &G7   &    5.8 &  --2.9 & $24.25\pm0.14$ &     $>1.24$    &     $<3.46$    &  0.013 & --70 & \\
\heada{HE1104--1805}{I (mag)}{V--I (mag)}{\hphantom{I--H (mag)}}
 &S1   &  --3.6 & --15.1 & $17.51\pm0.10$ &  $1.36\pm0.03$ &    $<-5.42$    &        &      & \\
 &G1   &  --3.1 &    4.6 & $21.89\pm0.11$ &  $0.78\pm0.08$ &  $1.45\pm0.04$ &  0.048 & --38 & \\
 &T1   &   73.2 & --43.7 & $19.17\pm0.10$ &  $2.59\pm0.04$ &                &  0.013 & --59 & \\
 &T2   &  --4.3 & --59.5 & $19.40\pm0.10$ &  $0.82\pm0.03$ &                &  0.017 &    5 & \\
 &T3   & --28.5 & --42.5 & $19.79\pm0.10$ &  $1.29\pm0.05$ &                &  0.016 &   35 & \\
 &T4   &   22.9 & --42.1 & $20.41\pm0.10$ &  $0.74\pm0.04$ &                &  0.013 & --28 & \\
\heada{Q1208+1011}{I (mag)}{V--I (mag)}{I--H (mag)}
 &G1   &  --3.8 &   18.8 & $20.77\pm0.10$ &  $0.94\pm0.03$ &                &  0.037 & --11 & \\
 &G2   &  --0.5 &  --9.3 & $21.57\pm0.10$ &  $1.06\pm0.04$ &  $1.43\pm0.04$ &  0.054 &    3 & \\
 &G3   &  --6.9 &  --7.7 & $22.80\pm0.11$ &  $1.26\pm0.06$ &  $1.57\pm0.06$ &  0.028 &   43 & \\
 &G4   & --12.4 &    2.7 & $23.95\pm0.12$ &  $0.77\pm0.06$ &  $2.27\pm0.06$ &  0.013 & --77 & \\
 &G5   &    9.1 &   16.7 & $24.22\pm0.13$ &     $>1.26$    &                &  0.008 &   28 & \\
 &G6   &  --4.5 &   14.5 & $24.26\pm0.13$ &  $0.84\pm0.08$ &                &  0.009 & --17 & \\
 &H1   &  --8.6 &    0.3 &                &                &     $>1.58$    &  0.020 & --86 & H=$21.34\pm0.11$ \\
\heada{PKS1830--211}{H (mag)}{I--H (mag)}{H--K (mag)}
 &S1   &    0.1 &    0.5 & $17.43\pm0.10$ &  $2.41\pm0.03$ &  $0.55\pm0.03$ &        &      & \\
 &G1   &    2.9 & --13.6 & $17.83\pm0.10$ &  $2.61\pm0.03$ &  $0.84\pm0.03$ &  0.032 & --15 & \\
 &G3   & --11.6 &  --1.7 & $19.06\pm0.10$ &  $2.80\pm0.04$ &  $0.88\pm0.03$ &  0.022 &   84 & \\
 &G2   &    0.2 &  --2.5 & $19.48\pm0.10$ &  $1.82\pm0.04$ &  $0.70\pm0.04$ &  0.094 & --20 & \\
 &G4   & --12.0 &  --3.7 & $20.60\pm0.10$ &     $>2.27$    &  $0.76\pm0.06$ &  0.010 &   74 & \\
\hline
\enddata

\tablecomments{
  The coordinate origin is the ``A'' image of the lens.  
  The standard catalog of objects extends to
  a radius of 20\arcsec$\,$ from the lens, with objects
  labeled G\# for galaxies and S\# 
  for stellar objects respectively.
  Galaxies outside the 20\arcsec$\,$ 
  radius which could produce a tidal shears 
  exceeding 1\% are labeled by T\# in order.
  Objects labeled H\# were detected only in the infrared. 
  Tidal shear estimates are scaled to the lens galaxy 
  magnitude in the corresponding filter, after correcting 
  for the SExtractor magnitude offset discussed in~\S2. 
  For Q~1208+1011, where we did not detect the lens 
  galaxy, the tidal shear estimates were made assuming 
  lens magnitudes of $I=23$ and $H=21$. } 
\end{deluxetable}

\vspace*{-1.5in}
\hspace*{-1.5in}
\centerline{\psfig{figure=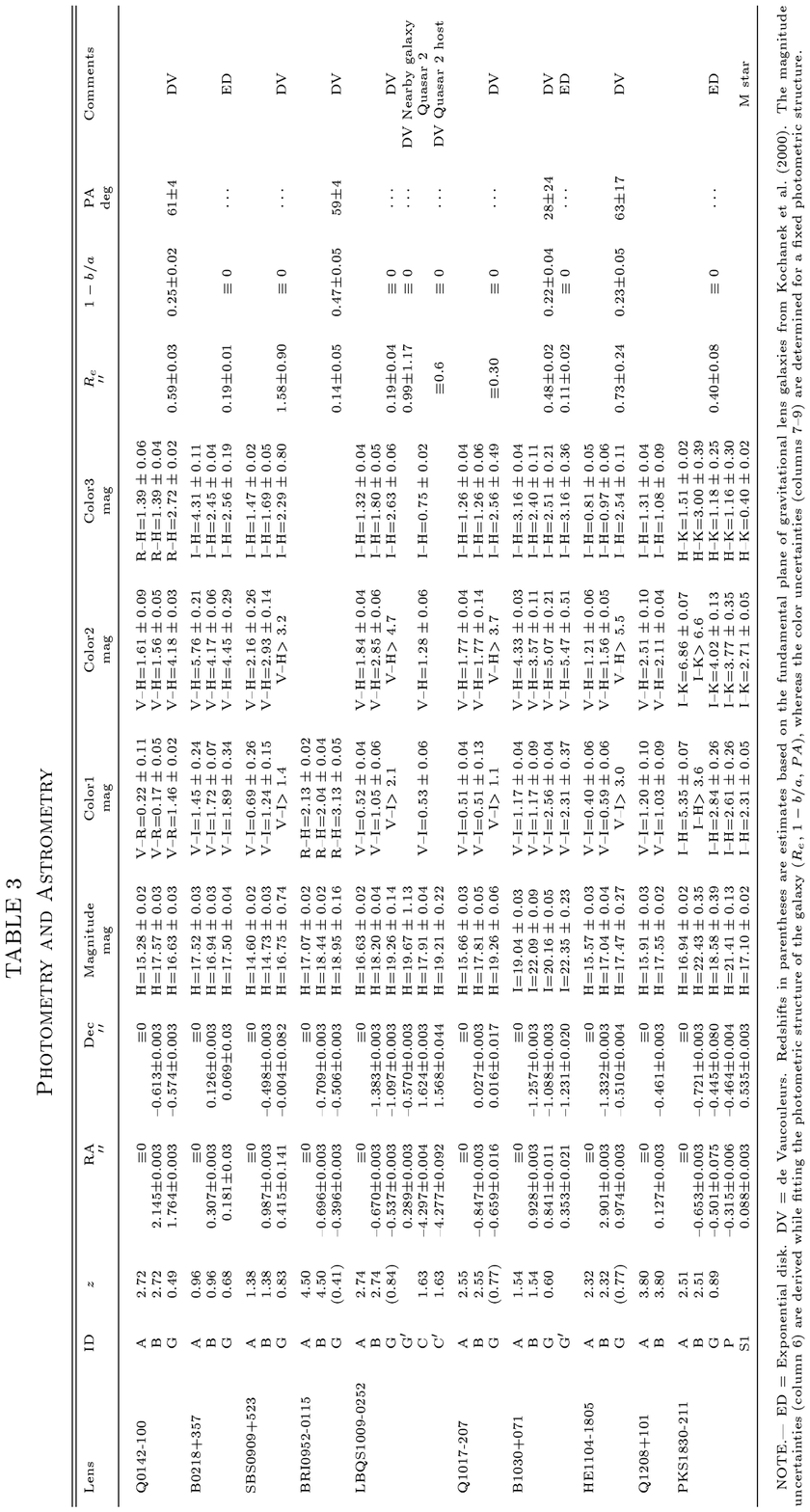,height=10.0in,angle=180}} 

\begin{deluxetable}{lccrccrccrcc}
\tablenum{4}
\label{tab-tidalshear}


\footnotesize
\tablecaption{Tidal Shear Estimates}
\tablewidth{0pt}
\tablehead{Lens &\multicolumn{3}{c}{Field Total} &\multicolumn{3}{c}{Largest} &\multicolumn{3}{c}{Remainder} &&Cosmic \nl
     &$\kappa_T$ &$\gamma_T$ &\multicolumn{1}{c}{PA ($^\circ$)}
     &$\kappa_T$ &$\gamma_T$ &\multicolumn{1}{c}{PA ($^\circ$)}
     &$\kappa_T$ &$\gamma_T$ &\multicolumn{1}{c}{PA ($^\circ$)} 
     &$\gamma_{M/L}$ & $\gamma_{LSS}$ }
\startdata
Q0142--100     &  0.138 &  0.072 &--24.6 &  0.079 &  0.079 & --6.8 &  0.060 &  0.047 &--64.6 &  0.011 & 0.04 \\
B0218+357      &  0.023 &  0.014 &  44.7 &  0.005 &  0.005 &  42.5 &  0.018 &  0.009 &  46.0 &  0.000 & 0.02 \\
SBS0909+523    &  0.075 &  0.057 &  62.6 &  0.037 &  0.037 &  51.3 &  0.038 &  0.027 &  78.8 &  0.003 & 0.03 \\
BRI0952--0115  &  0.081 &  0.027 &  72.3 &  0.026 &  0.026 &  87.2 &  0.055 &  0.014 &  37.3 &  0.002 & 0.05 \\
LBQS1009--0252 &  0.261 &  0.053 &--36.1 &  0.080 &  0.080 &--53.3 &  0.182 &  0.047 &  16.9 &  0.010 & 0.04 \\
Q1017--207     &  0.053 &  0.011 &  29.0 &  0.026 &  0.026 &  59.4 &  0.026 &  0.023 &--18.7 &  0.001 & 0.04 \\
B1030+071      &  0.131 &  0.093 &--39.3 &  0.044 &  0.044 &--48.9 &  0.087 &  0.053 &--31.4 &  0.006 & 0.03 \\
HE1104--1805   &  0.048 &  0.048 &--38.5 &  0.048 &  0.048 &--38.5 &  0.000 &  0.000 &   0.0 &  0.005 & 0.04 \\
Q1208+1011     &  0.169 &  0.072 &   5.1 &  0.055 &  0.055 &   3.5 &  0.114 &  0.017 &  10.0 &  0.008 & 0.05 \\
PKS1830--211   &  0.158 &  0.096 &--20.5 &  0.094 &  0.094 &--18.9 &  0.064 &  0.006 &--56.3 &  0.019 & 0.04 \\
\enddata

\tablecomments{The {\it field total} shear and convergence from
  all the objects within 20$\arcsec$ of the lens (see Table~3).  
  The total is broken down into the
  {\it largest} contribution from a single galaxy, 
  and the {\it remainder} from all other galaxies. We show the
  convergence $\kappa_T$, and 
  for the shear we give the magnitude $\gamma_T$ and its orientation $PA$.  
  These estimates assume SIS models for the galaxies,
  and that all galaxies lie at the lens redshift 
  and have the same mass-to-light ratio as the lens 
  (after applying the SExtractor magnitude shifts 
  of 0.1~mag for WFPC2 and 0.5~mag for NICMOS, see \S2).
  If the nearby galaxies have constant M/L, then the total shear, 
  $\gamma_{M/L}$, is much smaller.  
  The cosmological shear $\gamma_{LSS}$ estimates the variance
  in the shear produced by perturbations along the rays. 
  }
\end{deluxetable}

\vspace*{-2.8in}
\centerline{\psfig{figure=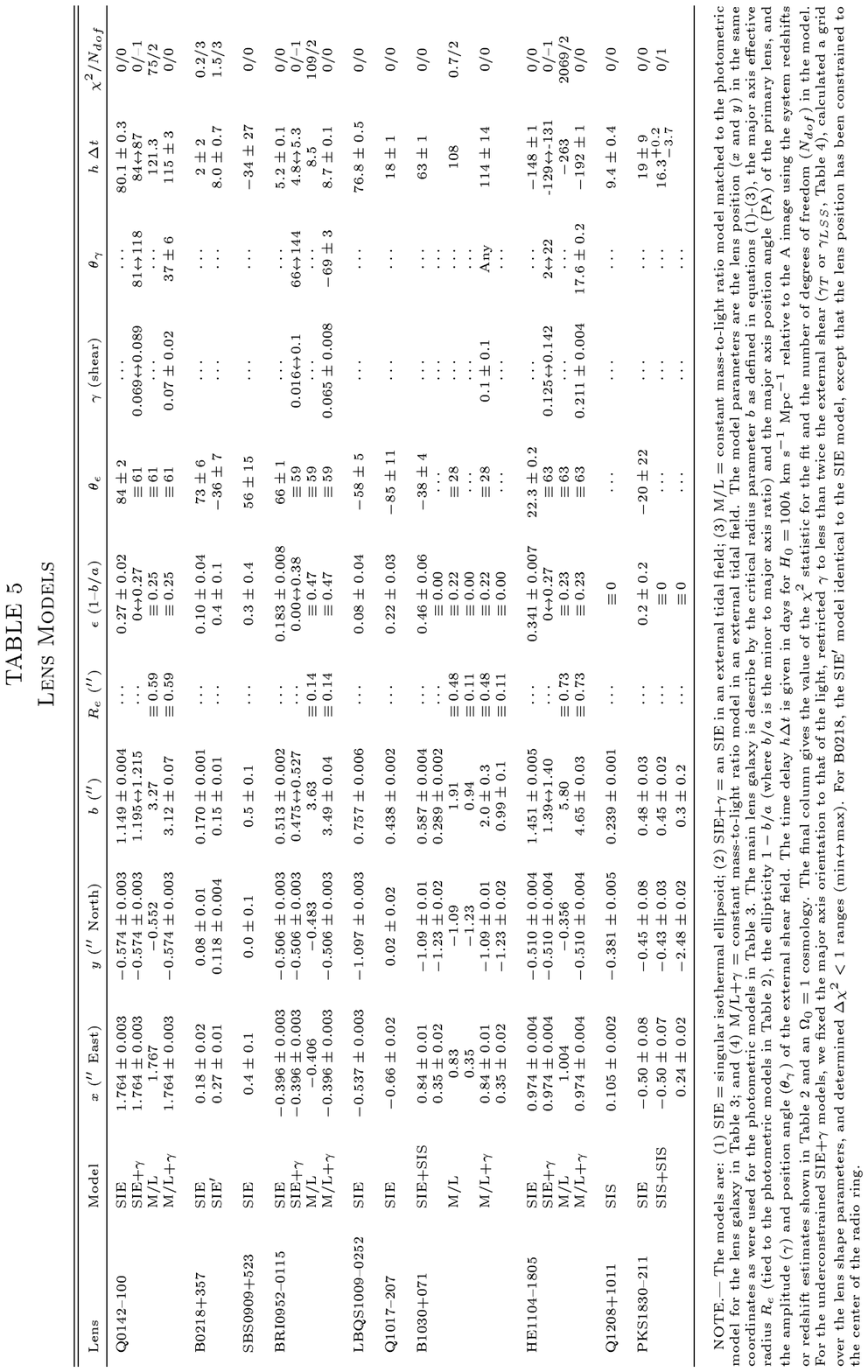,width=9.5in,angle=180}} 

\end{document}